\documentclass[aps,pra,reprint,showpacs,twocolumn,superscriptaddress]{revtex4-1}
\usepackage{graphicx}
\usepackage{amsmath}
\usepackage{amssymb}
\usepackage{amsfonts}
\usepackage{bbm}
\usepackage{bm}
\usepackage{braket}
\usepackage{verbatim}
\usepackage{cancel}
\usepackage{tikz}
\usepackage{wasysym}
\usepackage[bookmarks=false,colorlinks=true,urlcolor=blue,citecolor=blue,linkcolor=blue]{hyperref}
\usepackage[normalem]{ulem}
\usepackage[utf8]{inputenc}
\usepackage{float}
\usepackage{color}
\usepackage[all]{xy}
\usepackage[caption=false]{subfig}
\usepackage{bbold}

\graphicspath{{./Figures/}}
\allowdisplaybreaks

\definecolor{darkblue}{RGB}{0,0,127}

\definecolor{joe}{RGB}{127,0,127}

\definecolor{tom}{RGB}{0,127,0}

\newcommand{\lp}{\ensuremath \left(}
\newcommand{\rp}{\ensuremath \right)}
\newcommand{\lb}{\ensuremath \left[}
\newcommand{\rb}{\ensuremath \right]}

\newcommand{\XZ}{$x$-$z$~}
\newcommand{\YZ}{$y$-$z$~}
\newcommand{\XY}{$x$-$y$~}

\renewcommand{\L}{$\mathrm{L}$~}
\newcommand{\R}{$\mathrm{R}$~}
\newcommand{\B}{$\mathrm{B}$~}
\newcommand{\T}{$\mathrm{T}$~}

\newcommand{\TL}{$\mathrm{TL}$~}

\newcommand{\BR}{$\mathrm{BR}$~}
\newcommand{\BL}{$\mathrm{BL}$~}

\newcommand{\br}{\ensuremath \bm{r}}

\newcommand{\tphi}{\ensuremath \tilde{\phi}}
\newcommand{\ttheta}{\ensuremath \tilde{\theta}}

\newcommand{\sgn}{\text{sgn}}

\newcommand{\xhat}{\ensuremath \hat{\bm{x}}}
\newcommand{\yhat}{\ensuremath \hat{\bm{y}}}
\newcommand{\ahat}{\ensuremath \hat{\bm{a}}}

\begin{document}

\title{Planar p-String Condensation: Chiral Fracton Phases from Fractional Quantum Hall Layers and Beyond}

\author{Joseph Sullivan}
\affiliation{Department of Physics, Yale University, New Haven, Connecticut 06511, USA}

\author{Thomas Iadecola}
\affiliation{Department of Physics and Astronomy, Iowa State University, Ames, Iowa 50011, USA}

\author{Dominic J. Williamson}
\affiliation{Stanford Institute for Theoretical Physics, Stanford University, Stanford, California 94305, USA}

\date{\today}

\begin{abstract}
We present a coupled-wire construction of a model with chiral fracton topological order. 
The model combines the known construction of $\nu=1/m$ Laughlin fractional quantum Hall states with a planar p-string condensation mechanism. The bulk of the model supports gapped immobile fracton excitations that generate a hierarchy of mobile composite excitations. Open boundaries of the model are chiral and gapless, and can be used to demonstrate a fractional quantized Hall conductance where fracton composites act as charge carriers in the bulk. The planar p-string mechanism used to construct and analyze the model generalizes to a wide class of models including those based on layers supporting non-Abelian topological order. We describe this generalization and additionally provide concrete lattice-model realizations of the mechanism.
\end{abstract}

\maketitle

\tableofcontents

\section{Introduction}

Fracton phases are a new class of topological states of matter characterized by ``subdimensional" quasiparticles with emergent mobility restrictions (see Refs.~\onlinecite{nandkishore2018fractons,Pretko2020} for a recent review). 
Initially of interest for their glassy features~\cite{Chamon2005,CastelnovoPM2012} and utility as topological quantum memories~\cite{Haah,kim20123d,BravyiPRL2011,bravyi2013quantum,BravyiAOP2011,Brown2019,Devakul2020a} due to constrained quasiparticle dynamics~\cite{KimPRL2016,PremPRB2017}, the subject has grown to challenge the classification of topological phases of matter via topological quantum field theory~\cite{Shirley2017,Dua2019b,Wen2020,Aasen2020,JWang2020} and  demonstrated the possibility of heretofore unforeseen field  theories~\cite{PretkoPRB2017a,PretkoPRB2017b,PremPRB2018,ma2018fracton,PhysRevB.97.235112,PretkoPRL2018,Gromov2017,bulmash2018generalized,Gromov2018,SlaglePRB2017,Slagle2018a,PretkoGauge2018,Williamson2018a,Radzihovsky2019,Seiberg2019,Seiberg2020,Seiberg2020b,Fontana2020,Slagle2020}.

While many constructions of fracton phases have been proposed, a systematic understanding of all possible phases is yet to be rigorously established.  Most works so far have relied on constructing exactly solvable ``commuting-projector" models whose Hamiltonians are sums of commuting terms~\cite{Chamon2005,CastelnovoPM2012,Haah, kim20123d,YoshidaPRB2013,haah2014bifurcation, VijayPRB2015,VijayPRB2016,MaPRB2017,Vijay2017,VijayFu2017,HalaszPRL2017,HsiehPRB2017,Prem2018,Song2018,Prem2019,Bulmash2019,SSET,Dua2019a,Schmitz2019a,Williamson2020a}.  However, many topological phases, including \emph{chiral} phases with broken time-reversal symmetry, cannot be realized by such models~\cite{Kapustin19}. Such phases include many of the most famous (2+1)-dimensional topological orders, including fractional quantum Hall (FQH) phases~\cite{Tsui82,Laughlin83}. Different tools are thus required to build, study, and classify models of chiral fracton phases, the prospect of which has only recently been raised~\cite{GaugedLayers,Fuji19b}.

Chiral topological phases nevertheless admit analytically tractable microscopic models in the form of coupled-wire constructions. These constructions model topological phases as arrays of (1+1)-dimensional quantum wires with suitably chosen many-body interactions.  Coupled-wire constructions allow for the use of powerful techniques from (1+1)-dimensional systems, including bosonization and conformal field theory (CFT), to describe strongly interacting phases of matter in higher dimensions. They have been used to build and analyze numerous models of topological phases in (2+1)~\cite{Yakovenko91,Sondhi01,Kane02,Teo14,Neupert14,Gorohovsky15,Meng15,Huang16,Iadecola19,Imamura19}, (3+1)~\cite{Iadecola16,Fuji19a}, and higher dimensions~\cite{Iadecola16}, including both Abelian and non-Abelian examples. 

In this work, we show that the coupled-wire formalism can be applied to realize new chiral fracton phases in (3+1) dimensions. We focus primarily on a model inspired by the wire construction of the $\nu=1/m$ Laughlin FQH states~\cite{Kane02,Teo14} that realizes subdimensional excitations with anyonic statistics inherited from those of Laughlin quasiparticles. The models we consider have a useful interpretation in terms of anyon condensation, wherein stacks of $\nu=1/m$ Laughlin phases on \XZ and \YZ planes are coupled by condensing ``p-strings" composed of Laughlin quasiparticles at the lines of intersection of each pair of planes. This planar p-string condensation mechanism allows for the rapid determination of broad classes of new fracton phases, including examples based on non-Abelian topological orders.

The manuscript is laid out as follows. In Section~\ref{sec:Coupled-wire model} we introduce our coupled-wire model and study the topological properties of its bulk and boundary. 
In Section~\ref{sec:Planar p-String Condensation Mechanism} we provide a high level description of planar p-string condensation. 
In Appendix~\ref{sec: Coupled-wire Appendix} we provide detailed calculations on the coupled-wire model that complement the discussion in the main text. 
In Appendix~\ref{sec: Appendix G} we generalize our coupled-wire construction to a non-Abelian model. 
In Appendix~\ref{sec: Appendix H} we present further details about planar p-string condensation, including examples and relations to existing mechanisms to generate fracton topological order. 

\begin{figure}[t!]
\begin{center}
\includegraphics[width=.65\columnwidth]{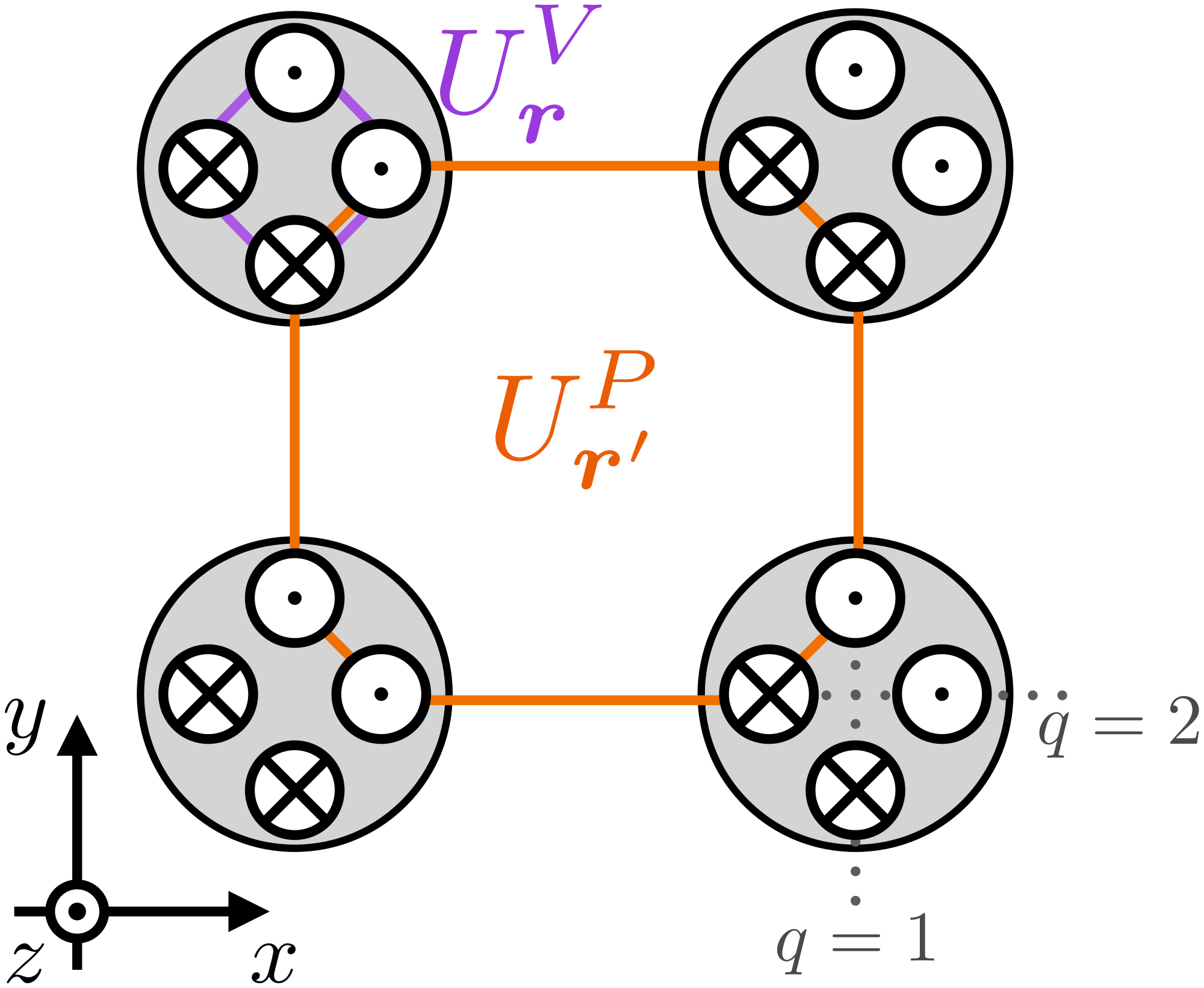}\\
\caption{
Schematic depiction of the coupled-wire array with vertex and plaquette terms $U^V_{\br}$ (purple) and $U^P_{\br}$ (orange). Gray circles represent the wires, and crossed and dotted circles represent right and left movers, respectively. The labels $q=1,2$ arise from viewing each wire as the intersection of a vertical and horizontal plane, respectively.
}
\label{fig:FQHModel}
\end{center}
\end{figure}

\section{Chiral Fracton Phase in a Laughlin Coupled-Wire Model}
\label{sec:Coupled-wire model}
In this section we introduce and analyze a coupled wire model that realizes a chiral fracton phase. In Sec.~\ref{sec:Model Definition} we define the model; starting from a decoupled array of two-component Luttinger liquids and then introducing two types of sine-Gordon terms to produce a strongly coupled phase. In Sec.~\ref{sec:Planar p-String Condensation Interpretation} we explore an interpretation of the model through the lens of the planar p-string condensation mechanism. The bulk excitations are discussed in the Sec.~\ref{sec:Bulk Excitations of the Coupled-Wire Model}. In Sec.~\ref{sec:Surface Theory} we discuss the surface theory of the model. Finally, the topological degeneracy is addressed in Sec.~\ref{sec:Topological Degeneracy}.

\subsection{Model Definition}
\label{sec:Model Definition}

We consider a set of (1+1)-dimensional quantum wires oriented along the $z$-axis and placed on the vertices $\br=(x,y)$ of an $L_x\times L_y$ square lattice $\Lambda$ in the \XY plane (see Fig.~\ref{fig:FQHModel}).  Each vertex contains two quantum wires labeled by $q=1,2$, and each wire contains two chiral channels labeled by $\eta=L,R$.  The wires consist of free fermions, which we write in bosonized form as $\Psi^{q}_{\eta,\br}\sim e^{i\phi^{q}_{\eta,\br}}$, where the chiral bosonic fields $\phi^{q}_{\eta,\br}(z)$ obey the equal-time canonical commutation relations
\begin{align}
    [\phi^{q}_{L/R,\br}(z),\phi^{q'}_{L/R,\br'}(z')]\!=\!\pm\delta_{q,q'}\delta_{\br,\br'}\,i\pi\, \sgn(z\!-\!z'),
\end{align}
where we associate the signs $+$ and $-$ with $\eta=L$ and $R$, respectively.
To specify the couplings between wires, we define new fields and commutation relations
\begin{align}
\label{eq:PhiTildeDef}
    \tphi^{q}_{L/R,\br}=[(\phi^{q}_{L,\br}+\phi^{q}_{R,\br})&\pm m(\phi^{q}_{L,\br}-\phi^{q}_{R,\br})]/2 \\
    [\tphi^{q}_{L/R,\br}(z),\tphi^{q'}_{L/R,\br'}(z')]\!=\!&\pm\delta_{q,q'}\delta_{\br,\br'}\, i\pi m\,  \sgn(z\!-\!z'),\nonumber
\end{align}
with $m$ an odd integer,
which are appropriate for describing Laughlin quasiparticles at filling $\nu=1/m$.  In these new variables, the local vertex operator $e^{i\tphi^{q}_{\eta,\br}}$ defines a chiral quasiparticle with fermionic statistics, while the nonlocal vertex operator $e^{i\tphi^{q}_{\eta,\br}/m}$ defines a chiral Laughlin quasiparticle with anyonic statistical angle $2\pi/m$.  

We now couple the wires with interactions ${H_{\rm int}=-\sum_{\br}\int^{L_z}_0\mathrm dz\, \lp\lambda_V\, U^V_{\br}+\lambda_P\, U^P_{\br}\rp}$, where
\begin{align}
\label{eq:AbelianHint}
    U^V_{\br}\!&=\cos\!\lb\frac{1}{m}\!\lp\tphi^{1}_{L,\br} \!-\! \tphi^{1}_{R,\br} \!+\! \tphi^{2}_{L,\br} \!-\! \tphi^{2}_{R,\br}\rp\rb \!\equiv\! \cos(\theta^V_{\br})\\
    U^P_{\br}\!&=\cos\!\lb2\!\lp\ttheta^{1}_{\br,\yhat} \!+\! \ttheta^{2}_{\br+\yhat,\xhat} \!-\! \ttheta^{1}_{\br+\xhat,\yhat} \!-\! \ttheta^{2}_{\br,\xhat}\rp\rb \!\equiv\! \cos(\theta^P_{\bm r}),\nonumber
\end{align}
and $2\ttheta^q_{\br,\ahat}=\tphi^q_{L,\br}-\tphi^q_{R,\br+\ahat}$, with $\ahat=\xhat,\yhat$ the unit vectors in the $x$ and $y$ directions. Importantly, these interactions are local when written in terms of the ``fundamental" fermions $\Psi^{q}_{\eta,\br}$; this can be checked explicitly using Eq.~\eqref{eq:PhiTildeDef} (see also Ref.~\cite{Teo14}).  Furthermore, it is straightforward to check using Eq.~\eqref{eq:PhiTildeDef} that the interaction terms $U^V_{\bm r}$ and $U^P_{\bm r}$ commute among themselves and can therefore be simultaneously diagonalized.  In the strong-coupling limit $\lambda_P,\lambda_V\to\infty$, the ground state manifold is obtained by pinning the arguments of $U^V_{\bm r}$ and $U^P_{\bm r}$ to integer multiples of $2\pi$.

We remark that, strictly speaking, the vertex terms $U^V_{\br}$ appearing in Eq.~\eqref{eq:AbelianHint} are not translation invariant in the $z$-direction when written in terms of the original fermions due to the presence of oscillatory factors $\sim e^{i\, 4k_F\, z}$, where $k_F$ is the Fermi wavenumber. These factors can be removed by making a global change of variables $\tilde{\phi}^2_{\eta,\br}\to-\tilde{\phi}^2_{\eta,\br}$, which amounts to choosing different bosonization conventions depending on the index $q=1,2$. This effectively redefines $k_F\to -k_F$ for the $q=2$ layers, leading to the pairwise cancellation of the $2k_F$ factors giving rise to the oscillations, while maintaining commutativity of the vertex and plaquette terms and preserving the canonical commutation relations. This transformation does not affect any of the properties of the model considered here, so we continue to use the original convention of Eq.~\eqref{eq:AbelianHint}.


\subsection{Planar p-String Condensation Interpretation}
\label{sec:Planar p-String Condensation Interpretation}

The interactions~\eqref{eq:AbelianHint} have an appealing interpretation in terms of coupled layers.
Consider a system of initially decoupled $\nu=1/m$ Laughlin FQH systems stacked along \YZ and \XZ planes of the cubic lattice. We assign the labels $q=1,2$ to \YZ and \XZ planes, respectively. Now define a square lattice in the \XY plane whose vertices $\br$ are the locations of the lines of intersection of pairs of \XZ and \YZ planes.  We can now couple pairs of Laughlin planes where they intersect by condensing a bosonic bound state of local excitations.  The simplest nontrivial object we can condense is a bound state of two quasiparticle-quasihole pairs, one pair for each $q=1,2$.  A microscopic model for this setup is obtained by representing each Laughlin plane using a coupled-wire construction.  Within this construction, the local operator that creates a Laughlin quasiparticle-quasihole pair within a layer is given by~\cite{Teo14}
\begin{align}
\label{eq:LaughlinPair}
    Q^q_{\br}=e^{i\frac{1}{m}(\tphi^{q}_{L,\br}-\tphi^{q}_{R,\br})}.
\end{align}
Adding $-\lambda_V\sum_{\br}\int^{L_z}_0\mathrm dz\, U^V_{\br}$ to the Hamiltonian and taking $\lambda_V\to\infty$ thus condenses the bound state of two such quasiparticle-quasihole pairs for all $z$ along the intersection line $\bm r$. This four-body composite can be viewed as a small loop composed of Laughlin quasiparticles (i.e., a p-string, see below), which fluctuates in the presence of condensation terms $U^V_{\br'}$ located at lattice sites $\br'\neq\br$. The plaquette terms $U^P_{\bm r}$ emerge by performing degenerate perturbation theory in the coupling $g$ that couples the wires within a given \XZ or \YZ plane to form the Laughlin FQH layer building blocks [see Appendix~\ref{sec:Perturbation theory}]. 

The coupled-wire model with interactions \eqref{eq:AbelianHint} is thus one simple instance of a large class of models obtained by coupling two interpenetrating stacks of (2+1)-dimensional topological phases.  The condensation process implemented by the vertex terms $U^V_{\br}$ is an example of \emph{p-string} condensation~\cite{MaPRB2017,Vijay2017,Prem2018}, because it proliferates closed loops composed of Laughlin quasiparticles. This distinguishes p-string condensation from standard anyon condensation~\cite{bais2009}, which proliferates pointlike anyon composites. Anyon condensation has also been used to build 3D topological orders from 2D building blocks~\cite{Jian14,Fuji19a,Fuji19b}, but such constructions do not yield totally immobile fracton excitations. In contrast, p-string constructions generically lead to fractons (for a summary, see Sec.~\ref{sec:Planar p-String Condensation Mechanism}).

The present class of models is distinguished from other p-string condensation constructions~\cite{MaPRB2017,Vijay2017,Prem2018,GaugedLayers} by the fact that p-strings are only allowed to fluctuate within \XY planes.  This new restriction enables the condensation of p-strings composed of anyons with nontrivial mutual statistics---the Laughlin p-strings defined above being a simple example.  In prior p-string constructions, such condensates cannot be consistently defined due to the nontrivial braiding of p-strings from intersecting planes. The construction introduced here thus enables a host of new fracton phases not obtainable by other means that are explored further in Sec.~\ref{sec:Planar p-String Condensation Mechanism} and Appendix~\ref{sec: Appendix H}.

\begin{figure}[t!]
\begin{center}
\includegraphics[width=\columnwidth]{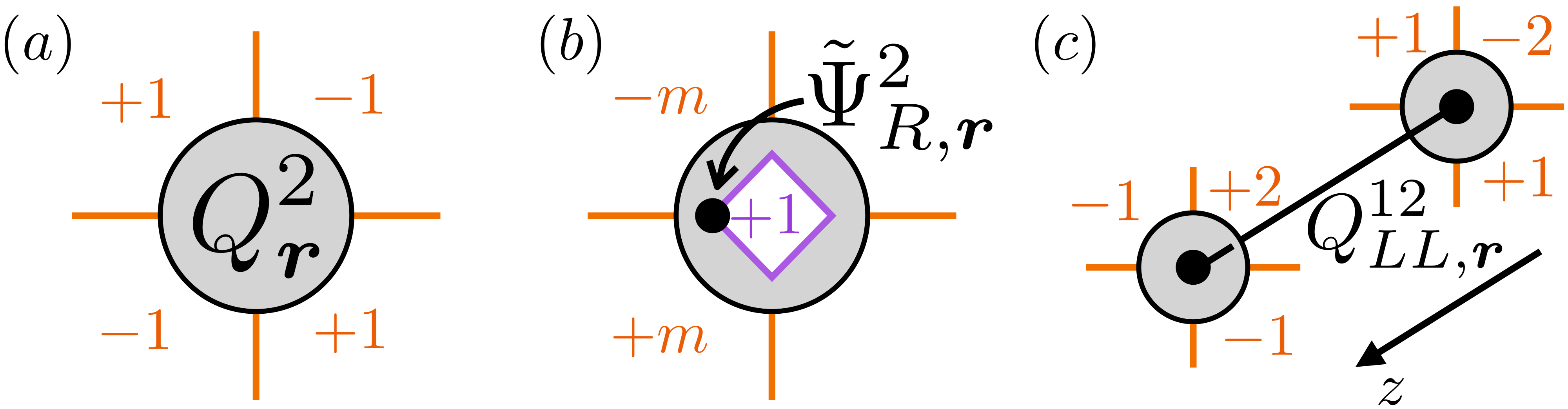}\\
\caption{
Action of (a) the Laughlin quasiparticle operator~\eqref{eq:LaughlinPair}, (b) the chiral electron operator~\eqref{eq:ChiralEl}, and (c) the composite chiral quasiparticle operator $Q^{12}_{LL,\br}$~\eqref{eq:ChiralLaughlin}. Integer charges of the vertex and plaquette solitons are indicated in purple and orange, respectively.
}
\label{fig:LocalOps}
\end{center}
\end{figure}

\subsection{Bulk Excitations of the Coupled-Wire Model}
\label{sec:Bulk Excitations of the Coupled-Wire Model}

The coupled-wire array supports two kinds of excitations: charged solitons, i.e., abrupt jumps of the pinned fields $\theta^V_{\br},\theta^P_{\br}$ between integer multiples of $2\pi$, and neutral Gaussian fluctuations of these fields around their minima. The solitons constitute gapped topological excitations of the theory. The Gaussian fluctuations become gapless in the thermodynamic limit $L_x,L_y\to\infty$~\cite{JoeCWFracton} [see Appendix~\ref{sec:Scaling of the gap}]. However, they are topologically trivial and do not contribute to the charge response at the level of Eq.~\eqref{eq:AbelianHint}. Furthermore, numerical results indicate a scaling limit $L_x=L_y\to\infty$, $U\gtrsim L_x^{2.8}$ in which they are gapped [see Appendix~\ref{sec:Scaling of the gap}]. We defer further analysis of the Gaussian fluctuations and their stability to future work.

Pointlike charged excitations of the coupled-wire array are identified with solitons:
\begin{align}
    \partial_z\theta^{V,P}_{\bm r}\to\partial_z\theta^{V,P}_{\bm r}+2\pi n\, \delta(z-z_0),
\end{align} 
for $n\in\mathbb Z$ and some $0\leq z_0<L_z$. A basis for these excitations is obtained by considering the action of all local vertex operators in a given wire. The vertex operators at our disposal are the Laughlin quasiparticle-quasihole pair operator $Q^q_{\br}(z)$ [Eq.~\eqref{eq:LaughlinPair}], the chiral ``electron" operator
\begin{align}
\label{eq:ChiralEl}
    \tilde{\Psi}^q_{\eta,\br}=e^{i\tphi^q_{\eta,\br}},
\end{align}
and the chiral operator
\begin{align}
\label{eq:ChiralLaughlin}
    Q^{q}_{\eta,\br}(z_1,z_2)=\exp\lp \frac{i}{m}\int^{z_2}_{z_1}\mathrm dz\ \partial_z\tphi^q_{\eta,\br}\rp,
\end{align}
which moves a Laughlin quasiparticle from $z=z_1$ to $z_2$.
(Note that $Q^{2}_{\br} = e^{i\theta^V_{\br}}Q^{1\dagger}_{\br}$, so that $Q^{1\dagger}_{\br}\simeq Q^2_{\br}$ when acting on the ground state, where $\theta^V_{\br}$ is pinned.)
Of these three operator types, the first two create genuine integer solitons in $U^{V,P}_{\bm r}$, see Fig.~\ref{fig:LocalOps}(a) and (b). $Q^{q}_{\eta,\br}(z_1,z_2)$ creates pairs of integer solitons in $U^{P}_{\bm r}$, but a pair of \textit{fractional solitons} in $U^{V}_{\bm r}$. This constitutes a \textit{linelike} excitation, because it shifts $U^{V}_{\bm r}$ by a noninteger multiple of $2\pi$ in the region between $z_{1,2}$. However, one can build composite operators 
\begin{align}
\label{eq:CompositeChiral}
Q^{qq'}_{\eta\eta',\bm r}(z_1,z_2)=Q^{q}_{\eta,\br}(z_1,z_2)[Q^{q'}_{\eta',\br}(z_1,z_2)]^\dagger
\end{align}
for which such linelike excitations cancel; for example, $Q^{12}_{LL,\br}(z_1,z_2)$ creates a pair of three integer plaquette solitons separated by $z_2-z_1$ along a wire [see Fig.~\ref{fig:LocalOps}(c)].

\begin{figure}[t!]
\begin{center}
\includegraphics[width=\columnwidth]{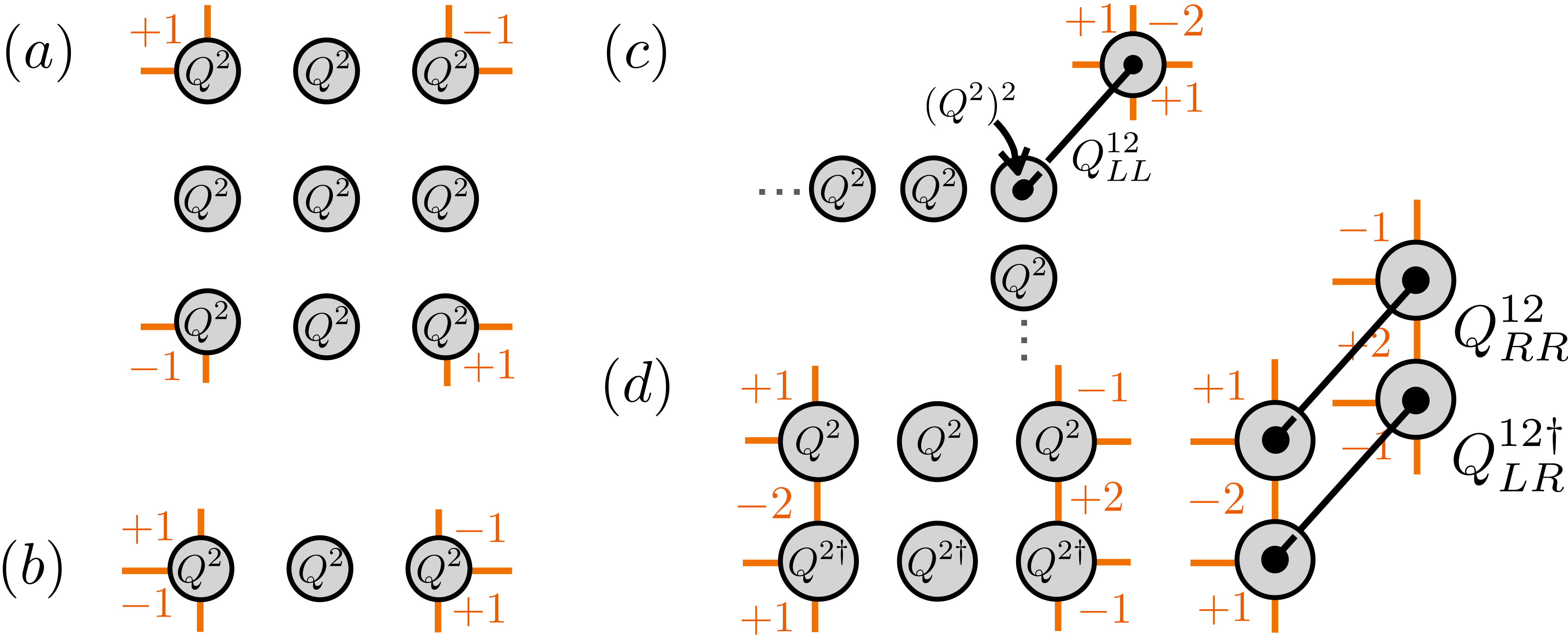}\\
\caption{
Subdimensional excitations of the coupled-wire array include (a) immobile fractons, (b) lineons mobile only along fixed lattice directions, and (d) planons mobile only within 2D planes. Panel (c) depicts the fusion of $x$ and $y$ lineons into a $z$ lineon, while (d) depicts a pair of operators that can be multiplied to allow a planon to ``turn a corner" between the $x$- and $z$-directions.
}
\label{fig:Excitations}
\end{center}
\end{figure}

A hierarchy of quasiparticle mobility restrictions follows from the observation that a single plaquette excitation cannot be moved by a local operator.
This mobility restriction in the \XY plane is visible in Fig.~\ref{fig:LocalOps}, which demonstrates that plaquette solitons of strength $\pm 1$ are created in groups of at least four. Immobility of the plaquette solitons in the $z$-direction can be deduced from a ``Gauss's-law" constraint; for any compact region $M$ in the \XY plane, we have
\begin{align}
\label{eq:Gauss}
    \sum_{\br\in M}\theta^P_{\br}(z) = \sum_{(\br,\hat{\bm a})\in\partial M}2\, \ttheta^q_{\br,\hat{\bm a}}(z),
\end{align}
where the sum on the right-hand side runs over bonds contained in the boundary $\partial M$ of $M$. [Note that $q$ and $\hat{\bm a}$ are correlated in this sum: $\hat{\bm a}=\xhat$ ($\yhat$) implies $q=2$ ($1$).] Suppose there exists a local operator $\mathcal O$ that moves a single $\pm 1$ plaquette soliton from $z_1$ to $z_2$. Such an operator shifts the left-hand side of Eq.~\eqref{eq:Gauss} by $\pm 2\pi$. However, because $\mathcal O$ is local, we can always choose $M$ larger than $\mathcal O$'s support; hence, $\mathcal O$ commutes with the right-hand side of \eqref{eq:Gauss}, a contradiction. Thus all local operators must create plaquette solitons in pairs of charge $\pm 1$ at fixed $z$, establishing charge-neutrality as a necessary (but not sufficient) condition for mobility in $z$.

The immobility of a single plaquette soliton implies that these excitations are fractons. Since they are created by the same operators $Q^q_{\br}$ that create Laughlin quasiparticle-quasihole pairs in the $\nu=1/m$ FQH state, we conclude that the condensation transition described in the coupled-layer picture transmutes quasiparticles with planar mobility into immobile subdimensional excitations. A group of four fractons created by applying, e.g., $Q^2_{\br}$ to the ground state can be separated from one another by sequential application of $Q^2$ operators on contiguous vertices, creating a rectangular membrane operator with fractons at its corners [see Fig.~\ref{fig:Excitations}(a)]. Alternatively, $Q^2_{\br}$ can be used to propagate \textit{pairs} of fractons in the $x$- or $y$-direction, indicating that such pairs become ``lineons," i.e., quasiparticles mobile only along one-dimensional submanifolds of the wire array [see Fig.~\ref{fig:Excitations}(b)]. As shown in Fig.~\ref{fig:Excitations}(c), lineons mobile in the $x$- and $y$-directions can ``fuse" to become a lineon mobile in the $z$-direction. Fig.~\ref{fig:Excitations}(d) shows how pairs of lineons can be combined to form ``planons" with mobility in, e.g., \XZ planes. This hierarchy of quasiparticle mobility is a familiar feature of many fracton models and also follows directly from the planar p-string condensation interpretation of the coupled-wire model [see Section~\ref{sec:Planar p-String Condensation Mechanism}].

Finally, we note that the braiding statistics of these subdimensional quasiparticles reveals their fractionalized nature. The notion of mutual and self-statistics of fractons, lineons, and planons has been defined~\cite{Bulmash2018b,Pai19} and follows from the phase acquired upon exchanging the membrane and string operators used to propagate the corresponding excitations. In the present case, this exchange phase follows from the commutation relations \eqref{eq:PhiTildeDef} and reflects the relationship between fractons and Laughlin quasiparticles.  For example, the statistical angle obtained from braiding two lineons [e.g., those depicted in Figs.~\ref{fig:LocalOps}(c) and \ref{fig:Excitations}(b)], or two planons in vertically offset \XZ planes [Fig.~\ref{fig:Excitations}(d)], is $\pm 2\pi/m$.

\subsection{Surface Theory}
\label{sec:Surface Theory}

We now show that the coupled-wire model with interactions \eqref{eq:AbelianHint} possesses chiral surface states that are gapless at any system size, evoking a (3+1)-dimensional generalization of FQH physics. To see this, we consider a square lattice with $L_x L_y$ vertices, each containing four chiral modes, and place periodic boundary conditions (PBC) in the $z$-direction and open boundary conditions (OBC) in the $x$- and $y$-directions. This leaves a 2D boundary with the topology of a 2-torus. Next, we apply a counting argument due to Haldane~\cite{Haldane95} to determine how many of the $4 L_x L_y$ chiral modes are gapped by the interactions \eqref{eq:AbelianHint}.  Recalling that a single cosine term gaps out two modes with opposite chirality in the strong-coupling limit, and noting that there are $L_x L_y$ vertex terms $U^V_{\br}$ and $(L_x-1)(L_y-1)$ plaquette terms $U^P_{\br}$, we conclude that the interactions \eqref{eq:AbelianHint} are sufficient to gap out all but $N=2(L_x+L_y-1)$ modes in the array. For the remainder of this paper we assume a fixed finite $L_x,L_y$.

In Appendix~\ref{sec:Teo Kane Laughlin}, we show that these $N$ gapless modes are strictly localized on the 2D surface of the array by explicitly identifying a set of surface modes that commute with the bulk couplings. Here, we summarize several notable features of these modes. First, they are chiral, with modes localized on opposite faces of the wire array having opposite chirality. Second, they are spatially overlapping and have nontrivial commutation relations that follow directly from Eq.~\eqref{eq:PhiTildeDef} and can be encoded in an $N\times N$ integer matrix $K$. Third, the $K$-matrix for modes living on the same face of the array closely resembles that of a stack of fractional quantum Hall states coupled by long-range Coulomb interactions~\cite{Qiu90,Naud00,Naud01,Ma20}. 

\begin{figure}[t!]
\begin{center}
\includegraphics[width=.6\columnwidth]{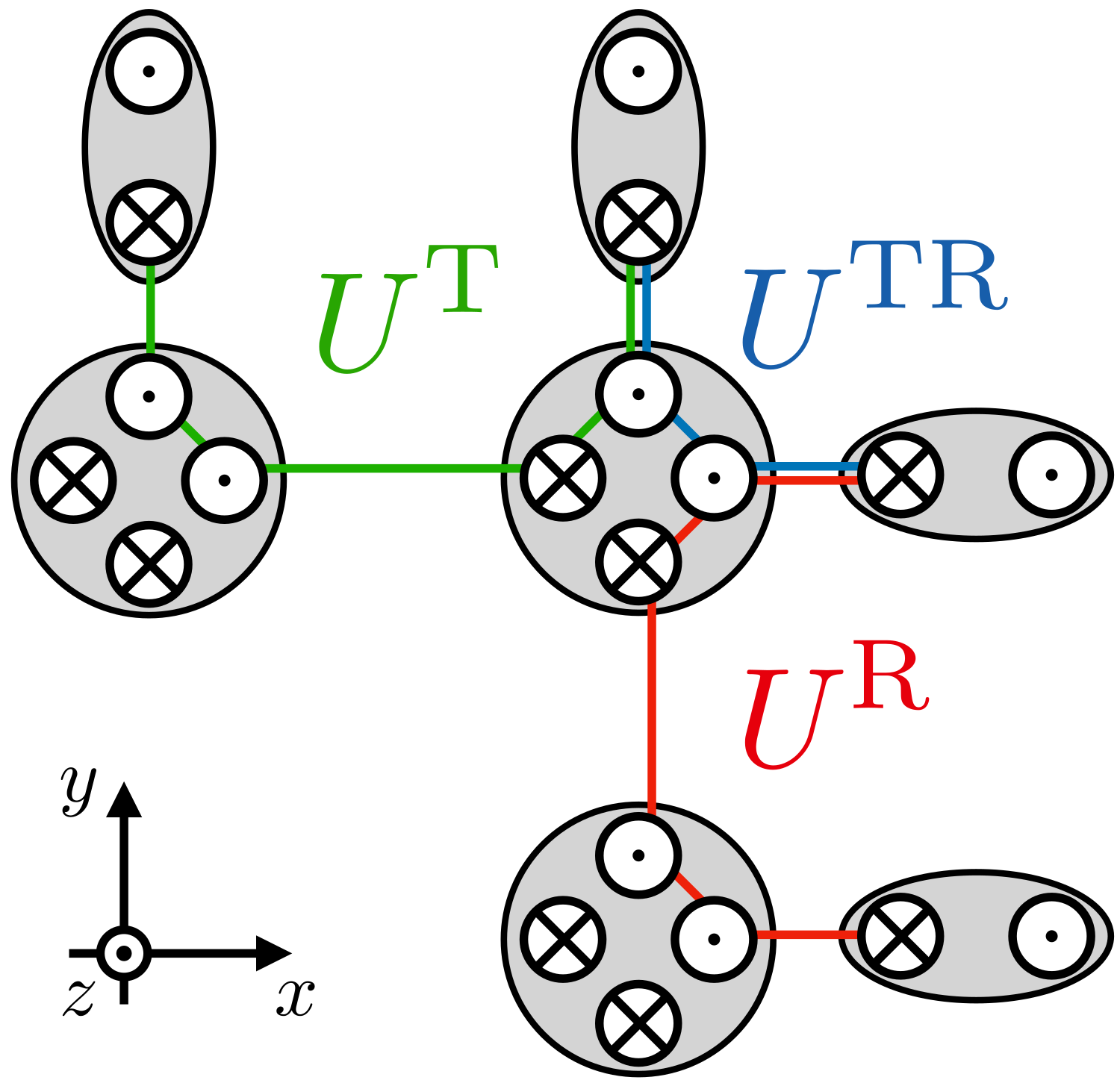}\\
\caption{
Schematic of the surface termination obtained by adding auxiliary boundary wires (gray ovals) and new boundary couplings (green, blue, and red). Dangling chiral gapless modes on the top and right surfaces are clearly visible.
}
\label{fig:Boundary}
\end{center}
\end{figure}

 The spatially overlapping and noncommuting chiral gapless surface modes discussed above can be disentangled by coupling additional boundary wires into the array as shown in Fig.~\ref{fig:Boundary}. We add $2(L_x+L_y)$ wires, each carrying one right- and one left-mover governed by the commutation relation \eqref{eq:PhiTildeDef}. The number of gapless modes in the array then increases to $N+4(L_x+L_y)=3N+4$. We now add additional strong commuting interaction terms to the Hamiltonian until $N+2$ gapless chiral modes remain. First we introduce $2(L_x-1)+2(L_y-1)$ truncated plaquette terms along the left, right, top, and bottom ($\mathrm{L,R,T,B}$) faces of the array. For example, on the $\mathrm{T}$ face we add the couplings [compare to Eq.~\eqref{eq:AbelianHint}]
 \begin{align}
     U^\mathrm{T}_{(x,L_y)}
     \!&=\cos\!\lb2\!\lp\ttheta^{1}_{(x,L_y),\yhat} \!-\! \ttheta^{1}_{(x+1,L_y),\yhat} \!-\! \ttheta^{2}_{(x,L_y),\xhat}\rp\rb 
 \end{align}
 for $x=1,\dots,L_x-1$, and likewise for the remaining three faces. Here, $2\ttheta^1_{(x,L_y),\yhat}=\tphi^1_{L,(x,L_y)}-\tphi^1_{R,(x,L_y+1)}$, where $\tphi^1_{R,(x,L_y+1)}$ is one of the additional boundary fields. This removes $2N-4$ gapless modes, leaving $6$ outstanding gapless modes. To dispose of these modes, it suffices to add truncated plaquette terms to three ``corners" of the array; for example, on the top-left ($\mathrm{TL}$) corner we add the coupling [compare to Eq.~\eqref{eq:AbelianHint}]
 \begin{align}
     U^\mathrm{TL}_{(1,L_y)}
     \!&=\cos\!\lb2\!\lp \ttheta^{1}_{(1,L_y),\yhat} + \ttheta^{2}_{(0,L_y),\xhat}\rp\rb,
\end{align}
and likewise for the $\mathrm{TR}$ and $\mathrm{BR}$ corners.
Here $\ttheta^{2}_{(0,L_y),\xhat}=\tphi^2_{L,(0,L_y)}-\tphi^2_{R,(1,L_y+1)}$, where $\tphi^2_{L,(0,L_y)}$ is one of the additional boundary fields. Note that we could also add a truncated plaquette term to the $\mathrm{BL}$ corner, but the argument of this plaquette term is linearly dependent with the other plaquette terms in the array and hence is not necessary for the construction.

The modified surface theory constructed above is that of $N+2=2(L_x+L_y)$  \textit{commuting} chiral gapless modes---precisely what one would obtain for a stack of $L_x+L_y$ decoupled $\nu=1/m$ Laughlin states. 
In fact, the surface theory defined above arises perturbatively within the coupled-layer interpretation of the model when the underlying Laughlin layers are arranged such that their chiral edges do not undergo p-string condensation [see Appendix~\ref{sec:Perturbation theory}].

One advantage of this alternative surface termination is that it makes the system's nontrivial Hall response transparent. Since the low-energy theory consists of $N+2$ decoupled chiral gapless modes that are identical to $\nu=1/m$ Laughlin edge states, inserting a vector potential in the $z$-direction corresponding to one flux quantum pumps a fractional charge $e L_x/m$ ($e L_y/m$) between the $\mathrm{T/B}$ ($\mathrm{L/R}$) faces of the array~\cite{Bardyn19}, corresponding to quantized fractional Hall conductivity $\sigma_{yz}$ ($\sigma_{xz}$)~\cite{Laughlin81,Halperin82}. This response is ultimately mediated by the bulk fractons, which descend from Laughlin quasiparticles and whose bound states are the only charged bulk excitations.

\subsection{Topological Degeneracy}
\label{sec:Topological Degeneracy}

In order to calculate the topological degeneracy of the ground-state manifold, we add additional couplings to remove the remaining chiral gapless surface modes. Starting from the surface termination shown in Fig.~\ref{fig:Boundary}, we add $N/2+1=L_x+L_y$ strong interaction terms of the form $\cos(2\ttheta^q_{\br,\hat{\bm a}})$ that couple gapless chiral modes on opposing faces of the array. The resulting model can be viewed as a three-torus containing two intersecting surface defects, each with the topology of a two-torus, on which the final interaction terms reside. This unusual boundary condition simplifies the analysis relative to the case of standard PBCs without auxiliary boundary wires [see Appendix~\ref{sec:Periodic boundary conditions}].

We calculate (in Appendix~\ref{sec:Topological groud-state degeneracy}) the topological degeneracy using
the method of Ref.~\cite{Imamura19}, starting from the set of strong-coupling ground states labeled by the values of the pinned bulk fields $\theta^{V,P}_{\br}\in 2\pi\mathbb{Z}$ and their boundary counterparts. Naively, this implies an infinite-dimensional ground state manifold; however, many of these ground states are equivalent since the bosonic fields $\phi^q_{\eta,\br}$ are only defined modulo $2\pi$. Accounting for this redundancy, we find that the ground-state manifold has dimension $m^{L_x+L_y}$. This subextensive ground-state degeneracy is a hallmark of ``type-I" fracton phases. We remark that the model also exhibits a subextensive number of superselection sectors with standard PBC, as shown in Appendix~\ref{sec:Periodic boundary conditions}. In Appendix~\ref{sec:Z_N lattice} we define an exactly solvable lattice model containing a chiral sector whose bulk topological excitations and ground-state degeneracy exactly match those of the coupled-wire model.

\section{Planar p-String Condensation Mechanism}
\label{sec:Planar p-String Condensation Mechanism}

In this Section we expound upon the planar p-string condensation mechanism for constructing fracton phases of matter. This was introduced in the context of the coupled-wire model in the previous section, but here we study the mechanism more abstractly from the point of view of coupled layers of topological orders that support nontrivial anyonic excitations. In Sec.~\ref{sec:Anyon theory description}, we discuss how to perform the planar p-string condensation procedure at the level of the anyon theory of the underlying 2D layers. 
In Sec.~\ref{sec:Z_N abstract}, we apply this condensation procedure to the example of chiral $\mathbb{Z}_N$ anyon layers, providing a high-level description of the topological fracton sectors in the coupled-wire model from the previous section.
A lattice model that is foliated equivalent~\cite{Shirley2017} to this chiral fracton theory is described in Appendix~\ref{sec: Appendix H} alongside further examples, as well as the connection between planar p-string condensation, gauging planar subsystem symmetries~\cite{GaugedLayers} and topological defect networks~\cite{Aasen2020}.

To date, several constructions of fracton models from coupled layers have appeared in the literature. 
They can all be understood as some form of p-string condensation on a stack of 2D layers with topological order. 
First we review Refs.~\cite{MaPRB2017,Vijay2017,Prem2018}, where the authors consider stacking topological layers along $xy$, $yz$, and $xz$ planes of the cubic lattice. 
These layers must support a group (under fusion) of Abelian bosons $\mathcal{A}$. 
The authors consider abstract $\mathcal{A}$-net configurations, which correspond to general stringlike objects that satisfy $\mathcal{A}$ fusion rules (for $\mathbb{Z}_2$ these are simply loops). 
Composite p-string excitations, with fusion rules given by $\mathcal{A}$, are formed along these $\mathcal{A}$-nets in three dimensional space by pinning the appropriate Abelian $g$ boson in a topological layer to its intersection point with a string segment labelled by $g$ in the $\mathcal{A}$-net. 
These p-string excitations are then condensed to form a cage-net fracton phase. 
This is achieved by adding local perturbations to the edges where layers intersect that fluctuate small loops of the p-string excitations, by creating particle-antiparticle pairs of $g$ bosons in both intersecting layers. 
The effect of the condensation is to confine any particles that braid nontrivially with the p-strings, and to promote the defect appearing at the open end of a p-string into a deconfined fracton excitation. Particles that braid trivially with the p-strings remain deconfined planons, and the $\mathcal{A}$ bosons in particular become equivalent to a pair of fractons (they can be viewed as a small segment of p-string). 
Particles that braid nontrivially with the p-string can be paired up across different layers to form deconfined lineons (for perpendicular layers) or planons (for parallel layers). 

More recently, in Ref.~\cite{GaugedLayers}, a construction was presented for a single stack of topological layers along the $xy$ planes of a cubic lattice, also supporting a group $\mathcal{A}$ of Abelian bosons or fermions. 
Once again p-string excitations  with fusion rules given by $\mathcal{A}$ can be constructed. 
However, in this construction, the p-strings are only condensed within $yz$ and $xz$ planes of the cubic lattice (this can be done simultaneously as the p-strings braid trivially). 
This renders particles that braid nontrivially with the p-strings immobile fractons, as their movement in the $\hat{x}$ and $\hat{y}$ directions becomes confined. 
Pairs of such fractons, separated along $\hat{x}$ or $\hat{y}$, are equivalent to charges under the condensing p-strings, and have planon mobility. 
The defects that appear at the ends of p-strings become lineons. 
Again particles that braid trivially with the p-strings remain deconfined planons. 
The $\mathcal{A}$ particles in particular are equivalent to a pair of lineons, as they can be viewed as small segments of p-string. 

Here, we describe yet another p-string construction. 
We consider topological layers stacked along the $xz$ and $yz$ planes of the cubic lattice that support a group $\mathcal{A}$ of Abelian anyons that do not need to be bosons or fermions (to construct a consistent lattice model they must have on-site string operators, which excludes semions in particular). 
We then consider condensing p-strings, made up of $\mathcal{A}$ excitations, within $xy$ planes of the cubic lattice only. This promotes particles in the layers that braid nontrivially with the Abelian $\mathcal{A}$ anyons into lineons. 
The defect at the open end of a p-string is promoted to a fracton. 
Particles in the layers that braid trivially with the $\mathcal{A}$ anyons remain planons. 
In particular, an $\mathcal{A}$ anyon in an $xz$ or $yz$ plane is equivalent to a pair of fractons, as it can be viewed as a small segment of p-string.

\begin{figure}[t!]
    \centering
    \subfloat[\label{fig:smallploop}]{\includegraphics[width=0.3\columnwidth]{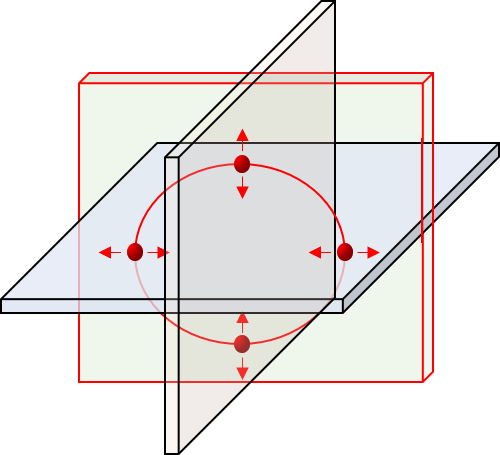}}
    \hspace{.1cm}
    \subfloat[\label{fig:longploop}]{\includegraphics[width=0.3\columnwidth]{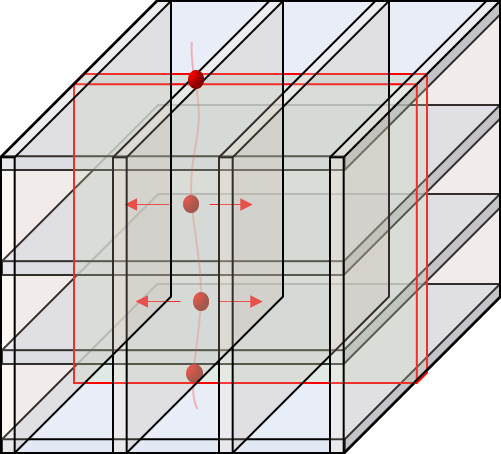}}
    \hspace{.1cm}
    \subfloat[\label{fig:ploopcondensed}]{\includegraphics[width=0.3\columnwidth]{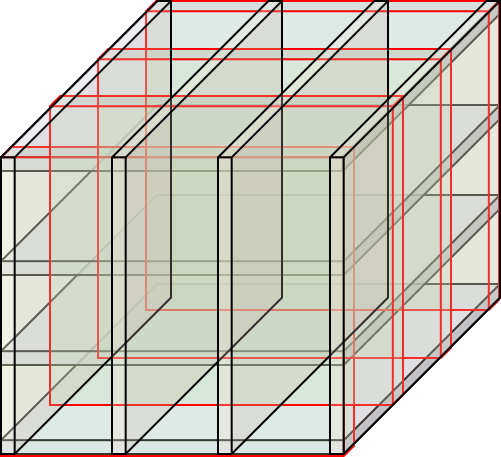}}
    \caption{
    (a) A loop of p-string excitation, confined to the $xy$ plane (green), at the junction of two topological layers. Anyons (red) are pinned to the p-string where it pierces through a topological layer. The p-string, and attached anyons, fluctuates over the $xy$ plane (shown by red arrows). 
    \\
    (b) A system of topological layers in $xz$ and $yz$ planes with an extended p-string excitation fluctuating over an $xy$ plane (green), as indicated by red arrows. 
    \\
    (c) A fracton model obtained from topological layers in $xz$ and $yz$ planes via p-string condensation within $xy$ layers (green). 
}
\end{figure}

\subsection{Anyon theory description}
\label{sec:Anyon theory description}

We consider layers described by an emergent anyon theory (modular tensor category) $\mathcal{M}$~\cite{Moore89b,Kitaev06a} that contains a group $\mathcal{A}$ of Abelian anyons. 
The anyon theory can be broken up into a direct sum according to the braiding phases of the anyons with the Abelian particles in $\mathcal{A}$, which form characters $\chi$ of the group  $\mathcal{A}$, 
\begin{align}
    \mathcal{M} = \bigoplus_{\chi \in \widehat{\mathcal{A}}} \mathcal{C}_\chi \, .
\end{align}
Specifically, an anyon $a_\chi \in \mathcal{C}_\chi$ and an Abelian anyon $g\in\mathcal{A}$ have $S$-matrix element  $S_{a_\chi,g}|S_{a_\chi,g}|^{-1}=\chi(g)$. 

We start from a system of decoupled topological layers, stacked along the $xz$ and $yz$ planes of a cubic lattice, that support anyon theories denoted by $\mathcal M_{xz}$ and $\mathcal M_{yz}$ respectively (The anyon theories need not be the same in every layer, so long as each supports a subgroup of Abelian anyons isomorphic to $\mathcal{A}$). 
We utilize the cubic lattice to label anyons by their position, where $a_{x \tilde{y} z}$ denotes an anyon in the $yz$ layer at coordinate $x$, located between $xz$ layers at $y$ and $y+1$, and contained within the $xy$ plane at coordinate $z$. 
Since the $a_{x \tilde{y} z}$ anyon is free to move throughout the $yz$ layer with coordinate $x$ (before the layers are coupled) we also utilize the notation $a_{x }$ to indicate the anyon is located somewhere in that layer. 
Similarly, we use the notation $a_{x \tilde{z}}$ to denote that the anyon is located in a strip of the $yz$ layer with coordinate $x$, between the $xy$ planes at $z$ and $z+1$. 
We employ similar notation throughout this section and Appendix~\ref{sec: Appendix H}. 

Next, we add coupling terms to the decoupled-layer Hamiltonians, at every triple intersection point of an $xz$ and $yz$ layer with an $xy$ plane, that simultaneously create $g$-$\overline{g}$ pairs, for $g\in\mathcal{A}$, in both $\mathcal{M}_{xz}$ and $\mathcal{M}_{yz}$, see Fig.~\ref{fig:smallploop}. 
In the limit of infinitely strong coupling this induces p-string excitations, formed by composites of $\mathcal{A}$ anyons, to fluctuate and condense in the $xy$ planes, see Fig.~\ref{fig:longploop}. 
For the planar p-string condensate to lead to a consistent gapped phase the $F$-symbols restricted to $\mathcal{A}$ must be trivial. 
The limit of the inter-plane spacing along $\hat{z}$ going to zero is particularly relevant for the coupled-wire construction from the main text.

\begin{figure}[t!]
    \centering
    \subfloat[\label{fig:ploopfracton}]{\includegraphics[width=0.3\columnwidth]{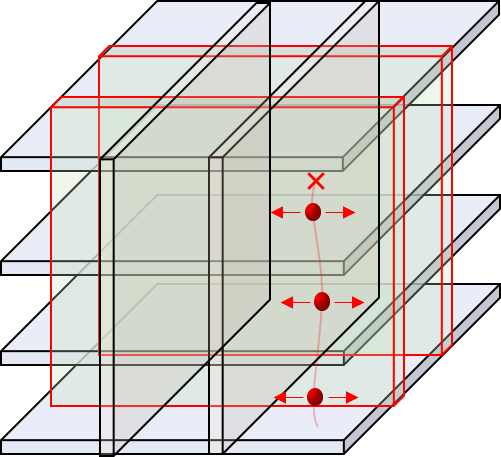}}
    \hspace{.1cm}
    \subfloat[\label{fig:plooplineonplanon2}]{\includegraphics[width=0.3\columnwidth]{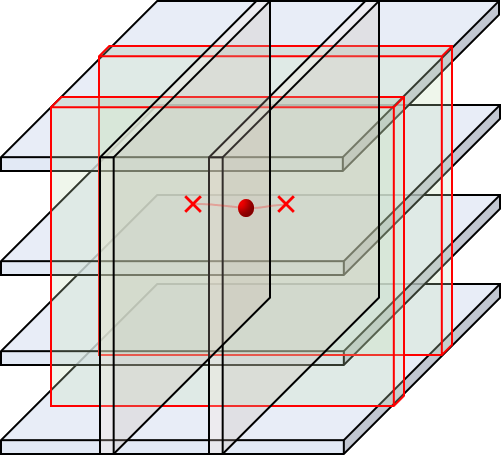}}
    \hspace{.1cm}
    \subfloat[\label{fig:plooplineonplanon1}]{\includegraphics[width=0.3\columnwidth]{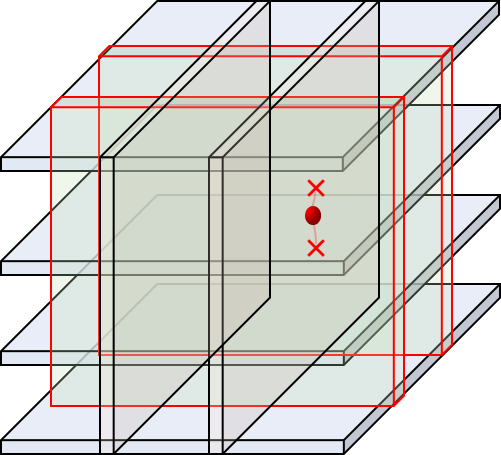}}
    \caption{
    (a) The red \textcolor{red}{$\times$} depicts a fracton excitation in the planar p-string condensed layer model. It appears at the end of a p-string that has been condensed in an $xy$-plane (indicated by red arrows) and so does not incur an energy penalty except at the open endpoint. Hence the excitation is pointlike.  
    \\
    (b) A fracton dipole oriented along $\hat{x}$. This is equivalent to an open p-string piercing a single $yz$ layer, which pins a single Abelian anyon (red sphere).  For bosonic (fermionic) p-strings this composite excitation is a $yz$ planon. For p-strings consisting of Abelian anyons with a modular braiding this excitation is a $\hat{y}$ lineon as it cannot pass through the p-string condensates on the $xy$ planes without incurring an energy penalty. 
    \\
    (c) A fracton dipole oriented along $\hat{y}$. Similar to (b) for bosonic (fermionic) p-string condensation this is an $xz$ planon. For modular Abelian anyon p-strings this is an $\hat{x}$ lineon.
}
\end{figure}

\begin{figure}[t!]
    \centering
    \subfloat[\label{fig:plooplineon1}]{\includegraphics[width=0.3\columnwidth]{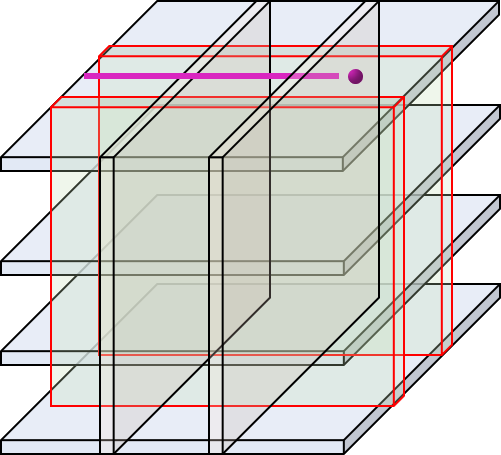}}
    \hspace{.1cm}
    \subfloat[\label{fig:plooplineon2}]{\includegraphics[width=0.3\columnwidth]{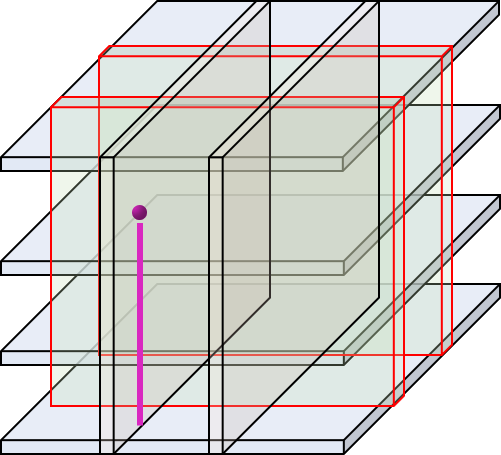}}
    \hspace{.1cm}
    \subfloat[\label{fig:plooplineon3}]{\includegraphics[width=0.3\columnwidth]{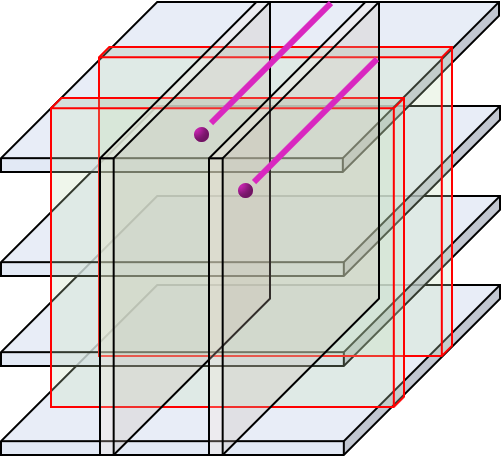}}
    \\
    \subfloat[\label{fig:ploopplanon2}]{\includegraphics[width=0.3\columnwidth]{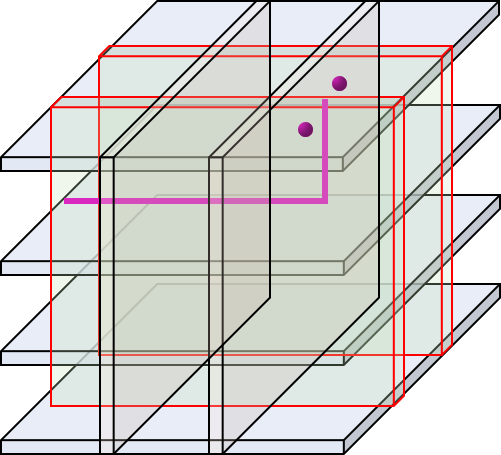}}
    \hspace{.1cm}
    \subfloat[\label{fig:ploopplanon3}]{\includegraphics[width=0.3\columnwidth]{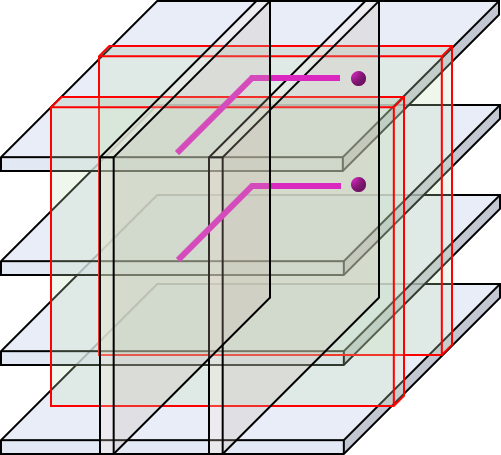}}
    \hspace{.1cm}
    \subfloat[\label{fig:ploopplanon4}]{\includegraphics[width=0.3\columnwidth]{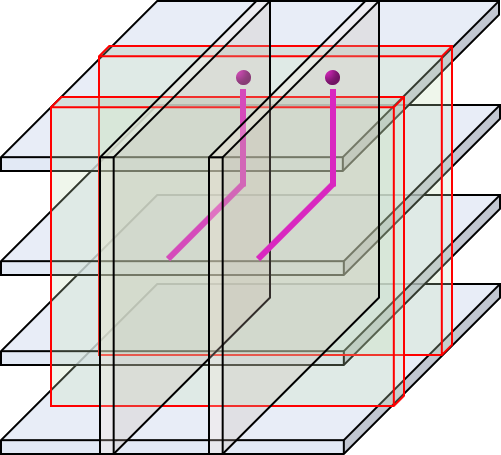}}
    \caption{
    (a) An anyon in an $xz$ layer that braids nontrivially with the p-strings becomes an $\hat{x}$ lineon after condensation.
    \\
    (b) Similarly an anyon in a $yz$ layer that braids nontrivially with the p-strings becomes a $\hat{y}$ lineon. 
    \\
    (c) A composite of $\hat{x}$ and $\hat{y}$ lineons that, together, braid trivially with the p-strings is a $\hat{z}$ lineon.
    \\
    (d) A dipole of $\hat{x}$ (or $\hat{y}$) lineons 
    that can be created by a local string operator oriented along $\hat{z}$, is an $xy$ planeon.
    \\ 
    (e) A dipole of $\hat{x}$ lineons separated along $\hat{y}$ that, together, braid trivially with the p-strings is an $xz$ planeon. 
    \\
    (f) Similarly a dipole of $\hat{y}$ lineons separated along $\hat{x}$ that, together, braid trivially with the p-strings is a $yz$ planeon.
}
\end{figure}

In the p-string condensed phase, the topological excitations are generated by fusion products of: 
\begin{itemize}
    \item Fractons $f^{g}_{\tilde x\tilde y z}\sim \prod_{ a < \tilde x} g_{a \tilde{y} z}$, with $\mathcal{A}$ fusion rules, that appear on $xy$ plaquettes of the cubic lattice. Where $\tilde{x}$ denotes the point between layers $x$ and $x+1$, and similarly for $\tilde{y}$. These fractons appear at the open endpoints of p-strings that consist of a line of $g$ anyons, such as the fusion product of all $g_{a \tilde{y} z}$ with $a < \tilde x$, see Fig.~\ref{fig:ploopfracton}. 
    
    \item Lineons  along $\hat{x}$ given by $(a_\chi)_{x \tilde{z}}$ and similarly along $\hat{y}$ given by $(b_{\chi})_{y \tilde{z}}$, for anyons $a_\chi,b_\chi\in\mathcal{C}_\chi$, see Figs.~\ref{fig:plooplineon1},~\ref{fig:plooplineon2}. 
    There are also composite lineons along $\hat{z}$, given by $\ell^{ab}_{x y}\sim (a_\chi)_{x \tilde{z}} (b_{\overline \chi})_{y \tilde{z}}$, see Fig.~\ref{fig:plooplineon3}. 
    
    \item Planons $(a_1)_{x/y}$. For bosonic or fermionic p-string excitations this includes any nontrivial $(g_1)_{x}=f^{g}_{\tilde{x} \tilde{y} z}f^{\overline{g}}_{(\tilde{x}-1) \tilde{y} z}$, see Fig.~\ref{fig:plooplineonplanon2}, and similarly for $y$, see Fig.~\ref{fig:plooplineonplanon1}. 
    For p-strings formed by more general Abelian anyons with a modular braiding the fracton composites $(g_\chi)_{x\tilde{z}}=f^{g}_{\tilde{x} \tilde{y} z}f^{\overline{g}}_{(\tilde{x}-1) \tilde{y} z}$ are in fact $\hat{y}$ lineons, and similarly $(g_\chi)_{y\tilde{z}}$ are $\hat{x}$ lineons. 
    (Since modular subtheories of an anyon model factor out~\cite{Mueger2002}, this case corresponds to the chiral $\mathbb{Z}_N$ fracton model described below stacked with some other layers.) 
    There are also composite planons $(a_\chi)_x (b_{\overline \chi})_{(x-1)}$, see Fig.~\ref{fig:ploopplanon4}, and similarly for $y$ and $z$, see Figs.~\ref{fig:ploopplanon3},~\ref{fig:ploopplanon2}. 
\end{itemize}
In this class of models the fractons are Abelian, while the lineons may be non-Abelian. 
The  braidings of the quasiparticles are inherited from the $\mathcal{M}$ layers. In particular, the fracton-composite planons may have nontrivial mutual braidings. Similarly the lineon-composite planons may have nontrivial mutual braidings. Single lineons and planons may also have nontrivial topological spin~\cite{Pai19}. 

In the limit that the inter-plane spacing goes to zero there is no space for excitations supported between layers. However, all the topological excitations can be pushed onto p-string layers and these representatives survive the limit. This is required to match the excitations with those arising in the coupled-wire model from the previous section.

\subsection{Example: $\mathbb{Z}_N$ anyon layers}
\label{sec:Z_N abstract}


We now present an example of the construction outlined above that reproduces the topological fracton sectors found in the Abelian coupled-wire construction from Section~\ref{sec:Bulk Excitations of the Coupled-Wire Model}. 
We consider Abelian chiral topological layers with $\mathbb{Z}_N^{(n)}$ anyons, in the notation of Ref.~\cite{Bonderson2007}, for $N$ an odd integer and $n$ coprime to $N$. 
For $n=2,N=m,$ this describes the anyon theory of a Laughlin FQH state at filling fraction $\nu=\frac{1}{m}$, modulo the physical fermion. 
The topological charges, and their fusion, are described by the cyclic group $\mathbb{Z}_N$ under addition. 
The $S$-matrix and topological spins of the anyons are
\begin{align}
S_{a,b} &= \frac{1}{\sqrt{N}} e^{i \frac{4 \pi n}{N} a b} \, ,
&&
\theta_a = e^{i \frac{2 \pi n}{N} a^2} \, ,
\end{align}
while the quantum dimensions and $F$ symbols are trivial. 
The obvious $\mathbb{Z}_N$ grading on the anyons is induced by braiding with the 1 anyon that generates the $\mathbb{Z}_N$ under fusion with itself (using additive notation for the composition rule i.e. 0 denotes the vacuum). 
The 1 anyon is not a boson or fermion as it has topological spin $e^{i \frac{2 \pi n}{N}}$. 

A fracton model is constructed by driving a $\mathbb{Z}_N$ p-string condensation transition within the $xy$ planes of a stack of $\mathbb{Z}_N$ anyon theories along the $xz$ and $yz$ planes of a cubic lattice. 
This case is slightly degenerate and unusual in an interesting way, since there are no nontrivial particles in the trivial sector (i.e. the sector containing the particles that braid trivially with 1), as $n$ is coprime to $N$. 
Even the generating particle 1 braids nontrivially with itself. 
The resulting model contains topological charges with a hierarchy of subdimensional topological excitations generated by: 
\begin{itemize}
    \item $\mathbb{Z}_N$ fractons that appear on the open ends of condensed p-strings. These fractons are more exotic than the usual fractons appearing in p-string condensation, as the p-strings are formed by anyons and impart a vestige of the anyonic statistics onto the fractons. 
    \item $\hat{x}$ lineons from nontrivial anyons in an $xz$ layer, trapped between p-string planes. Similarly there are $\hat{y}$ lineons from the $yz$ layers. There are also $\hat{z}$ lineons from composites of an $\hat{x}$ and $\hat{y}$ lineon trapped between the same p-string planes. 
    A pair of fractons adjacent on either side of a p-string plane is equivalent to a single p-string condensed anyon, and hence is also a lineon in this example. These lineons can be obtained by moving an $\hat{x}$ or $\hat{y}$ lineon onto a p-string plane. 
    We remark that this behavior is due to the anyonic nature of the p-strings, and does not occur for bosonic or fermionic p-string condensations. 
    \item Planons that arise from pair composites of lineons, which are themselves composites of fractons, that have opposite braiding phases with the condensed p-strings. In this example a composite formed by only a pair of fractons is not a planon due to the anyonic nature of the p-strings. 
\end{itemize}

In this example we have considered a planar p-string condensation involving anyonic p-strings. 
Attempting to apply the conventional 3D p-string condensation to these anyonic p-strings would not succeed in producing a gapped phase due to their nontrivial braidings. 
This is even true for planar p-string condensation with intersecting planes, again due to the nontrivial braidings. 
However, as we have only considered the anyonic p-string planes to be nonoverlapping, and the $F$-symbols are trivial, there is no inconsistency in the above construction. This inconsistency can be formulated as the anomaly of a subsystem symmetry in the gauging formulation of planar p-string condensation~\cite{GaugedLayers}, see Appendix~\ref{sec: Appendix H}. 

A lattice model with fracton topological order that is foliated equivalent to the example in this section is presented in Appendix~\ref{sec:Z_N lattice} where it is used to calculate the ground space degeneracy of the current example for various boundary conditions.

\section{Conclusion}
\label{sec:Conclusion}

In this work we have introduced a coupled-wire construction for a family of chiral fracton phases in (3+1) dimensions with chiral gapless boundary modes. This construction inspired a planar p-string condensation mechanism which we elaborated upon in the main text, as well as Appendix~\ref{sec: Appendix H}, where it is shown to yield a wide variety of fracton models, including ones based upon non-Abelian layers. In Appendix~\ref{sec: Appendix G}, we also propose a coupled-wire realization of such non-Abelian models.

The models constructed here motivate further exploration of coupled-wire constructions for fracton phases and new potential paths towards experimental realizations of fracton physics. They also raise the challenge of developing a deeper theoretical understanding of chiral fracton phases. In particular, further analysis of the gapless Gaussian fluctuations of the model is necessary---for example, it may be possible to add interactions to gap these fluctuations. We expect that such interactions exist, since one can construct a gapped lattice model realizing the same topological order, see Appendix~\ref{sec: Appendix H}.

Finally, this work points towards intriguing field theories obtained by taking the continuum limit in the remaining two directions that were left discrete in the current work. 
This brings to the forefront challenging technical issues surrounding the continuum limit of a system with an exponentially scaling topological degeneracy. 

\begin{acknowledgments}
We thank Dave Aasen for inspiring discussions at the conception of this project, and Danny Bulmash for valuable feedback during its early stages. 
We especially thank Meng Cheng for important clarifying discussions about the Gaussian fluctuations. 
J.S. thanks Arpit Dua for illuminating discussions on related work.
D.W. acknowledges support from the Simons Foundation.
T.I. and D.W. acknowledge hospitality courtesy of the Kavli Institute for Theoretical Physics (supported in part by the National Science Foundation under Grant No.~NSF PHY-1748958), where this work was started.
\end{acknowledgments}


\appendix 

\begin{widetext}

\section{Details on the coupled-wire model}
\label{sec: Coupled-wire Appendix}

In this Appendix we provide some of the details for calculations relating to the wire model. First, in Sec \ref{sec:Teo Kane Laughlin} we review the 2D coupled wire construction for the Laughlin state. In Sec \ref{sec:Perturbation theory} we analyze, via perturbation theory, a model of intersecting Laughlin layers with pairs of Laughlin quasiparticles condensed at vertices and show that the leading order effective Hamiltonian for this model is the one presented in Sec.~\ref{sec:Coupled-wire model}. The gap to neutral Gaussian fluctuations is discussed in Sec \ref{sec:Scaling of the gap}. We give a thorough discussion of the surface theory in Sec \ref{sec:Details on surface theory}. Lastly, we calculate the ground state degeneracy of the model with the alternative boundary conditions considered in the body of the text (Sec  \ref{sec:Topological groud-state degeneracy}) and with fully periodic boundary conditions (Sec \ref{sec:Periodic boundary conditions}) 

\subsection{Coupled-wire construction of the Laughlin state}
\label{sec:Teo Kane Laughlin}

\setcounter{equation}{0}
\renewcommand{\theequation}{A\arabic{equation}}

Here we review the coupled-wire construction of the $\nu = 1/m$ Laughlin state, first presented in \cite{Kane02,Teo14}.
The building blocks of this construction are a collection of 1D quantum wires of free fermions oriented along the z direction and labelled by a site index $j$. Upon bosonization; one has $\Psi_{\eta,j}(z)\sim e^{i \phi_{\eta,j}(z)}$ for sites j and chirality $\eta=L,R$. This construction involves only a single species of fermion on each wire so there is no need for the additional superscript $q$ used in the main text. Any local vertex operator must be decomposable into an integer combination of $\phi_{\eta,i}$. The chiral bosons have the familiar commutation relations:

\begin{equation}
[\phi_{L/R,j}(z),\phi_{L/R,i}(z^\prime)]=\pm i \pi\delta_{ij}\sgn(z-z^\prime) ~~~ \text{and} ~~~ [\phi_{L,j}(z),\phi_{R,i}(z^\prime)] = 0
 \end{equation}
 
 \noindent Next, for $m \in 2\mathbb{Z}+1$, consider the new composite fields

 \begin{equation}
 \label{eq:TeoKane-comm}
    \tilde{\phi}_{L/R,j} = \frac{\left(\phi_{L,j}+\phi_{R,j}\right) \pm m\left(\phi_{L,j} - \phi_{R,j}\right)}{2} ~~~ \text{with} ~~~ [\tilde{\phi}_{L/R,j}(z),\tilde{\phi}_{L/R,i}(z^\prime)]=\pm i m \pi\delta_{ij}\sgn(z-z^\prime)
    \end{equation}
Note that because $m$ is odd $e^{i \tilde{\phi}_{\eta}}$ is a local operator and creates a chiral quasiparticle with fermionic statistics. The non-local vertex operator  $e^{i \tilde{\phi}_{\eta}/m}$ creates an anyonic quasiparticle with self-statistics $2\pi/m$. Finally consider the field defined by 

\begin{equation}
    2 \tilde{\theta}_{j,\hat{\bm{a}}} = \tilde{\phi}_{L,j} - \tilde{\phi}_{R,j+\hat{\bm{a}}}
\end{equation}
where $\hat{a}$ is a unit vector indexing the lattice site direction. In this situation one can equivalently think of the site $j+\hat{\bm{a}}$ as $j+1$. More generally, and in the case of the main text, one may consider unit vectors $\hat{\bm{a}}$ in more than one direction and so the notation used here is chosen so as to reinforce the notation used in the main text.
With all of the relevant fields defined consider a collection of evenly spaced parallel wires with the following Hamiltonian:

\begin{equation}
H = H_0 + H_{\text{int}}= \sum_{j} \int_0^{L_z} \mathrm dz \left[ \left ( \partial_z \phi_{L,j}\right)^2 + \left (\partial_z \phi_{R,j}\right)^2\right] - \lambda\sum_j \int_0^{L_z} \mathrm dz \cos{(2\tilde{\theta}_{j,\hat{\bm{a}}})}
\end{equation}
In the $\lambda \to \infty$ limit, the interaction term condenses and $2\tilde{\theta}_{j,\hat{\bm{a}}}=0$ $\forall ~ i$. In this regime the model is gapped and has an $m$-fold ground state degeneracy. The gapped excitations correspond to solitons in the condensed fields: $2\tilde{\theta}_{j,\hat{\bm{a}}}(z) \rightarrow 2\tilde{\theta}_{j,\hat{\bm{a}}}(z) + 2\pi n \Theta(z-z_0)$. The model has anyonic excitations corresponding to minimal strength solitons (kinks of size $2\pi$) which are fully mobile in 2D. The anyons are moved along the wire direction by an operator such as $e^{\frac{i}{m}\int_{z_1}^{z_2}\mathrm{d}z\, \partial_z \tilde{\phi}_{\eta,j}}$ and are moved along the lattice direction by the operator $e^{i(\tilde{\phi}_{L,j}-\tilde{\phi}_{R,j})/m}= e^{i(\phi_{L,j}-\phi_{R,j})}$. Using these quasiparticle translation operators one can compute the anyonic braiding statistics of $2\pi/m$ by considering, e.g., the commutator
\begin{align}
    \left(\prod_j e^{\frac{i}{m}(\tilde{\phi}_{L,j}-\tilde{\phi}_{R,j})}\right)
    e^{\frac{i}{m}\int^L_0\mathrm{d}z\, \partial_z \tilde{\phi}_{L,i}}
    =
    e^{\frac{i}{m}\int^L_0\mathrm{d}z\, \partial_z \tilde{\phi}_{L,i}}
    \left(\prod_j e^{\frac{i}{m}(\tilde{\phi}_{L,j}-\tilde{\phi}_{R,j})}\right)
    e^{-2\pi i/m},
\end{align}
which follows directly from Eq.~\eqref{eq:TeoKane-comm}.


\subsection{Perturbation theory}
\label{sec:Perturbation theory}

\indent We claim that the model discussed in the main text emerges as the lowest (fourth) order term in perturbation theory in a model of interpenetrating stacks of $\nu=1/m$ Laughlin planes coupled by condensing Laughlin quasiparticles using the vertex terms $U^V_{\br}$. Heuristically one can see that the plaquette terms are related to products of four Laughlin interaction terms, $\cos(2\tilde{\theta}^q_{{\bm{r},\hat{\bm{a}}}})$ (see Appendix~\ref{sec:Teo Kane Laughlin}).

Let's see how this emerges. The Hamiltonian can be decomposed into two parts:
\begin{subequations}
\begin{align}
H_0&= H_{\text{kin}} - \lambda_V\sum_{\br}\int_0^{L_z}\mathrm{d}z\, \cos{(2\theta_{\br}^{1}+2\theta_{\br}^{2})} \\
H_1&= g\sum_{\br}\int_0^{L_z}\mathrm{d}z\,\left[\cos{\big(2\tilde{\theta}^{1}_{\bm{r},\hat{\bm{y}}}\big)} + \cos{\big(2\tilde{\theta}^{2}_{\bm{r},\hat{\bm{x}}}\big)}\right],
\end{align}
\end{subequations}
where $H_{\text{kin}}$ is the kinetic term for the wires in the absence of the  couplings $\lambda_V,g$.

\begin{figure}[t!]
\begin{center}
\includegraphics[width=.85\columnwidth]{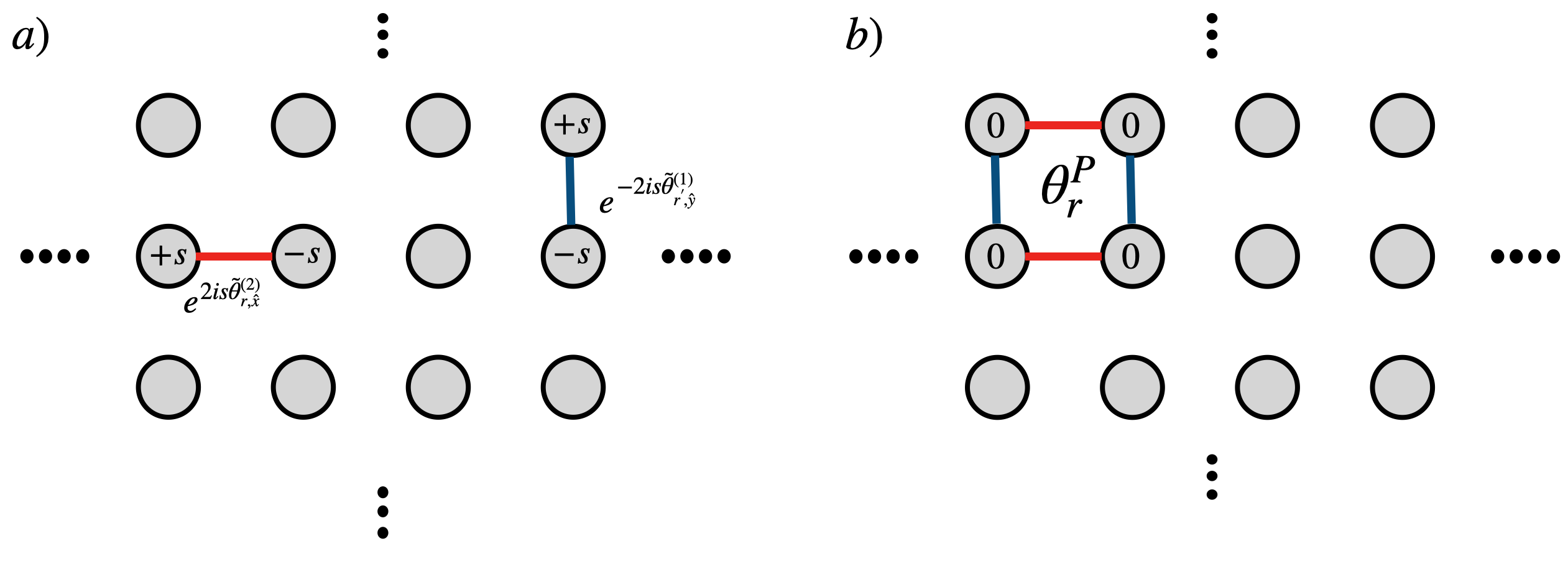}
\caption{(a) Dipole configurations are created in the vertex terms by the Laughlin interaction operators $\exp(2is\tilde{\theta})$. (b) To avoid being projected out of the low-energy subspace, these operators must enter the effective Hamiltonian in the form of a plaquette term. In such configurations, the vertex solitons generated by different Laughlin interaction operators cancel.
}
\label{fig:PertTh}
\end{center}
\end{figure}

In the limit $\lambda_V\to\infty$, solitons in the argument of the vertex term in $H_0$ do not lie in the spectrum of low-energy states. The aim is to find an effective theory which describes the physics in this low-energy subspace. This can be accomplished using Wigner-Brillouin perturbation theory: we introduce the operator $P$ which projects onto the ground-state subspace of $H_0$ and in particular throws out any states with solitons in the $\lambda_V$ terms. Our effective Hamiltonian is then given by $H_0+H_{\text{eff}}$, where 
\begin{equation}
    H_{\text{eff}} = PH_1\sum_{n=0}^{\infty} \bigg\{\big[E_0 - (1-P)H_0(1-P)\big]^{-1}H_1 \bigg\}^n P
\end{equation}
In the limit $\lambda_V\to\infty$, eigenstates of $H_0$ can be labeled schematically by occupation numbers associated with solitons in each vertex term. Occupied states will be projected out, so we need to determine the lowest-order term in the series which does not excite any vertices. 
Suppressing the integral over $z$, we can express $H_1$ as $$H_1=\frac{g}{2}\sum_{\br,s=\pm} \left[\exp\left(2is\tilde{\theta}^{1}_{\bm{r},\hat{\bm{y}}}\right) +\exp\left(2is\tilde{\theta}^{2}_{\bm{r},\hat{\bm{x}}}\right)\right].$$ 
As one can see from Fig.~\ref{fig:PertTh}(a), $\exp(2is\tilde{\theta}^{1})$ creates a $\pm s$ dipole of vertex solitons in the $y$-direction while $\exp(2is\tilde{\theta}^{2})$ creates an analogous dipole in the $x$-direction.
The terms in $H_{\text{eff}}$ correspond to products of these dipoles. One needs to go to $4^{\text{th}}$ order to create a dipole configuration that leaves behind no vertex solitons and avoids being projected out:
\begin{align}
\begin{split}
        H_{\text{eff}}^{(4)} &= P\sum \frac{g^4}{16\Delta_{\text{kink}}^3}\Bigg[ \exp\left(2is\tilde{\theta}^{1}_{\bm{r},\hat{\bm{y}}} + 2is^{\prime}\tilde{\theta}^{2}_{\bm{r}^\prime,\hat{\bm{x}}} + 2is^{\prime \prime}\tilde{\theta}^{1}_{\bm{r}^{\prime \prime},\hat{\bm{y}}} + 2is^{\prime\prime\prime}\tilde{\theta}^{2}_{\bm{r}^{\prime \prime \prime},\hat{\bm{x}}}\right)+\text{H.c.} \Bigg]P\\
        &~~~~~+ \text{terms w/ unequal numbers of $x$- and $y$-dipoles that get projected out},
\end{split}
\end{align}
where $\Delta_\mathrm{kink}$ is the energy gap to creating a vertex-soliton dipole.
Applying the projection operators one can see that the only terms that survive form a unit square with $s=-s^\prime=-s^{\prime\prime}=s^{\prime\prime\prime}$, see Fig.~\ref{fig:PertTh}(b). This yields
\begin{align}
     &H_{\text{eff}} \sim \frac{g^4}{\Delta_{\text{kink}}^3}\sum_{\br} \cos\left(2\tilde{\theta}^{1}_{\bm{r},\hat{\bm{y}}}-2\tilde{\theta}^{2}_{\bm{r},\hat{\bm{x}}}- 2\tilde{\theta}^{1}_{\bm{r} + \hat{\bm{x}},\hat{\bm{y}}}+2\tilde{\theta}^{2}_{\bm{r}+\hat{\bm{y}},\hat{\bm{x}}}\right)+ \text{higher-order terms},
\end{align}
which is, up to subleading corrections, precisely the plaquette term defined in Eq.~\eqref{eq:AbelianHint} in the main text.

The above analysis can be extended to describe the case where strong vertex terms are only turned on in a bounded subregion, which we take to be an $L_x\times L_y$ rectangle for simplicity, of the full lattice of wires. In the perturbative treatment, strong vertex terms on the boundary of this subregion generate truncated plaquette operators which contain the only edges of a full plaquette that touch one of the strong vertex terms. For example, vertex terms along the top boundary of the subregion generate the interaction $\cos\lb2\!\lp\ttheta^{1}_{(x,L_y),\yhat} \!-\! \ttheta^{1}_{(x+1,L_y),\yhat} \!-\! \ttheta^{2}_{(x,L_y),\xhat}\rp\rb$, while the vertex term in the top-right corner of the subregion generates the interaction $\cos\lb 2\!\lp\ttheta^{1}_{(L_x,L_y),\yhat} \!-\! \ttheta^{2}_{(L_x,L_y),\xhat}\rp\rb$. These are precisely the boundary interactions used in the surface termination discussed in the main text, which can be viewed as a minimal example in which the topmost and bottommost $q=2$ Laughlin layers and the leftmost and rightmost $q=1$ Laughlin layers are omitted.

Given this perturbative analysis, we can try to understand the excitations of the model from the perspective of p-string condensation, as discussed in the main text. The term $\lambda_V$ condenses two pairs of anyons created by $\exp{2i\theta_{xy}^{1}}$ and $\exp{2i\theta_{xy}^{2}}$. This condensate proliferates in the limit $\lambda_V\rightarrow \infty$. Low energy excitations must commute with the condensate and so the emergent mobility restrictions can be understood as stemming from this constraint. For more on this perspective, we refer the reader to Appendix~\ref{sec: Appendix H}.


\subsection{Scaling of the gap}
\label{sec:Scaling of the gap}

\begin{figure}[t!]
 \centering
    \includegraphics[width=9cm]{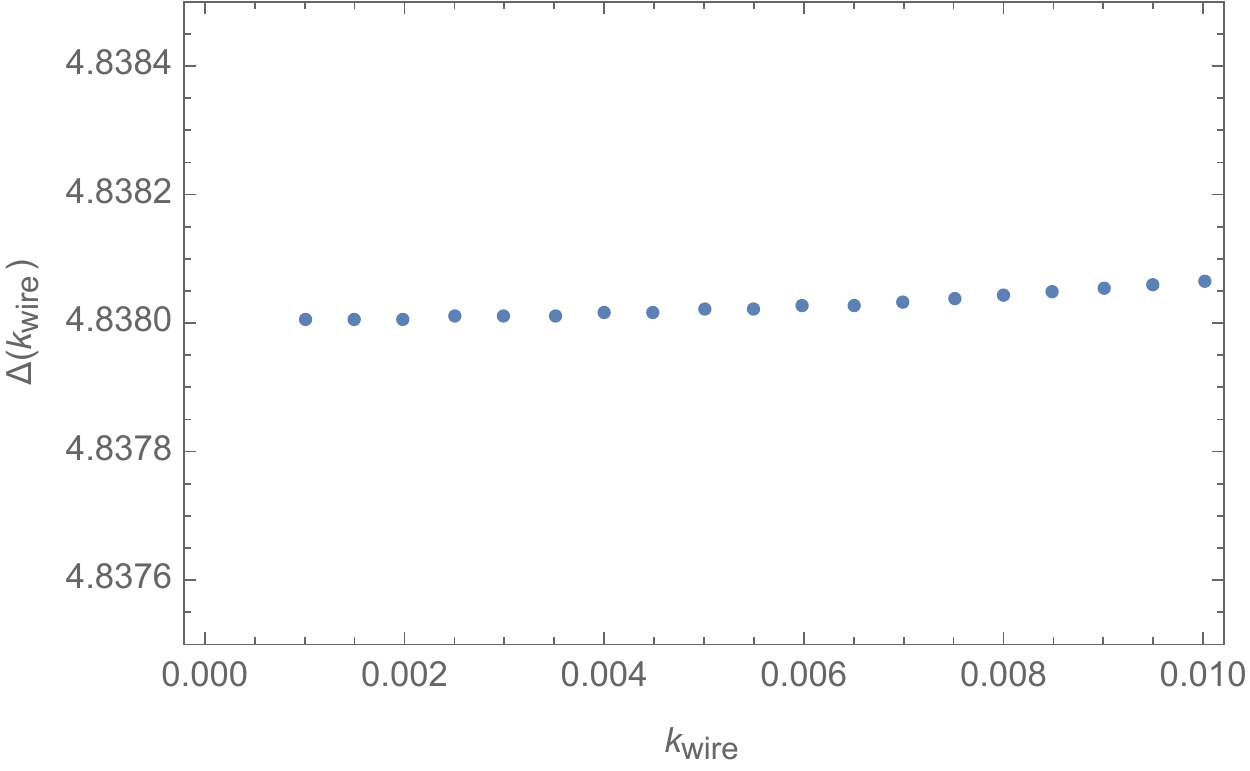}
    \caption{A plot of the smallest eigenvalue at a given $k_z$ along the wire direction with $U=100$ and system size $L=15$. One can see that the gap $\Delta_{k_z\to 0} \rightarrow 4.8380 + \epsilon$ ($\epsilon\ll 1$) reflecting the fact that the model is gapped at finite system size.}
    \label{fig:kscaling}
  \end{figure}

While gapped to charged topological excitations (solitons), the model studied in the main text possesses gapless neutral excitations in the thermodynamic limit. These neutral excitations, which we refer to as ``phonons," correspond to Gaussian density fluctuations and do not participate in the transport of charge. 
 
In this Appendix we present numerical results on the phonon spectrum for PBC in the $z$-direction and the boundary conditions in the $x$-$y$ plane used to calculate the topological degeneracy in the main text and in Appendix~\ref{sec:Topological groud-state degeneracy}. For these calculations, we set $L_x=L_y\equiv L$ and consider a vertex coupling $\lambda_V=100\, U$, with all other couplings including $\lambda_P$ and the boundary and corner couplings set to $U$; in turn, we take $U\gg v$, where $v$ is the kinetic energy scale of the decoupled wires. We set $v=1$ unless specified otherwise. This hierarchy of energy scales is consistent with the perturbative treatment of Appendix~\ref{sec:Perturbation theory}. We analyze the scaling of the phonon gap with the momentum $k_z$ along the wire direction, the system size $L$, and the strong coupling $U$. Our results indicate that the phonons are gapped at finite system size, and in a particular scaling limit in which $U$ scales at least as a sufficiently large power of $L$ as $L \rightarrow \infty$.

\begin{figure}[t!]
    \centering
    \includegraphics[width=9cm]{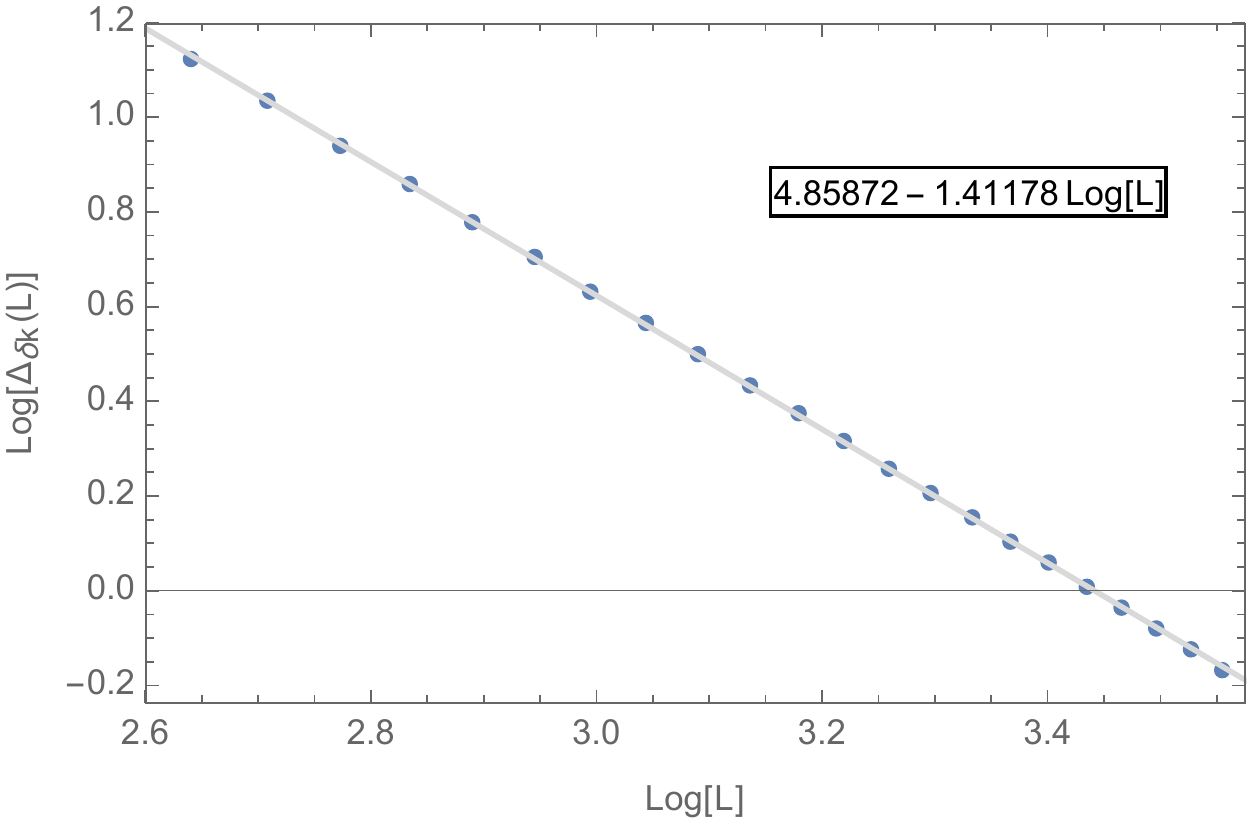}
    \caption{Log-Log plot of $\Delta_{\delta k_z}(L)$ vs $L$ at $U=100$. We see that the scaling relation $\Delta_{\delta k_z}(L) \sim L^\alpha$ with $\alpha \approx -1.4118$. }
    \label{fig:Lscaling}
 \end{figure}
 
 \begin{figure}[t!]
    \centering
    \includegraphics[width=9cm]{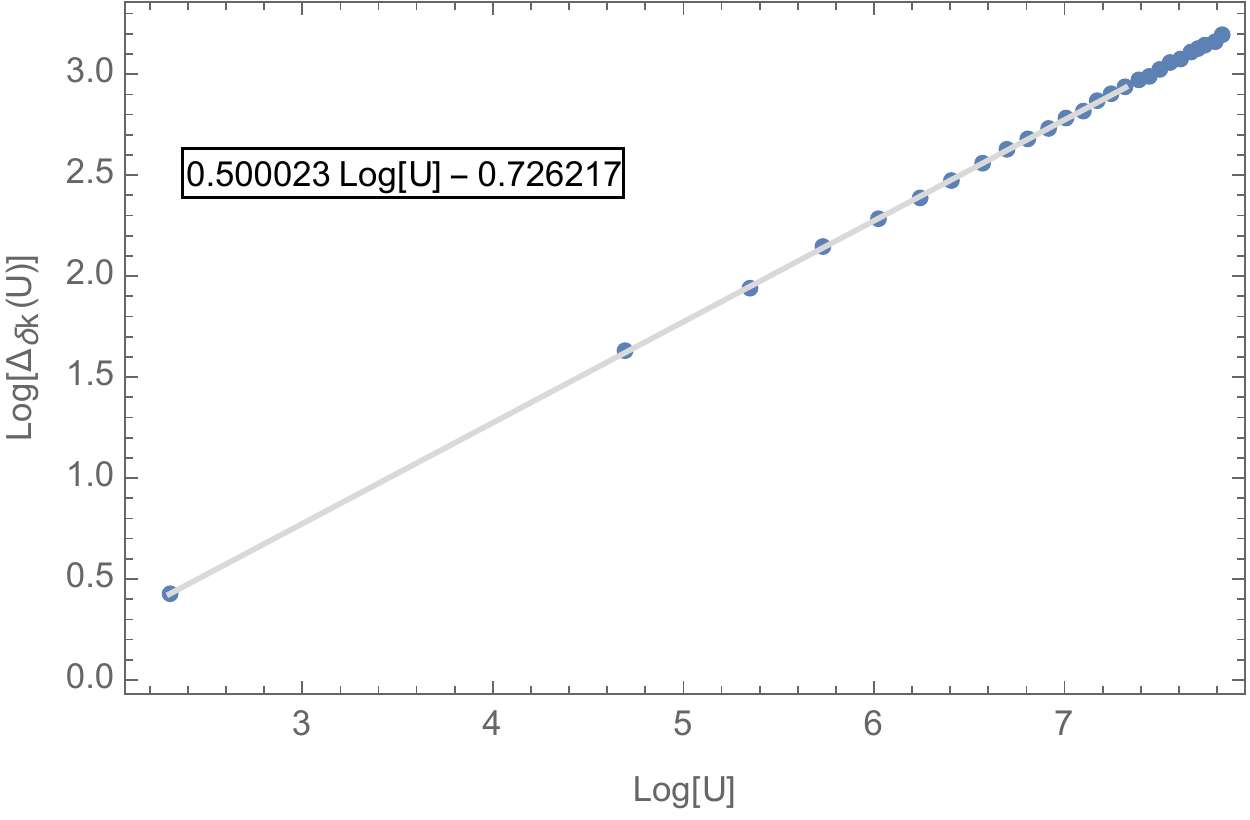}
    \caption{Log-Log plot of $\Delta_{\delta k_z}(U)$ vs $U$ at $L=20$. We see that the scaling relation $\Delta_{\delta k_z}(U) \sim U^\beta$ with $\beta \approx .500023$. This points to a scaling $\Delta_{\delta k_z}(U) \sim \sqrt{U}$.}
    \label{fig:Uscaling}
\end{figure}

Following~\cite{JoeCWFracton}, for each pinning term we make the replacement $\cos(\Lambda \cdot \Phi) \sim 1-\frac{(\Lambda \cdot \Phi)^2}{2}$ and analyze the corresponding Bogoliubov-de-Gennes mean field theory. The alternate boundary conditions break translation symmetry in the discrete directions $x,y$ so the model must be solved using a mixed basis $\{a^\dagger_{\br}(k_z),a_{\br}(k_z)\}$, which create/destroy bosonic fluctuations of momentum $k_z$ on wire $\br=(x,y)$. Since $k_z$ is a well defined quantum number, let us define $\Delta_{k_z}(U,L)$ to be the smallest energy eigenvalue with momentum $k_z$ at system size $L$ and coupling strength $U$ (note that we identify this quantity with the phonon gap since the Hamiltonian is positive semidefinite). We suppress the arguments $U$ and $L$ when convenient.

First let us consider the behavior of  $\Delta_{k_z}$ as we let $k_z \rightarrow 0.$ Since the real space calculation is done numerically one cannot actually evaluate the $\Delta_{k_z}$ at $k_z=0$ because the BdG Hamiltonian will involve terms proportional to $\frac{1}{|k_z|}$ coming from factors $\phi^2$ and $\theta^2$ (see Ref.~\cite{JoeCWFracton}). An analytical expression for the eigenenergies is prohibitively complicated but should be a function $f(vk_z,vU)$ in order to the prevent divergence as $k_z \rightarrow 0$. Fig.~\ref{fig:kscaling} shows $\Delta_{k_z}$ at system size $L=15$. Evidently, at this system size $\Delta_{k_z} \rightarrow \Delta_0 \in [4.380 - \epsilon,4.380+\epsilon]$ with $\epsilon\ll 1$, reflecting the fact that the model is gapped at finite $L$. In other words, at finite $L$, $\Delta_{k_z}(L) \rightarrow \Delta_0(L)>0$ as $k_z \rightarrow 0$. With this justification, we henceforth approximate $\Delta_{0}$ by $\Delta_{\delta k_z}$ with a small $\delta k_z=10^{-5}$.

 We now consider the scaling of the gap with $L$ and $U$. Determining how $\Delta_{0}(L)$ depends on $L$ enables us to check if the gap persists in the thermodynamic limit. We find that $\Delta_{\delta k_z}(L) \rightarrow 0$ as $L \rightarrow \infty$. Thus, while gapped at finite $L$, the phonons become gapless in the limit of infinite system size. In Fig.~\ref{fig:Lscaling} we see a power-law dependence of the form $\Delta_{\delta k_z}(L)\sim L^\alpha$ with $\alpha \sim - 1.412$. Lastly, the scaling of $\Delta_{\delta k_z}(U)$ with $U$ is shown in Fig \ref{fig:Uscaling}: $\Delta_{\delta k_z} \sim U^{.500023}.$ Since we use a $\delta k$ of order $10^{-5}$, our numerical results are consistent with the dependence $\Delta_{k_z\to 0}\sim\sqrt{U}$, which is also found in Ref.~\cite{JoeCWFracton}'s analysis of translation-invariant coupled-wire models with gapless fluctuations. Putting these dependences together, we conclude that $\Delta_{k_z\to 0}\sim \sqrt{U}L^\alpha$.

One interesting take-away from this analysis is that there exists a scaling limit, in which the strong-coupling limit $U\to \infty$ is taken alongside the thermodynamic limit $L\to\infty$ such that $U\gtrsim L^{-2\alpha}$, in which the fluctuations are gapped. In this limit, one has $\Delta_0\gtrsim L^{\frac{1}{2}(-2\alpha)}L^\alpha \sim O(1)$. At first glance this may seem a somewhat artificial limit to take. Recall though that the $U \rightarrow \infty$ limit has been assumed throughout in our discussion of both the charged and neutral sectors of the theory. Thus the scaling limit merely demands that the $U\to\infty$ limit be taken sufficiently ``fast" compared to the $L\to\infty$ limit.

    


\subsection{Details on surface theory}
\label{sec:Details on surface theory}

\begin{figure}[h!]
\begin{center}
\includegraphics[width=.35\columnwidth]{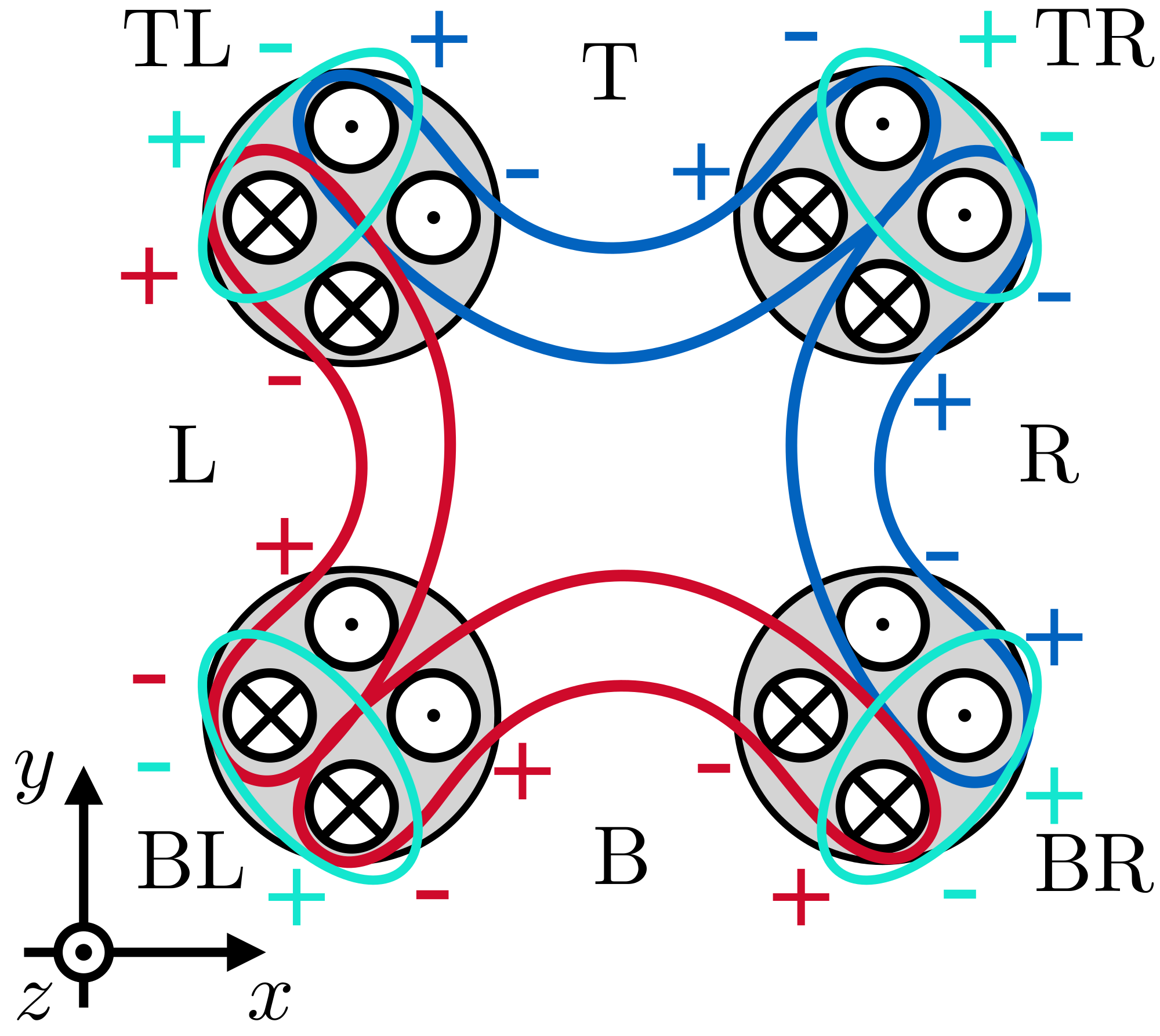}\\
\caption{
Pictorial definition of the gapless surface ($\mathrm{L,R,T,B}$) and corner ($\mathrm{TL,TR,BL,BR}$) modes with open boundary conditions in the $x$- and $y$-directions. Chiral modes belonging to the same surface or corner mode are encircled, and $\pm$ indicates the relative sign with which each chiral mode appears [see, e.g., Eq.~\eqref{eq:SurfMode}].
}
\label{fig:Boundary-modes}
\end{center}
\end{figure}

In this Appendix we provide further details on the construction of the surface theory for the case of open boundary conditions in $x$ and $y$ and periodic boundary conditions in $z$. In particular, we construct the $2(L_x+L_y)$-dimensional $K$-matrix of the surface theory, which encodes the commutation relations among the gapless surface modes,
and show that it has two zero modes that correspond to bulk degrees of freedom that are gapped in the strong coupling limit. This leaves $N=2(L_x+L_y-1)$ gapless modes residing on the surface, as expected from the counting argument presented in the main text.

We identify the chiral gapless boundary modes by finding linear combinations of bosonic fields on the surface that commute with the bulk interaction terms \eqref{eq:AbelianHint}.  The boundary of the coupled-wire array can be divided into left, right, top, and bottom ($\mathrm{L,R,T,B}$) faces. The \L and \R faces each contain $L_y-1$ bonds, and the \T and \B faces each contain $L_x-1$ bonds.  Each such bond is associated with a chiral mode that commutes with the bulk interactions.  For example, on the \L surface [$\br=(1,y),$ $y=1,\dots,L_y-1$], one finds using Eq.~\eqref{eq:PhiTildeDef} that the $L_y-1$ chiral modes
\begin{align}
\label{eq:SurfMode}
\begin{split}
        \hat{\phi}^{\rm L}_{R,\br}&=\tilde{\phi}^2_{R,\br}-\tilde{\phi}^1_{R,\br}+\tilde{\phi}^1_{L,\br-\yhat}-\tilde{\phi}^2_{R,\br-\yhat}
\end{split}
\end{align}
all commute with the interactions \eqref{eq:AbelianHint}. Analogous definitions for the $\mathrm{R,T,B}$ surfaces can be read off from Fig.~\ref{fig:Boundary-modes}. In addition to the $2(L_x+L_y-2)$ chiral modes from the $\mathrm{L,R,T,B}$ surfaces, there are four gapless modes associated with the $\mathrm{TL,TR,BL,BR}$ corners of the array; for example, at the \BL corner we have
\begin{align}
\label{eq:CornerMode}
    \hat{\phi}^{\rm BL}_{R}=\tilde{\phi}^2_{R,(1,1)}-\tilde{\phi}^1_{R,(1,1)},
\end{align}
and similar expressions for the other corners can be read off from Fig.~\ref{fig:Boundary-modes}.  We thus find a total of $2(L_x+L_y)=N+2$ boundary modes that commute with the bulk interactions.  Of these, all but the modes at the \TL and \BR corners are chiral, i.e., contain an excess of right- or left-movers.

Each of the surface modes identified above has a nontrivial commutation relation both with itself and with its immediate neighbors; these commutation relations follow directly from Eq.~\eqref{eq:PhiTildeDef}. We organize these commutation relations into an $(N+2)$-dimensional square matrix $K$ defined by
\begin{align}
\label{eq:KDef}
[\hat\phi_\alpha(z),\hat\phi_\beta(z')]=i\pi\, K_{\alpha\beta}\, \sgn(z-z'),
\end{align}
where $\alpha,\beta=1,\dots,N+2$ label the surface modes identified above. We compute $K$ as an $8\times8$ block matrix whose diagonal blocks describe the $\mathrm{L,R,T,B}$ faces and the $\mathrm{TL,TR,BL,BR}$ corners, and whose off-diagonal blocks encode nontrivial commutation relations among the corners and faces.

We first focus on the block-diagonal part of $K$. Each of the $\mathrm{L,R,T,B}$ faces has an $(L_a-1)$-dimensional ($a=x,y$) $K$-matrix for the modes identified in Eq.~\eqref{eq:SurfMode} that is proportional to
\begin{align}
\label{eq:SurfaceKOBC}
    K_a
    =
    \begin{pmatrix}
    -2m & m & 0 & \dots & 0 & 0\\
    m & -2m & m & 0 & \dots & \\
    0& m & -2m & m & 0 & \dots \\ 
    \vdots & \vdots & \vdots & \vdots & \vdots & \vdots\\
    0 & 0 & \dots & 0 & m & -2m
    \end{pmatrix}.
\end{align}
This $K$ matrix is (up to an unimportant sign on the diagonal entries) proportional to that of the so-called ``121" phase of a stack of fractional quantum Hall layers identified in Ref.~\cite{Naud01}.
The $K$-matrices for the $\mathrm{T,B,L,}$ and $\mathrm{R}$ surfaces are
\begin{align}
\label{eq:SideKs}
    K_{\rm T}=-K_x,\indent K_{\rm B}=K_x,\indent K_{\rm L}=K_y,\indent K_{\rm R}=-K_y,
\end{align}
indicating that opposite surfaces have opposite chiralities, as expected. The remaining diagonal blocks of $K$ describe the self-commutation relations of the $\mathrm{TL,TR,BL,BR}$ corner modes identified in Eq.~\eqref{eq:CornerMode}; they are
\begin{align}
\label{eq:CornerKs}
    K_{\mathrm{TL}}=0,\indent K_{\mathrm{TR}}=2m,\indent K_{\mathrm{BR}}=0,\indent K_{\mathrm{BL}}=-2m.
\end{align}
Note that $K_{\mathrm{TL}}=K_{\mathrm{BL}}=0$ because these modes are nonchiral.

We now determine the off-diagonal blocks of $K$. Each of the corner modes has a nontrivial algebra with neighboring modes from the two surfaces it touches.  This algebra is encoded in the $1\times (L_x-1)$ $K$-matrices
\begin{subequations}
\label{eq:CouplingKs}
\begin{align}
    K_{\rm TLT}=\begin{pmatrix} -m & 0 &\dots &0 \end{pmatrix}=-K_{\rm BRB},
\end{align}
the $(L_x-1)\times 1$ $K$-matrices
\begin{align}
    K_{\rm TTR}=\begin{pmatrix} 0 \\ \vdots \\ 0 \\ -m \end{pmatrix}=-K_{\rm BBL},
\end{align}
the $1\times (L_y-1)$ $K$-matrices
\begin{align}
    K_{\rm TRR}=\begin{pmatrix} m & 0 &\dots &0 \end{pmatrix}=-K_{\rm BLL},
\end{align}
and the $(L_y-1)\times 1$ $K$-matrices
\begin{align}
    K_{\rm RBR}=\begin{pmatrix} 0 \\ \vdots \\ 0 \\ m \end{pmatrix}=-K_{\rm TLL}^\mathsf{T}.
\end{align}
\end{subequations}
Combining Eqs.~\eqref{eq:SurfaceKOBC}, \eqref{eq:SideKs}, \eqref{eq:CornerKs}, and \eqref{eq:CouplingKs}, we can write the full $2(L_x+L_y)$-dimensional $K$-matrix in block form as
\begin{align}
\label{eq:KAllOBC}
    K
    =
    \begin{pmatrix}
    K_{\rm TL} & K_{\rm TLT} & 0 & 0 & 0 & 0 & 0 & K_{\rm TLL}\\
    K_{\rm TLT}^\mathsf{T} & K_{\rm T} & K_{\rm TTR} & 0 & 0 & 0 & 0 & 0\\
    0 & K_{\rm TTR}^\mathsf{T} & K_{\rm TR} & K_{\rm TRR} & 0 & 0 & 0 & 0\\
    0 & 0 & K_{\rm TRR}^\mathsf{T} & K_{\rm R} & K_{\rm RBR} & 0 & 0 & 0\\
    0 & 0 & 0 & K_{\rm RBR}^\mathsf{T} & K_{\rm BR} & K_{\rm BRB} & 0 & 0\\
    0 & 0 & 0 & 0 & K_{\rm BRB}^\mathsf{T} & K_{\rm B} & K_{\rm BBL} & 0 \\
    0 & 0 & 0 & 0 & 0 & K_{\rm BBL}^\mathsf{T} & K_{\rm BL} & K_{\rm BLL}\\
    K_{\rm TLL}^\mathsf{T} & 0 & 0 & 0 & 0 & 0 & K_{\rm BLL}^\mathsf{T} & K_{\rm L}
    \end{pmatrix}.
\end{align}

The existence of the $2(L_x+L_y)$-dimensional $K$-matrix~\eqref{eq:KAllOBC} seems to imply the existence of $2(L_x+L_y)=N+2$ gapless modes that commute with the bulk interaction terms, rather than the $N$ modes expected based on counting the bulk interaction terms.  However, it is possible that some linear combinations of these surface modes can be rewritten in terms of pinned bulk fields, since these bulk fields also commute with the interaction Hamiltonian.  Indeed, we find that the $K$-matrix \eqref{eq:KAllOBC} has two zero modes at any system size, and that these zero modes correspond to linear combinations of pinned bulk fields.  The first zero mode can be written compactly in $(N+2)$-dimensional vector form as
\begin{align}
    \Upsilon_+=\begin{pmatrix} -1 & -\bm 1_{L_x-1} & -1 & +\bm 1_{L_y-1} & -1 & -\bm 1_{L_x-1} & -1 & +\bm 1_{L_y-1} \end{pmatrix}^\mathsf{T},
\end{align}
where $\bm 1_d$ is a $d$-dimensional vector with unit entries.
The second is 
\begin{subequations}
\begin{align}
    \Upsilon_-=\begin{pmatrix} \frac{L_x+L_y}{2} & C_{+-} & \frac{L_y-L_x}{2} & C_{-+} & -\frac{L_x+L_y}{2} & -C_{+-} & \frac{L_x-L_y}{2} & -C_{-+} \end{pmatrix}^\mathsf{T},
\end{align}
where the $(L_x-1)$-dimensional vector
\begin{align}
    C_{+-}=\begin{pmatrix} \frac{L_x+L_y}{2}-1 & \frac{L_x+L_y}{2} - 2 & \dots & \frac{L_y-L_x}{2} + 1  \end{pmatrix}
\end{align}
and the $(L_y-1)$-dimensional vector
\begin{align}
    C_{-+}=\begin{pmatrix} \frac{L_x-L_y}{2}+1 & \frac{L_x+L_y}{2} + 2 & \dots & \frac{L_x+L_y}{2} - 1  \end{pmatrix}.
\end{align}
\end{subequations}
Writing these zero modes as linear combinations of the underlying bosonic surface fields defined by Eqs.~\eqref{eq:SurfMode} and \eqref{eq:CornerMode} and their analogs, we find that they can be reexpressed as linear combinations of the pinned bulk vertex and plaquette fields.  For example,
\begin{align}
    \Upsilon_+ = \sum_{\br\in\Lambda} \theta^P_{\br},
\end{align}
where the sum runs over all vertices $\br$ in the square lattice $\Lambda$ with OBC and where $\theta^P_{\br}$ is the pinned (i.e., gapped) plaquette field. $\Upsilon_-$ can also be expressed as a linear combination of pinned vertex and plaquette fields $\theta^V_{\br}$ and $\theta^P_{\br}$, respectively, but the expression is more complicated (in particular, it depends on system size) and we omit it here. Pictorial examples of the expressions for $\Upsilon_\pm$ in terms of pinned bulk fields for $L_x=L_y=4$ are shown in Fig.~\ref{fig:ZeroModes}. In summary, while a naive identification of surface modes commuting with the bulk interaction terms finds $N+2$ such modes, a closer look shows that two of these modes can be reexpressed as linear combinations of pinned bulk fields.  We thus find $N$ gapless surface modes, as expected from the counting of bulk interaction terms.

\begin{figure}[t!]
\begin{center}
\includegraphics[width=.6\columnwidth]{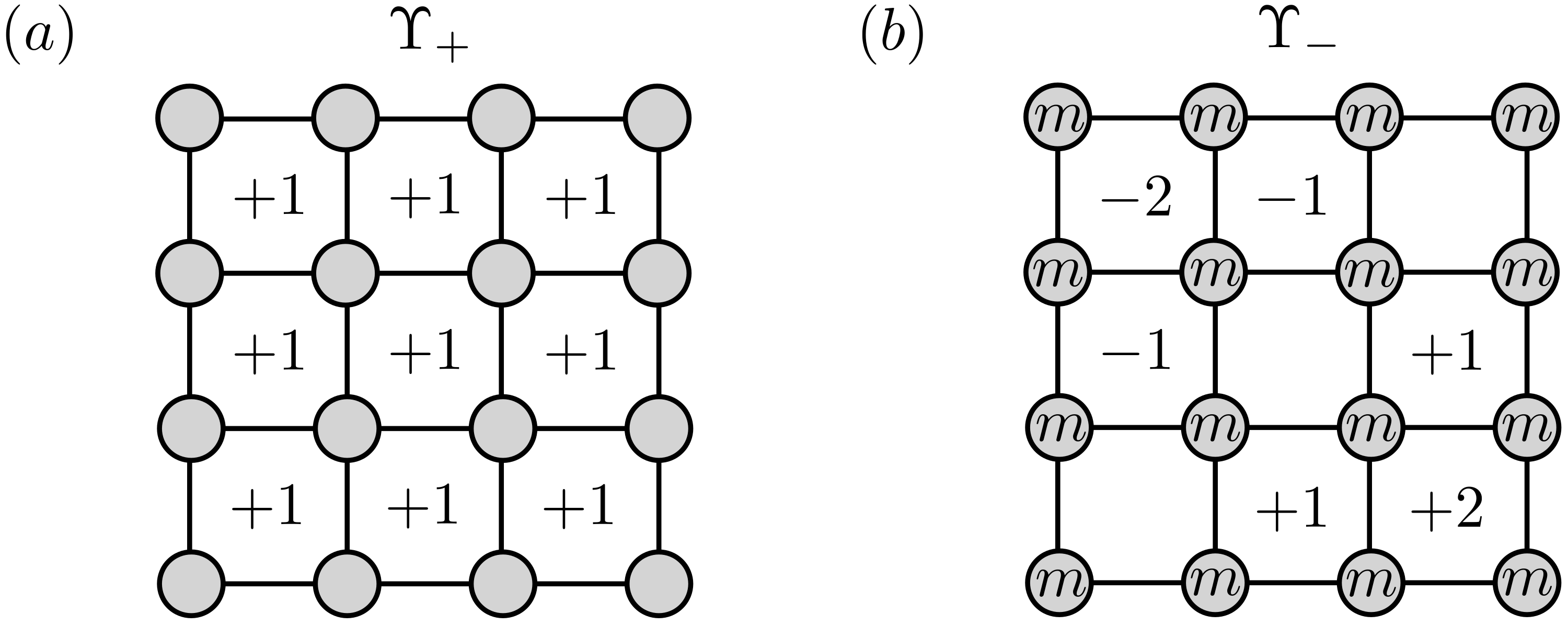}\\
\caption{
Schematic depiction of the expressions for the zero modes $\Upsilon_+$ (a) and $\Upsilon_-$ (b) in terms of pinned bulk fields $\theta^{V,P}_{\br}$ at system size $L_x=L_y=4$.  Integers appearing in a plaquette or vertex of the square lattice signify the coefficient with which the corresponding $\theta^P_{\br}$ or $\theta^V_{\br}$ field (respectively) enters the expression of $\Upsilon_\pm$ as a linear combination of these fields.
}
\label{fig:ZeroModes}
\end{center}
\end{figure}

We note in passing that the Lagrangian for the surface theory constructed in this Appendix can be written as
\begin{align}
\label{eq:Lbdy}
    L_{\rm bdy}=\int^{L_z}_0\frac{\mathrm dz}{4\pi}\, [(\partial_t\hat{\bm\phi})^\mathsf{T} K^{+}(\partial_z\hat{\bm\phi})-(\partial_z\hat{\bm\phi})^\mathsf{T} V(\partial_z\hat{\bm\phi})],
\end{align}
wherein we have collected the boundary modes into an $(N+2)$-component vector $\hat{\bm\phi}$, and where the $(N+2)$-dimensional symmetric rational matrix $K^+$ is the pseudoinverse of $K$. The matrix $V$ encodes the kinetic energy of the boundary modes and depends on microscopics. This surface theory is unusual because $K^+$ is generically not sparse, and deserves further study.


\subsection{Topological ground-state degeneracy}
\label{sec:Topological groud-state degeneracy}

Here we compute the topological ground-state degeneracy (GSD) for the coupled-wire model with the alternative ``unusual" boundary conditions defined in the main text. The method used is presented in Ref.~\cite{Imamura19}. The closed model is shown in Fig.~\ref{fig:FullBC} a). Note that in the strong coupling limit any configuration in the ground state manifold is labeled by a set of integers defined by $\theta^{V,P}_{\bm{r}},2\tilde{\theta}^{q}_i\in 2\pi\mathbb{Z}$ where $\bm{r}\in\Lambda$ labels wires in the original 2D array and $i=1,\dots,N/2+1$ labels pairs of added boundary wires. Naively then, a general ground state can be labelled by assigning each plaquette, vertex, and oval in \ref{fig:FullBC}(a) a value in $\mathbb{Z}$. However, because of the compact nature of the degrees of freedom ($\phi^q_{\eta,\bm{r}} \equiv \phi^q_{\eta,\bm{r}} +2\pi$), many of these ground-state configurations should, in fact, be identified. Starting with an arbitrary configuration, one can ``clean" the set of ground-state labels by using local shifts $\phi^q_{\eta,\bm{r}} \rightarrow \phi^q_{\eta,\bm{r}} +2\pi$ to set extraneous labels to zero. The patterns of these local $2\pi$ shifts are the same as those produced by the application of vertex operators, some of which are displayed in Figs.~\ref{fig:LocalOps} and \ref{fig:Excitations}. Patterns of particular use are displayed in Fig.~\ref{fig:FullBC}(b); we refer to the patterns therein by the labels $A$-$E$ in the cleaning argument below. 

\begin{figure}[t!]
\centering
   \includegraphics[width=\columnwidth]{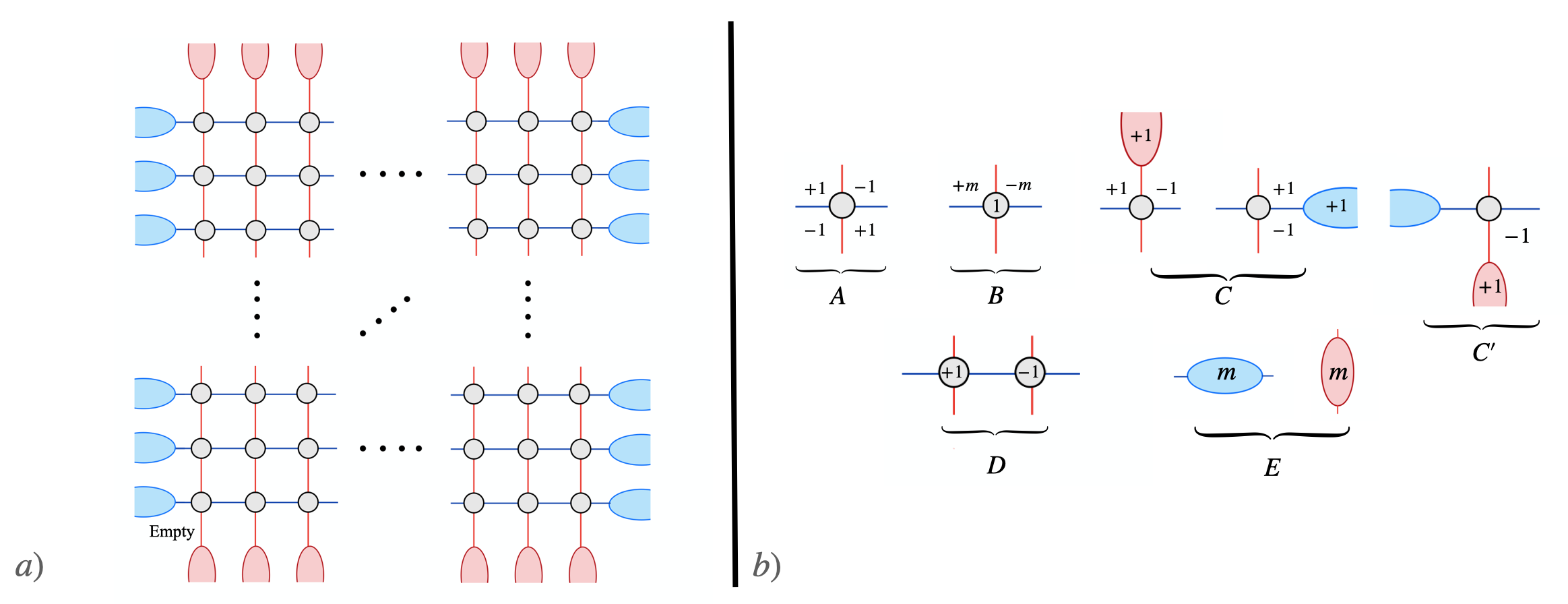}
    \caption{(a) A representation of the model with the alternative boundary conditions discussed in the main text. The grey circles correspond to the vertex terms $\theta^V_{\bm{r}}$ and the squares (both complete and partially complete) correspond to the plaquette terms $\theta^P_{\bm{r}}$. The blue and red ovals, which are shared between the left/right and top/bottom faces, respectively, correspond to the argument of the Laughlin interaction term, $2\tilde{\theta}^{1, 2}$. Note that the bottom-left corner does not have a plaquette term, as discussed in the main text. (b) Here we provide examples of some useful phase shift patterns which are employed to clean a general configuration in the ground state manifold. Note $C^\prime$ in particular, which is a special case of pattern $C$ in the bottom left-corner where the plaquette term is absent. This will be key for cleaning the entire configuration of plaquettes.}
    \label{fig:FullBC}
   \end{figure}
  
First we address the configuration of the vertex terms which are represented in Fig \ref{fig:FullBC}(a) by the grey circles. We claim that any configuration of vertex terms is trivial and can by cleaned so that $\theta^V_{\bm{r}}=0$ at each $\bm{r}$ while the other terms remain unchanged. First note that, by repeated application of $D$, all vertex terms except one can be set to zero. Suppose the remaining nonzero vertex term is in the top edge and has value $\theta^V=q$. Then, by combining $B$ and $C$, we can set $\theta^V=0$ at the expense of shifting the red oval directly above, corresponding to $\tilde{\theta}^{1}$, by $mq$. Finally, using $E$, this factor of $mq$ can be cleaned. 
  
Next we consider the plaquette terms $\theta^P_{\bm{r}}$. First we assign some value in $\mathbb{Z}$ to each square in \ref{fig:FullBC}(a) except for the bottom-left corner, which is assumed to be free of a truncated plaquette term as discussed in the main text. Observe that by using pattern $A$ one can set all plaquettes to 0 except those in an ``L"-shaped region on the perimeter. Suppose all nonzero plaquettes are confined to the top and right edge. By applying $C$ from left to right and then top to bottom, the values of the plaquettes can be shifted onto the red or blue ovals. This procedure leaves one remaining nonzero plaquette in the bottom-right corner. For this term, we can apply $C$ to shift its value to the left until it reaches the bottom-left corner, where it can be removed using $C^\prime$. 

At this point all plaquettes and vertices have been set to zero, leaving only the integers corresponding to interaction terms represented by the red and blue ovals in Fig.~\ref{fig:FullBC}. Using $E$ one can now clean the red and blue ovals modulo $m$. This analysis yields 
\begin{equation}
\label{eq:GSD}
\text{GSD}=m^{L_x+L_y}
\end{equation}
where $L_x\times L_y$ is the size of the square lattice $\Lambda$ of vertices in the array. Another method for computing topological ground-state degeneracies for coupled-wire models was introduced in \cite{Ganeshan16}. Applying it here produces the same answer.


\subsection{Periodic boundary conditions}
\label{sec:Periodic boundary conditions}

We now briefly comment on the coupled-wire model when the natural periodic boundary conditions (PBC) are imposed in all directions, so that the coupled-wire array has the topology of a three-torus. In this case, the Gauss law defined in Eq.~\eqref{eq:Gauss} in the main text can be applied with $M=\Lambda$, leading to the global constraint
\begin{align}
\label{eq:Relation}
    \sum_{\br\in\Lambda}\theta^P_{\br}=0.
\end{align}
Now we can apply the Haldane counting argument summarized in the main text. The number of chiral gapless modes in the array is $4L_xL_y$, and with PBC there are $L_x L_y$ vertex terms and $L_xL_y$ plaquette terms. Since each interaction term gaps a pair of chiral modes, the number of interaction terms at first appears sufficient to fully gap all chiral modes. However, the constraint \eqref{eq:Relation} reduces by one the number of linearly independent interaction terms in the Hamiltonian. We thus arrive at the conclusion that the model retains one pair of gapless chiral modes with opposite chirality when PBC are imposed in all directions. We remark that that the model with open boundary conditions (OBC) also has chiral gapless modes, but, as shown in Appendix~\ref{sec:Details on surface theory}, these modes are associated entirely with the surface. We expect that local bulk properties of the coupled-wire array---e.g., the energy gap for \textit{local} excitations---cannot depend on boundary conditions. We therefore expect that the remaining chiral gapless modes cannot be excited by any local operator.
We attribute these modes to an infinite ground-state degeneracy that occurs when the continuum limit of a closely related fracton lattice model (see Appendix~\ref{sec: Appendix H}) is taken in the wire direction. 
This subtlety, which is also present in Ref.~\cite{Iadecola16} but was overlooked there, will be investigated in future work.

Under the plausible assumption of an energy gap to all topological excitations created by local operators, the topological ground-state degeneracy with PBC can be computed using the techniques of Appendix~\ref{sec:Topological groud-state degeneracy}. This calculation finds that the ground-state manifold has dimension 
\begin{align}
\text{GSD}=m^{L_x+L_y-2}\ \text{gcd}(L_x,L_y).
\end{align}
This differs from Eq.~\eqref{eq:GSD} by the factor $\text{gcd}(L_x,L_y)/m^2$. The numerator of this factor comes from the fact that the final vertex in the vertex-cleaning procedure described in Appendix~\ref{sec:Topological groud-state degeneracy} can no longer be eliminated in the case of PBC, and the minimal shift of this remaining vertex is $2\pi\ \text{gcd}(L_x,L_y)$. The denominator of this factor comes from relations among the plaquette terms that only arise for PBC. One factor of $m$ comes from the fact that cleaning all plaquettes into an ``L" shape as in Appendix~\ref{sec:Topological groud-state degeneracy} yields only $L_x+L_y-1$ unique plaquette terms for PBC. The remaining factor of $m$ comes from the global constraint~\eqref{eq:Relation}, which reduces by one the number of independent plaquette terms.


\section{Generalization to non-Abelian coupled-wire models}
\label{sec: Appendix G}

\setcounter{equation}{0}
\renewcommand{\theequation}{B\arabic{equation}}

We now propose a direct generalization of the coupled-wire construction depicted in Fig.~\ref{fig:FQHModel} that yields models with non-Abelian excitations.  The idea is to promote each Luttinger-liquid wire in Fig.~\ref{fig:FQHModel} to a rational conformal field theory (CFT), and to couple these CFTs in such a way that chiral non-Abelian topological phases (rather than Abelian Laughlin $\nu=1/m$ phases) reside on \XZ and \YZ planes of the square lattice.  We then couple the planes by adding strong local interactions at the vertices where they intersect; similar to the Laughlin construction, these interactions condense p-strings consisting of fractionalized excitations from different planes.  This condensation process generates plaquette terms that give rise to subdimensional non-Abelian excitations. For a high-level analysis of related models, we refer the reader to Appendix~\ref{sec: Appendix H}.

A relatively simple and very interesting class of examples uses $SU(2)_k$ CFTs as building blocks.  These CFTs can be realized using spin chains at criticality or fermionic wires with $2k$ fermion species.  They can be coupled using current-current interactions within the \XZ and \YZ planes of the square lattice to yield chiral $SU(2)_k$ topological phases in each plane.  Specifically, we define the $SU(2)_k$ current operators using the decomposition $SU(2)_k=\mathbb{Z}_k\times U(1)_k$ as~\cite{Iadecola19}
\begin{align}
\label{eq:Currents}
    J^{q,+}_{\eta,\bm r}&=\sqrt{k}\, \psi^{q}_{\eta,\bm r}\, e^{+i\tilde\phi^q_{\eta,\bm r}/k}\\
    J^{q,-}_{\eta,\bm r}&=\sqrt{k}\, \psi^{q\, \dagger}_{\eta,\bm r}\, e^{-i\tilde\phi^q_{\eta,\bm r}/k}\\
    J^{q,z}_{\eta,\bm r}&=i\frac{\sqrt{k}}{2}\, \partial_z\phi^q_{\eta,\bm r},
\end{align}
where $\bm r\equiv(x,y)$, $q=1,2$ labels whether the CFT belongs to a vertical or horizontal plane, respectively, and $\eta=L,R$ labels the chirality.  Here, $\psi^{q}_{\eta,xy}$ and $\psi^{q \,\dagger}_{\eta,xy}$ are parafermion operators that are the simple currents in the $\mathbb Z_k$ CFT, and $\phi^q_{\eta,xy}$ are chiral boson operators from the $U(1)_{k}$ CFT.  The parafermion operators in a given wire obey the exchange algebra
\begin{align}
    \psi^{q}_{L/R,\bm r}(z)\psi^{q'}_{L/R,\bm r'}(z')&=\psi^{q'}_{L/R,\bm r'}(z')\psi^{q}_{L/R,\bm r}(z)\, e^{\pm i\frac{\pi}{k}\delta_{q,q'}\delta_{\bm r,\bm r'}\, \sgn(z-z')}\\
    \psi^{q\,\dagger}_{L/R,\bm r}(z)\psi^{q'\,\dagger}_{L/R,\bm r'}(z')&=\psi^{q'\,\dagger}_{L/R,\bm r'}(z')\psi^{q\,\dagger}_{L/R,\bm r}(z)\, e^{\pm i\frac{\pi}{k}\delta_{q,q'}\delta_{\bm r,\bm r'}\, \sgn(z-z')}\\
    \psi^{q}_{L/R,\bm r}(z)\psi^{q'\,\dagger}_{L/R,\bm r'}(z')&=\psi^{q'\,\dagger}_{L/R,\bm r'}(z')\psi^{q}_{L/R,\bm r}(z)\, e^{\mp i\frac{\pi}{k}\delta_{q,q'}\delta_{\bm r,\bm r'}\, \sgn(z-z')},
\end{align}
while the chiral bosons obey the algebra
\begin{align}
\label{eq:AbelianCommutator}
    [\tilde\phi^{q}_{L/R,\bm r}(z),\tilde\phi^{q'}_{L/R,\bm r'}(z')]=\pm i\pi\,  k\, \delta_{q,q'}\delta_{\bm r,\bm r'}\, \sgn(z-z').
\end{align}
We can then couple the wires using current-current interactions of the form
\begin{align}
\label{eq:nonAbelianPlane}
    H=\lambda\sum_{x,y}\sum^{3}_{a=1}\left(U^1_{\bm r,\bm r+\yhat}+U^2_{\bm r,\bm r+\xhat}\right),
\end{align}
where
\begin{align}
\label{eq:CurrentCurrent}
  U^q_{\bm r,\bm r'}&=J^{q,+}_{L,\bm r}J^{q,-}_{R,\bm r'}+J^{q,-}_{L,\bm r}J^{q,+}_{R,\bm r'}\\
  &=k\Big[\psi^{q}_{L,\bm r}\psi^{q\, \dagger}_{R,\bm r'}\, e^{i\left(\tilde{\phi}^{q}_{L,\bm r}-\tilde{\phi}^{q}_{R,\bm r'}\right)/k}+\text{H.c.}\Big].
\end{align}
The interactions \eqref{eq:nonAbelianPlane} are marginally relevant under the renormalization group, so the coupling constant $\lambda$ flows to infinity when it has the appropriate sign.  Each plane then enters a gapped phase with $SU(2)_k$ non-Abelian topological order~\cite{Huang16,Iadecola19}.

Next we seek vertex terms that couple intersecting \XZ and \YZ planes.  The simplest local operators are products of primary operators $\Phi^{q,(\ell)}_{\bm r}$ in each CFT, from which we can construct Hamiltonian terms
\begin{align}
\label{eq:GeneralVertex}
    U^{V,\ell\ell'}_{\bm r}=\Phi^{1,(\ell)}_{\bm r}\Phi^{2,(\ell')}_{\bm r}+\text{H.c.}
\end{align}
Here, the index $\ell=0,\dots,k$ labels the $k+1$ primary fields of the $SU(2)_k$ CFT. These primary operators can be expressed in terms of operators in the $\mathbb Z_k$ and $U(1)_k$ CFTs as (suppressing the labels $q$ and $\bm r$ for compactness)~\cite{Zamolodchikov85}
\begin{align}
\label{eq:Primaries}
    \Phi^{(\ell)}=\Phi^\ell_{L}\Phi^\ell_{R}\  e^{i\frac{\ell^2}{4k}\tilde{\phi}_L}\,e^{-i\frac{\ell^2}{4k}\tilde{\phi}_R},
\end{align}
where $\Phi^\ell_{L}\Phi^\ell_{R}$ is a primary operator in the $\mathbb Z_k$ CFT. (Note that $\Phi^0_{L}\Phi^0_{R}=\Phi^k_{L}\Phi^k_{R}=\mathbbm{1}$.) The primary operators $\Phi^{(\ell)}$ are in one-to-one correspondence with anyons in the gapped bulk of the coupled-wire array, creating quasiparticle-quasihole pairs consisting of anyons with label $\ell$. 
The vertex terms \eqref{eq:GeneralVertex} thus create bound states of anyonic excitations in the intersecting layers; adding such a term to the Hamiltonian and manually imposing a large coupling $\lambda_V\gg\lambda$ leads to condensation of p-strings composed of these anyons. The natural anyon to condense in this fashion is the one with label $\ell=k$. This anyon is always Abelian and carries topological spin $e^{i\frac{\pi k}{2}}$---thus, it is a boson when $k=0\ \text{mod}\ 4$, a fermion when $k=2\ \text{mod}\ 4$, and a semion or antisemion when $k=1$ or $3\mod 4$. To condense p-strings composed of these anyons, we choose $\ell=\ell'=k$ in Eq.~\eqref{eq:GeneralVertex}.

Implementing p-string condensation using the vertex terms $U^{V,kk}_{\br}$ leads to modified couplings between wires arising from perturbation theory in $\lambda/\lambda_V$. Based on our understanding of the Abelian case, it is clear that the current-current couplings \eqref{eq:CurrentCurrent} generically excite the vertex terms \eqref{eq:GeneralVertex}---this can be seen, for example, by inspection of the Abelian components of the current operators \eqref{eq:Currents} and the primary operators \eqref{eq:Primaries}, whose commutation is governed by Eq.~\eqref{eq:AbelianCommutator}.  Thus, perturbation theory generates products of the current-current interactions \eqref{eq:CurrentCurrent}. An example of a term generated at fourth order is
\begin{align}
\label{eq:NonAbelianPlaquette}
    U^P_{\bm r}\sim J^{1,+}_{L,\bm r}J^{1,-}_{R,\bm r+\yhat}J^{2,+}_{L,\bm r+\yhat}J^{2,-}_{R,\bm r+\yhat+\xhat}J^{1,+}_{R,\bm r+\yhat+\xhat}J^{1,-}_{L,\bm r+\xhat}J^{2,+}_{R,\bm r+\xhat}J^{2,-}_{L,\bm r}+\text{H.c.},
\end{align}
whose sign structure mimics that of its Abelian counterpart, see Eq.~\eqref{eq:AbelianHint}.  To see that this indeed commutes with $U^{V,kk}_{\br}$, we first consider the Abelian sector.  The commutator of the Abelian parts of the current operators in Eq.~\eqref{eq:NonAbelianPlaquette} with the Abelian part of $U^{V,kk}_{\br}$ can be shown to vanish using~\eqref{eq:AbelianCommutator}. Next, we consider the non-Abelian sector.  In order for the non-Abelian parts of Eqs.~\eqref{eq:NonAbelianPlaquette} and $U^{V,kk}_{\br}$ to commute, we must demand that the combination of $\mathbb{Z}_k$ primary operators entering Eq.~\eqref{eq:GeneralVertex} with $\ell=\ell'=k$ has trivial monodromy with the parafermion operators entering Eq.~\eqref{eq:NonAbelianPlaquette}.  For general $\ell,\ell'$, this is achieved when the following two relationships hold:
\begin{align}
    \Delta_{\Phi^{\ell}}+\Delta_{\psi}-\Delta_{\Phi^{\ell}\times\psi}&=-(\Delta_{\Phi^{\ell'}}+\Delta_{\psi^\dagger}-\Delta_{\Phi^{\ell'}\times\psi^\dagger})\ \text{mod}\ 1 \label{eq:NonAbelianConstraint1}\\
     \Delta_{\Phi^{\ell}}+\Delta_{\psi^\dagger}-\Delta_{\Phi^{\ell}\times\psi^\dagger}&=\Delta_{\Phi^{\ell'}}+\Delta_{\psi}-\Delta_{\Phi^{\ell'}\times\psi}\ \text{mod}\ 1 \label{eq:NonAbelianConstraint2},
\end{align}
where $\Delta_\mathcal{O}$ is the chiral scaling dimension of the operator $\mathcal O$ and $\mathcal O\times\mathcal O'$ denotes the fusion product of the operators $\mathcal O$ and $\mathcal O'$.  Using the data~\cite{Zamolodchikov85}
\begin{align}
    \Delta_{\Phi^\ell}&=\frac{\ell(\ell+2)}{4(k+2)}-\frac{\ell^2}{4k}\\
    \Delta_{\psi}&=-\frac{1}{k}\\
    \Delta_{\psi^\dagger}&=-\frac{(k-1)^2}{k}\\
    \Delta_{\Phi^\ell\times\psi}&=\frac{\ell(\ell+2)}{4(k+2)}-\frac{(\ell+2)^2}{4k}\\
    \Delta_{\Phi^\ell\times\psi^\dagger}&=\frac{\ell(\ell+2)}{4(k+2)}-\frac{(\ell+2k-2)^2}{4k}
\end{align}
we see that Eq.~\eqref{eq:NonAbelianConstraint1} reduces to $\ell=\ell'\ \text{mod}\ k$, while Eq.~\eqref{eq:NonAbelianConstraint2} reduces to $\ell=-\ell'\ \text{mod}\ k$.  The only nontrivial solution satisfying both constraints is $\ell=\ell'=k$.  

We thus arrive at the interesting conclusion that condensing p-strings composed of $\ell=k$ $SU(2)_k$ anyons yields a class of models that naturally generalizes the Abelian construction depicted in Fig.~\ref{fig:FQHModel}.  

Having constructed a nontrivial class of models, we now discuss how to determine the allowed quasiparticles after condensation.  Here we can take advantage of the connection between anyon condensation in topological quantum field theories (TQFTs) and chiral algebra extensions in CFTs~\cite{bais2009}.  Namely, when a primary operator in a CFT is ``condensed" (or, algebraically speaking, added to the representation of the vacuum sector) by adding a term of the form \eqref{eq:GeneralVertex} to the Hamiltonian, the operator content of the CFT reorganizes itself in a manner reminiscent of anyon condensation in TQFT.  In particular, the new primary operators in the ``extended" CFT are in one-to-one correspondence with deconfined quasiparticles after condensation in the TQFT.  Each new primary operator corresponds to a quasiparticle species, as we saw in the Abelian case where each local nonchiral product of vertex operators makes a bound state of fractons.  Non-Abelian excitations are created by primary operators in the extended CFT that have a nontrivial fusion algebra, i.e.~if their fusion with one or more other primary operators has multiple possible channels.

As an example, consider a wire construction based on $SU(2)_4$ CFTs coupled by current-current interactions \eqref{eq:CurrentCurrent} and vertex terms \eqref{eq:GeneralVertex} with $\ell=\ell'=4$.  The effect of adding the vertex terms at strong coupling can be understood heuristically by viewing the $p=1,2$ copies of $SU(2)_4$ as two separate $SU(2)_4\times \overline{SU(2)_4}$ topological orders, whose anyons we label by $(\ell,\ell')_p$ with $\ell,\ell'=0,\dots,4$ and $p=1,2$.  In this analogy, each anyon with integer topological spin in either of the two copies, including the ``diagonal" anyons $(\ell,\ell)_p$, corresponds to a local operator in the associated CFT.  These two topological orders are then coupled by condensing the anyon $(4,4)_1(4,4)_2$.  One can verify by explicit calculation along the lines of Ref.~\cite{bais2009} that the condensed theory hosts a pair of diagonal non-Abelian anyons that can be labeled by $(1,1)_1(0,0)_2$ and $(0,0)_1(1,1)_2$.  The corresponding local operators in the CFT thus create non-Abelian quasiparticles.

Although it is interesting that we can generate non-Abelian generalizations of the coupled-wire model studied in the main text, the analysis of this model is cumbersome to carry out at the level of the underlying CFTs.  In Appendix~\ref{sec: Appendix H}, we introduce a general algebraic prescription for carrying out the planar p-string condensation procedure discussed here. Applying this construction to $SU(2)_k$ layers allows for a much more rapid analysis of quasiparticle mobility in the resulting fracton models.

\section{Further aspects 
of planar p-string condensation}
\label{sec: Appendix H}
 
\setcounter{equation}{0}
\renewcommand{\theequation}{C\arabic{equation}}

In this Appendix we present further details about examples of and connections between planar p-string condensation and existing mechanisms that generate fracton topological order.

In Sec.~\ref{sec:FurtherExamples}, we present further high-level examples of planar p-string condensation, one of which is non-Abelian. 
In Sec.~\ref{sec:Lattice models}, we introduce several spin lattice models constructed via planar p-string condensation, including a model that is foliated equivalent~\cite{Shirley2017} to the chiral fracton theory that emerges in the bulk of the coupled-wire model introduced in the main text. We go on to discuss in Sec.~\ref{sec:Gauging} how this mechanism is related to gauging planar subsystem symmetries~\cite{GaugedLayers}, and in Sec.~\ref{sec:Defect Network} how it fits into the recently developed framework of topological defect networks~\cite{Aasen2020}. 

\subsection{Further high-level examples}
\label{sec:FurtherExamples}

In this section we present a pair of examples, the first generalizing the $\mathbb{Z}_N$ layer examples to even $N$, the second realizing non-Abelian fracton sectors that emerge from the coupled-wire construction in Appendix~\ref{sec: Appendix G}. 

\subsubsection{Semion layers} 

There is a closely related family of examples to those presented in Section~\ref{sec:Planar p-String Condensation Mechanism} with $N$ even. These correspond to the semion theory (and related theories). Again the topological charges and their fusion is given by $\mathbb{Z}_N$ with $N$ an even integer in this case. 
The $S$-matrix and topological spins are
\begin{align}
S_{a,b} &= \frac{1}{\sqrt{N}} e^{i \frac{2 \pi}{N} a b} \, ,
&&
\theta_a = e^{i \frac{ \pi}{N} a^2} \, ,
\end{align}
and the quantum dimensions are all 1. However, in this case the $F$ symbols are nontrivial, namely
\begin{align}
    F^{a b c}_{[a+b+c]} = e^{i \frac{\pi}{N} a( b + c - [b +c] )} \, ,
\end{align}
where $[~]$ denotes addition modulo $N$. 

There is an obvious $\mathbb{Z}_N$ grading generated by braiding with the $1$ anyon, which is semionic (or a generalization thereof). 
The planar p-string condensation construction of a fracton model can be formally followed through exactly as above. However, in this case the string operators for the anyonic p-strings being condensed in disjoint layers do in fact have a nontrivial anomaly due to the nontrivial $F$-symbol. 
This $F$-symbol implies that the string operators making up the membrane operator that creates a p-string cannot be realized as on-site operators, and hence cannot be condensed in a consistent way, see Sec.~\ref{sec:Gauging}. 
Another way to say this is that the p-strings cannot be condensed into the vacuum of a gapped phase, as that would allow a vacuum to vacuum process involving the creation and annihilation of p-strings resulting in the vacuum state being equal to minus itself due to the nontrivial $F$-symbol, which takes values $\pm 1$.

\subsubsection{$SU(2)_k$ anyon layers}
\label{sec:SU(2)_k}

For an example that is related to the non-Abelian coupled-wire construction proposed in Appendix~\ref{sec: Appendix G}, we consider chiral topological layers supporting $SU(2)_k$ anyons. 
The topological charges of the $SU(2)_k$ anyon theory are labelled by half integers $\{0,\frac{1}{2},\dots,\frac{k}{2}\}$. Their fusion rules, quantum dimensions, $S$-matrix and topological spins are 
\begin{align}
j_1 \times j_2 &= \sum_{j=|j_1-j_2|}^{\min(j_1+j_2,k-j_1-j_2)} j \, ,
&&
d_j = \frac{\sin \frac{(2j+1)\pi}{k+2}}{\sin \frac{\pi}{k+2}} \, , \nonumber
 \\
S_{j_1,j_2} &= \sqrt{\frac{2}{k+2}} \sin \frac{(2 j_1 + 1)(2 j_2 +1)\pi}{k+2} \, ,
&&
\theta_j = e^{2\pi i \frac{j(j+1)}{k+2}} \, ,
\end{align}
respectively. 
See Ref.~\cite{Bonderson2007} for a review of the $F$ and $R$ symbols of the $SU(2)_k$ anyon theory. 

The $\frac{k}{2}$ particle is an Abelian $\mathbb{Z}_2$ anyon; it is a boson for $k=0 \mod 4$, a semion for $k=1\mod 4$, a fermion for ${k=2 \mod 4}$, and an antisemion for $k=3 \mod 4$. 
The braiding phases $S_{j,\frac{k}{2}}|S_{j,\frac{k}{2}}|^{-1}=\pm 1$ with $\frac{k}{2}$ induce a $\mathbb{Z}_2$ grading on the topological charges, organizing them into integers and half-integers $\{0,1,\dots \}_+ \oplus \{\frac{1}{2},\frac{3}{2},\dots \}_-$. 
The half-integer $-1$ sector contains a non-Abelian anyon for any $k> 1$. 

We construct a fracton model by driving $\mathbb{Z}_2$ p-string condensation of $\frac{k}{2}$ anyons within $xy$ planes of a stack of $SU(2)_k$ anyon layers along the $xz$ and $yz$ planes of the cubic lattice. 
The resulting fracton model has a hierarchy of subdimensional topologial excitations generated by: 
\begin{itemize}
    \item Abelian $\mathbb{Z}_2$ fractons that appear on the open ends of condensed p-strings. 
    \item Non-Abelian (and Abelian) $\hat{x}$ lineons from the half-integer anyons in an $xz$ layer, trapped between p-string planes. Similarly there are $\hat{y}$ lineons from the $yz$ layers. There are also non-Abelian $\hat{z}$ lineons from composites of an $\hat{x}$ and $\hat{y}$ lineon trapped between the same p-string planes. 
    \item Planons, that may be non-Abelian, coming from the integer anyons in an $xz$ or $yz$ layer or from composites of fractons or lineons that have an overall trivial braiding with the p-strings. 
\end{itemize}

We remark that for odd $k$ there is an anomaly of the string operators preventing the p-strings from condensing to form a consistent condensate, due to the nontrivial $F$-symbols of the semions or antisemions. 
For this reason we do not expect that driving such a p-string condensation in odd-$k$ $SU(2)_k$ planes can lead to a gapped phase. 
In Sec.~\ref{sec:Gauging}, this is expressed as an anomaly of the planar subsystem symmetry that applies semion or antisemion string operators to layers intersected by the plane.

For $k/2$ an odd integer, the p-strings being condensed in layers are made up of emergent fermions. 
For conventional p-string condensation throughout the whole 3D bulk this would be anomalous, due to the nontrivial topological spin of the emergent fermions. 
However, as was shown in Ref.~\cite{GaugedLayers}, a planar subsystem symmetry generated by fermion string operators is not anomalous and can be gauged. 
This is equivalent to the condensation of p-strings consisting of emergent fermions, see Sec.~\ref{sec:Honeycomb} for a further discussion. 

\subsection{Lattice models}
\label{sec:Lattice models}

We now present several lattice-model constructions using the planar p-string condensation mechanism. We highlight in particular that the lattice model discussed in Sec.~\ref{sec:Z_N lattice} is closely related to the coupled-wire model studied in the main text and to the abstract model discussed in Sec.~\ref{sec:Z_N abstract}.

\subsubsection{X-cube from $\mathbb{Z}_2$ gauge theory layers}

As a warm up we consider 2D layers of toric code, i.e. $\mathbb{Z}_2$ lattice gauge theory, stacked along the $xz$ and $yz$ planes of the cubic lattice.
We introduce couplings that induce p-loop condensation of $m$ anyons on the $xy$ planes. 
The resulting planar p-string condensed model is simply the well-known X-cube model~\cite{VijayPRB2016}. 

The Hamiltonian governing the edge qubits in each 2D layer is
\begin{align}
    H_{\text{TC}} = -\sum_v \prod_{e \ni v} Z_e - \sum_{p} \prod_{e \in p} X_e 
    \, .
\end{align}
where $e \ni v$ denotes the edges $e$ containing a vertex $v$ and $p$ is used to denote plaquettes. 
The layers are stacked along the $xz$ and $yz$ planes of the cubic lattice, leading to a single qubit per edge in each $xy$ plane and two qubits per $\hat{z}$ edge $e$, which we label $e_{xz}$ and $e_{yz}$. 
The $m$ anyon p-string creation operators are given by $Z_{e_{xz}} Z_{e_{yz}}$. 
These couplings are introduced to the decoupled layer Hamiltonian 
\begin{align}
    H_\lambda = \sum_{\ell_{xz}} H_{\text{TC}}^{\ell_{xz}} +  \sum_{\ell_{yz}} H_{\text{TC}}^{\ell_{yz}}  - \lambda \sum_{e\perp \hat{z}} Z_{e_{xz}} Z_{e_{yz}} \, ,
\end{align}
where $\ell_{xz}$ and $\ell_{yz}$ denote $xz$ and $yz$ planes, respectively. 
In the limit of infinitely strong coupling $\lambda\rightarrow\infty$ the two qubit Hilbert space on each $xy$-plane edge is projected onto a single qubit described by the operators $Z_{e_{xz}}\sim Z_{e_{yz}}\mapsto Z_e$ and $X_{e_{xz}} X_{e_{yz}}\mapsto X_e$.
The resulting strongly coupled Hamiltonian has cube terms, given by products of four plaquette terms, arising at leading order in degenerate perturbation theory (higher order terms are not independent and hence simply shift the energetics of gapped excitations). 
This is simply the X-cube model at leading order:
\begin{align}
    H_{\text{condensed}} = -\sum_v \prod_{e \ni v,e \perp \hat{x}} Z_e +  \prod_{e \ni v,e \perp \hat{y}} Z_e - \sum_{c} \prod_{e \in c} X_e \, .
\end{align}

The anyons in the toric code layers have $\mathbb{Z}_2\times \mathbb{Z}_2$ fusion generated by the $\mathbb{Z}_2$ electric charge $e$, which are created by $X$ string operators along edges of the graph,  and magnetic flux $m$, which are created by $Z$ string operators along dual edges. 
Since the $m$ particles are planar p-string condensed in this example, the $e$ particles are promoted to lineons, while $m$ becomes a planon composite of a pair of fractons, as described in the general treatment above. 

We remark that this example extends directly to planar p-loop condensing $\mathbb{Z}_N$ lattice gauge theory layers to obtain the $\mathbb{Z}_N$ X-cube model. In the next example we introduce an alternate anyonic planar p-loop condensation transition that drives $\mathbb{Z}_N$ lattice gauge theory layers to a twisted $\mathbb{Z}_N$ X-cube model that is foliated equivalent to the chiral fracton model introduced in the main text and discussed in Sec.~\ref{sec:Z_N abstract}.

\subsubsection{Anomalous string operators in $\mathbb{Z}_N$ gauge theory}
\label{sec:Z_N lattice}

For our next example we consider anyonic planar p-string condensation in layers containing $\mathbb{Z}_N$ gauge theory. 
The resulting fracton model is equivalent to the coupled-wire fracton model introduced in the main text and Sec.~\ref{sec:Z_N abstract}, up to stacking with decoupled 2D layers (also known as foliated equivalence).

To describe the model we denote the $\mathbb{Z}_N$ clock and shift matrices by $X$ and $Z$. They satisfy the relations
\begin{align}
    X^N=Z^N=\mathbb{1} \, ,
    && 
    XZ=\overline{\omega} ZX \, ,
\end{align}
where $\omega$ is a primitive $N$th root of unity. 
The 2D layer Hamiltonians act on edge qubits of a square lattice via 
\begin{align}
    H_{2D}=-\sum_v A_v - \sum_p B_p + \text{H.c.}
\end{align}
where the vertex and star terms are given by 
\begin{align}
A_v =
\begin{array}{c}
\xymatrix@!0{%
& Z^\dagger
\\
Z &  \ar@{-}[u]  \ar@{-}[d]  \ar@{-}[l]  \ar@{-}[r]  & Z^\dagger
\\
& Z
}
\end{array},
&&
B_p=
\begin{array}{c}
\xymatrix@!0{%
  & X^\dagger \ar@{-}[l]  \ar@{-}[r]  & 
\\
 X^\dagger \ar@{-}[u]  \ar@{-}[d]   & & X \ar@{-}[u]  \ar@{-}[d]
\\
  & X \ar@{-}[l]  \ar@{-}[r] & 
}
\end{array}.
\label{eq:2DTC}
\end{align}
The above terms generate $Z$ string operators on the dual lattice, and $X$ string operators on the lattice. These string operators create emergent anyons corresponding to gauge flux and charge, which we denote by $m$ and $e$ respectively, that generate $\mathbb{Z}_N\times\mathbb{Z}_N$ fusion rules. 
The anyon theory describing these particles is formally denoted by the Drinfeld center  $\mathcal{Z}(\text{Vec}_{\mathbb{Z}_N})$, which is discussed further in the next section. 
The braiding $S$-matrix of this anyon theory is 
\begin{align}
    S_{e^im^j,e^{k}m^{\ell}} = \frac{1}{N}\omega^{i\ell +j k} \, .
\end{align}

For $N$ odd this theory can be decomposed into layers of opposite chirality, $\mathcal{Z}(\text{Vec}_{\mathbb{Z}_N})\cong \mathbb{Z}_N^{(n)} \boxtimes \mathbb{Z}_N^{(-n)}$, for $n=1,2,$ or any other integer coprime to $N$, where we have used the notation of Ref.~\cite{Bonderson2007}. 
The $\boxtimes$ notation we have used refers to the operation of stacking decoupled layers. 
The generating anyon for the $\mathbb{Z}_N^{(n)}$ chiral layer is given by $e^n m$, while the generator of the antichiral $\mathbb{Z}_N^{(-n)}$ layer is $e^n m^{N-1}$. Denoting anyons in terms of the chiral-antichiral generators as $(i,j)$ the $S$-matrix is then written 
\begin{align}
    S_{(i,j),(k,\ell)} = \frac{1}{N}\omega^{2n(ik-j \ell)} = S^{(n)}_{i,k} S^{(-n)}_{j,\ell} \, ,
\end{align}
where $S^{( n)}_{i,k},S^{(-n)}_{j,\ell},$ are the $S$-matrices of the chiral and antichiral layers respectively. 

The anyon theory describing the superselection sectors of the Laughlin state at filling fraction $\nu=\frac{1}{m}$, modulo the physical fermion, is $\mathbb{Z}_N^{(2)}$ for $N=m$, see  Ref.~\cite{Bonderson2007} for example. 
Below we utilize the embedding of the $\mathbb{Z}_N^{(2)}$ anyons into the $\mathcal{Z}(\text{Vec}_{\mathbb{Z}_N})$ anyons of the $\mathbb{Z}_N$ lattice gauge theory to construct a lattice model via p-string condensation that is foliated equivalent~\cite{Shirley2017} (i.e. equivalent up to stacking decoupled $\mathbb{Z}_N^{(-2)}$ layers) to the fracton model arising from the coupled-wire construction in the main text when a lattice cutoff is introduced in the wire direction. 
We explicitly consider $n=2$, which is relevant to the Laughlin case, but a more general family of models starting from hierarchy FQH states can be obtained using $n\neq 2$.

The string operators for the chiral Abelian anyons in a given 2D layer, as viewed from above, are of the form 
\begin{align}
\dots
    \begin{array}{c}
\xymatrix@!0{%
 Z & & Z & &  Z & &  Z & &  Z &
\\
  \ar@{-}[u]  \ar@{-}[r]  & X^2 & \ar@{-}[u] \ar@{-}[l]  \ar@{-}[r]  & X^2 & \ar@{-}[u] \ar@{-}[l]  \ar@{-}[r]  & X^2 &  \ar@{-}[u] \ar@{-}[l]  \ar@{-}[r]  & X^2 &  \ar@{-}[u] \ar@{-}[l]  \ar@{-}[r]  & X^2
}
\end{array} \dots 
\end{align}
For convenience we conjugate the model by the following local unitary circuit to simplify the horizontal string operators
\begin{align}
    U = \prod_{v} H_{v+\frac{\hat{x}}{2}} CX_{v+\frac{\hat{x}}{2},v+\frac{\hat{y}}{2}}^{2} H^\dagger_{v+\frac{\hat{x}}{2}}  \, ,
    \label{eq:CZCirc}
\end{align}
where $\hat{x}$ and $\hat{y}$ denote the axes of the square lattice depicted above. 
where $H_i$ is a generalized Hadamard matrix, or $\mathbb{Z}_N$ Fourier transform, which satisfies 
\begin{align}
    H X H^\dagger = Z^\dagger \, ,
    && 
    H Z H^\dagger = X \, ,
\end{align}
while $CX_{i,j}$ is a generalized controlled-X matrix, which satisfies 
\begin{align}
    CX (XI) CX^\dagger  = XX \, ,
    && 
    CX (IX) CX^\dagger  =  IX \, ,
    &&
    CX (ZI) CX^\dagger  = ZI \, ,
    &&
    CX (IZ) CX^\dagger  = Z^\dagger Z \, , 
\end{align}
where $XI$, $IX$ denote $X_i$, $X_j$, and similarly for $Z$. 
The conjugated string operator becomes 
\begin{align}
\dots
    \begin{array}{c}
\xymatrix@!0{%
 Z & & Z & &  Z & &  Z & &  Z &
\\
  \ar@{-}[u]  \ar@{-}[r]  &  & \ar@{-}[u] \ar@{-}[l]  \ar@{-}[r]  &  & \ar@{-}[u] \ar@{-}[l]  \ar@{-}[r]  &  &  \ar@{-}[u] \ar@{-}[l]  \ar@{-}[r]  &  &  \ar@{-}[u] \ar@{-}[l]  \ar@{-}[r]  &  
}
\end{array} \dots .
\end{align}
The $B_p$ terms in the Hamiltonian are left invariant under conjugation by $U$, while the $A_v$ terms become
\begin{align}
    U A_v U^\dagger =
\begin{array}{c}
\xymatrix@!0{%
X^{-2} & & Z^\dagger X^2
\\
\ar@{-}[u]  \ar@{-}[r] & Z &  \ar@{-}[u]  \ar@{-}[d]  \ar@{-}[l]  \ar@{-}[r]  & Z^\dagger X^2
\\
& & Z &
\\
& & \ar@{-}[u] \ar@{-}[r] & X^{-2}
}
\end{array}.
\label{eq:UVertexTerm}
\end{align}
To facilitate the coupled-layer construction we modify our choice of the Hamiltonian vertex terms by multiplying the above operators with plaquette terms, which preserves the topological phase, as follows:
\begin{align}
    \widetilde A_v =
\begin{array}{c}
\xymatrix@!0{%
X^2 & \ar@{-}[l] \ar@{-}[d] 
\\
& Z^\dagger 
\\
 Z X^{-2} &  \ar@{-}[u]  \ar@{-}[d]  \ar@{-}[l]  \ar@{-}[r]  & Z^\dagger X^2
\\
& Z & 
\\
&  \ar@{-}[u] \ar@{-}[r] & X^{-2}
}
\end{array}.
\label{eq:TildeVertexTerm}
\end{align}

The planar p-string coupled-layer model is obtained by stacking the 2D Hamiltonian along $xz$ and $yz$ planes of the cubic lattice, such that there are two qudits per $\hat{z}$ edge coming from the vertical edges of the intersecting 2D layers, and driving a phase transition with a strong uniform $ZZ^\dagger$ field applied to these edges, i.e.,
\begin{align}
    H(\alpha) = \sum_{xz,\, yz\, \text{planes}} H_{2D} - \alpha \sum_{e \parallel \hat{z}} ZZ^\dagger + \text{H.c.}
\end{align}
In the planar p-string condensed limit, as $\alpha \rightarrow \infty$, the low energy subspace has one effective qudit per $\hat{z}$ edge with logical operators $ZI \sim IZ \mapsto Z, XX \mapsto X$ 
and the leading order Hamiltonian on this Hilbert space is given by 
\begin{align}
    H = -\sum_v ( \widetilde{A}_v^{xz} + \widetilde{A}_v^{yz} )
    - \sum_c B_c + \text{H.c.},
    \label{eq:TXC}
\end{align}
where $\widetilde{A}_v^{xz}$ and $\widetilde{A}_v^{yz}$ are modified star terms that now act on a common set of qubits on the $\hat{z}$ edges, and 
\begin{align}
    B_c = \prod_{e\in c} X_e^{\sigma_e}
\end{align}
is the cage term of the $\mathbb{Z}_N$ X-cube model, where $\sigma_e=\pm1$ depending upon the edge. 
Hence we refer to the Hamiltonian~\eqref{eq:TXC} as a twisted $\mathbb{Z}_N$ X-cube model, though we remark that the way this model is twisted is distinct from previously considered generalizations of X-cube~\cite{MaPRB2017,Prem2018}. 

We now discuss the ground space degeneracy of the model~\eqref{eq:TXC} with periodic and open boundary conditions and describe how these results compare with the coupled-wire model discussed in the main text and Appendices~\ref{sec:Topological groud-state degeneracy} and \ref{sec:Periodic boundary conditions}.

\textbf{Periodic boundary conditions:}  On an $L_1 \times L_2 \times L_3$ torus there are an equal number of qubits and local stabilizer generator terms in the Hamiltonian and so we can compute the ground space degeneracy by counting the number of independent relations between the generators, i.e. nontrivial products of generators equal to the identity. 
The cube terms are identical to those in the X-cube model, where it is known that the product over any dual lattice plane gives rise to a relation. Furthermore, there are two redundancies in these relations as the product over all dual $xy$ planes is identical to the product over all dual $yz$ planes, and similarly for dual $xz$ planes. 
The vertex terms in Eq.~\eqref{eq:TXC} are twisted relative to those in the X-cube model, however a similar set of relations still hold: the product of the $\widetilde{A}_v^{xz}$ terms over an $xz$ plane gives a relation, and similarly for the product of the $\widetilde{A}_v^{yz}$ terms over a $yz$ plane. 
Finally, we can take the product of $\widetilde{A}_v^{xz}(\widetilde{A}_v^{yz})^\dagger$ over an $xy$ plane, leaving a product of Pauli $X^{\pm 2}$ operators over the edges in a pair of $xy$ planes, which can be cancelled out by multiplication with cube terms $B_c$ between the planes, provided $L_x,L_y,$ are multiples of $N$. 
There is one redundancy in these relations as the product of the $\widetilde{A}_v^{xz}$ relations over all $xz$ planes, and the $(\widetilde{A}_v^{yz})^\dagger$ relations over all $yz$ planes is identical to the product of the $\widetilde{A}_v^{xz}(\widetilde{A}_v^{yz})^\dagger$ relations over all $xy$ planes.  

The counting of the relations modulo redundancies gives a total ground space degeneracy of $N^{2(L_x+L_y+L_z)-3}$, matching that of the untwisted X-cube model. 
As explained above, the lattice model in this section is foliated-equivalent to the topological phase of the coupled-wire model, up to stacking with $L_x+L_y$ decoupled $\mathbb{Z}_N^{(-2)}$ layers, which are not affected by the p-string condensation. The degeneracy of the  decoupled antichiral layers after planar p-string condensation is $N^{L_x+L_y}$, leaving a degeneracy of $N^{ L_x+L_y+2L_z-3}$ associated to the chiral fracton model. 

To match with the degeneracy of the wire model we must take the continuum limit in the $\hat{z}$ direction, which sends the number of sites in that direction to infinity, i.e., $L_z\rightarrow \infty$. Hence we see from the lattice model that there is necessarily some infinite topological degeneracy due to the continuum limit along $\hat{z}$. This infinite degeneracy is quite subtle, although a similar phenomenon can already be seen to occur for the continuum limit of decoupled topological layers stacked along $\hat{z}$.  It is topologically protected in the sense that splitting by local operators is exponentially suppressed as a function of $L_x$ and $L_y$ but it can be lifted by local operators to give a nontrivial dispersion in the $\hat{z}$ direction within the exponentially suppressed window. We leave a more in-depth study of this limit to future work. In the continuum limit along $\hat{z}$, relations that limit to a product over a continuum are lumped into the infinite degeneracy, i.e. we separate out the component of the degeneracy that becomes infinite via $N^{2 L_z-2}\rightarrow \infty$. This leaves a degeneracy of $N^{L_x+L_y-1}$ which matches that of the coupled-wire model with PBC (see Appendix~\ref{sec:Periodic boundary conditions}) for $N=m$ and $L_x,L_y$ such that $\gcd (L_x,L_y)=N$.

\textbf{Open boundaries:}  
It is simple to modify the above example to match the alternative boundary conditions used to calculate the ground space degeneracy in the main text and in Appendix~\ref{sec:Topological groud-state degeneracy}. To see this, we first note that by picking gapped open boundary conditions at $x=0,L_x,$ for an $xz$ layer, and $y=0,L_y,$ for a $yz$ layer, 
we can induce the chiral layer to fold over and become the antichiral layer. Equivalently, with electric charge-condensing rough boundaries~\cite{Bravyi1998} we have that pairs of chiral and antichiral anyons condense at the boundary. 
Combining this with periodic boundary conditions in the $\hat{z}$ direction we have a system that can be viewed as decoupled tori on $xz$ and $yz$ planes supporting chiral $\mathbb{Z}_N^{(2)}$ anyons. 
Inducing p-string condensation on the chiral layers only, as described above, drives the chiral layers to enter the phase of the fracton model described in the main text, but with gapped boundary conditions where the $x=0$ and $x=L_x$ ($y=0$ and $y=L_y$) boundaries of the fracton model are connected via a stack of 2D antichiral layers.

To construct the lattice model we again start from decoupled 2D Hamiltonians that we write as 
\begin{align}
    H_{2D}^{RBC}=-\sum_v A_v - \sum_p B_p - \sum_{p\in L} B_p^L - \sum_{p\in R} B_p^R  + \text{H.c.},
\end{align}
where the $A_v$ terms are the same as above, and the $B_p$ terms on plaquettes not touching the left or right open boundaries (as viewed from above) are also the same. The plaquette terms touching the left, or right, boundaries (viewed from above) are given by 
\begin{align}
B_{p}^L = 
\begin{array}{c}
\xymatrix@!0{%
  X^\dagger  \ar@{-}[r]  & 
\\
  & X \ar@{-}[u]  \ar@{-}[d]
\\
   X  \ar@{-}[r] & 
}
\end{array}, 
&&
B_{p}^R=
\begin{array}{c}
\xymatrix@!0{%
  & X^\dagger \ar@{-}[l] 
\\
 X^\dagger \ar@{-}[u]  \ar@{-}[d]   & 
\\
  & X \ar@{-}[l] 
}
\end{array},
\label{eq:2DGBTC}
\end{align}
respectively. 
At this rough boundary, $e$ anyons condense as single vertex terms $A_v$ can be excited by an open string of $X$ operators ending on the boundary. 
In terms of the decomposition into chiral and antichiral layers $\mathbb{Z}_N^{(2)}\boxtimes \mathbb{Z}_N^{(-2)}$ generated by $e^2m$ and $e^2m^{N-1}$, respectively, this gapped boundary corresponds to a simple fold, since $e^2m\times e^2m^{N-1} =e^4$ generates the condensate there. 

As above, we apply the local unitary circuit from Eq.~\eqref{eq:CZCirc}, restricted to the vertices not on the boundaries. 
All but the leftmost vertex terms take the same form as in Eq.~\eqref{eq:UVertexTerm}. 
After a phase preserving redefinition of the vertex terms they are all brought into the form of Eq.~\eqref{eq:TildeVertexTerm} (including the leftmost vertex terms, by way of multiplication with the $B_p^L$ terms). 
The coupled-layer model is found by driving planar p-string condensation on the $xy$ planes of a stack of $\mathbb{Z}_N^{(2)}$ layers in $xz$ and $yz$ planes with rough boundary conditions on the $x=0,L_x,$ and $y=0,L_y,$ planes 
\begin{align}
    H(\alpha) = \sum_{xz,\, yz\, \text{planes}} H_{2D}^{RBC} - \alpha \sum_{e \parallel \hat{z}} ZZ^\dagger + \text{H.c.} 
\end{align}
Taking the planar p-string condensed limit, $\alpha \rightarrow \infty$, projects each edge into a single qudit subspace spanned by operators $ZI \sim IZ \mapsto Z, XX \mapsto X$. The Hamiltonian on this Hilbert space is given by 
\begin{align}
    H = -\sum_v ( \widetilde{A}_v^{xz} + \widetilde{A}_v^{yz} )
    - \sum_c B_c -\sum_{c\in \partial } B_c^\partial
    + \text{H.c.} 
\end{align}
where $\widetilde{A}_v^{xz}$, $\widetilde{A}_v^{yz}$ and $B_c$ are as above and 
\begin{align}
    B_c^\partial = \prod_{e\in c} X_e^{\sigma_e} \, ,
\end{align}
is a partial cage term of the X-cube model in the presence of a rough gapped boundary~\cite{Bulmash2018b}, where $c$ contains eight edges for a boundary term and five edges for a corner term, while $\sigma_e=\pm1$ depending upon the edge. 
To count the ground space degeneracy in the above stabilizer Hamiltonian we first note that with the gapped open boundary conditions there are $3L_x L_y+L_x+L_y$ edge qubits, $2L_xL_y$ star terms, and $L_xL_y+L_x+L_y+1$ cube terms (including truncated edge and corner cubes) per $xy$ layer. 
There are $(L_y+1)+(L_x+1)+L_z$ constraints from products of cube terms over dual $xz$, $yz$, and $xy$ planes that give identity. However there are two global redundancies between the product of the relations over all dual $xz$ and $yz$ planes, and all $xz$ and $xy$ planes, respectively. 
Combining the above contributions yields a ground state degeneracy of 
\begin{align}
N^{L_x+L_y} =N^{ (3L_x L_y+L_x+L_y)L_z 
- (3L_xL_y+L_x+L_y+1)L_z
+(L_x+1+L_y+1+L_z)
-2}
\end{align} 
which for $N=m$ matches the result quoted in the main text and derived in Appendix~\ref{sec:Topological groud-state degeneracy}.

\subsubsection{String-net layers} 

Finally, we consider a class of examples based on inducing planar p-string condensation on decoupled layers supporting models from the general class of 2D string-net Hamiltonians. 

Any nonchiral anyon theory that admits a gapped boundary to vacuum (technically Witt trivial~\cite{Davydov2013a}) can be realized by a string-net lattice model~\cite{Levin2005}. 
The starting point is a theory $\mathcal{C}$ consisting of a finite number of string types $\{s \}$, including the vacuum 1, together with a fusion operation described by coefficients $N_{a b}^c$ that is not strictly associative, which is captured by $F$-symbols. This mathematical object is formalised by a unitary fusion category (UFC)~\cite{etingof2005fusion}.  

The string-net Hamiltonian based on $\mathcal{C}$ is defined on a honeycomb lattice (and more generally on any directed trivalent planar graph)  
\begin{align}
    H_{\text{SN}}= - \sum_v A_v - \sum_p B_p \, ,
\end{align}
where the vertex term enforces the fusion rule at every vertex of the lattice 
\begin{align}
    A_v \Bigg |
    \begin{array}{c}
    \includegraphics{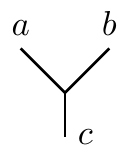}
    \end{array}
    \Bigg \rangle
    = \delta_{a b}^{c}
    \Bigg |
    \begin{array}{c}
    \includegraphics{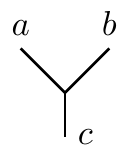}
    \end{array}
    \Bigg \rangle \, ,
\end{align}
with  
\begin{align}
    \delta_{ab}^{c} = 
    \begin{cases}
    0 & N_{ab}^{c}=0 \, ,
    \\
    1 & N_{ab}^{c}>0 \, ,
    \end{cases} 
\end{align}
and the plaquette term further decomposes as
\begin{align}
   B_p = \frac{1}{\mathcal{D}^2} \sum_{s\in \mathcal{C}} d_s B_p^s  \, ,
\end{align}
where $d_s$ is the quantum dimension of string type $s$, $\mathcal{D}^2=\sum_s d_s^2$ is the total quantum dimension of $\mathcal{C}$ and 
\begin{align}
    B_p^s \Bigg |
    \begin{array}{c}
    \includegraphics{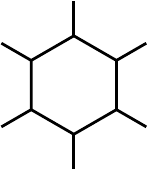}
    \end{array}
    \Bigg \rangle = 
    \Bigg |
    \begin{array}{c}
    \includegraphics{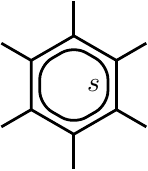}
    \end{array}
    \Bigg \rangle \, ,
\end{align}
inserts a loop of string type $s$ into the plaquette $p$, which is then fused into the lattice. 

The emergent anyons in the topological phase containing the string-net model based on the UFC $\mathcal{C}$ are described by the Drinfeld center $\mathcal{Z}(\mathcal{C})$. In the special case that $\mathcal{C}$ already describes an algebraic theory of anyons, known as a modular tensor category (MTC), the Drinfeld center is simply given by stacking the anyon theory with its time-reverse, $\mathcal{Z}(\mathcal{C})\cong  \mathcal{C} \boxtimes \overline{\mathcal{C}}$. 
Another important special case is where the string types are given by elements of a finite group $G$ and the $F$-symbols are trivial, denoted $\text{Vec}_G$, in which the emergent anyons $\mathcal{Z}(\text{Vec}_G)$ correspond to the charges, fluxes and dyons of $G$ gauge theory.

If the emergent anyon theory $\mathcal{Z}(\mathcal{C})$ contains a group $G$ of Abelian bosons that are closed under fusion, then the above lattice model can be constructed so as to have an on-site 1-form $G$ symmetry~\cite{gelaki2009centers,HeinrichPRB2016,cheng2016exactly,NewSETPaper2017}. 
This is achieved by taking an input UFC $\mathcal{C}_G$ that is $G$-graded, without loss of generality. 
This simply means that the string types decompose into nonempty $g$-sectors $\mathcal{C}_g$ containing string types we denote $\{ s_g\}$. The full UFC is recovered by a direct sum over these sectors, i.e. 
\begin{align}
    \mathcal{C}_G=\bigoplus_g \mathcal{C}_g \, ,
\end{align} and the fusion rule respects the grading, i.e. 
\begin{align}
    N_{a_g b_h}^{c_k} = \delta_{gh\bar{k}} N_{a_g b_h}^{c_k} \, ,
\end{align} where $\bar{k}=k^{-1}$. 
The plaquette terms of the string-net Hamiltonian can then be rearranged to form a projection onto the symmetric sector of a $G$ representation 
\begin{align}
   B_p = \frac{1}{|G|} \sum_g B_p^g  \, ,
\end{align}
where
\begin{align}
   B_p^g = \frac{1}{\mathcal{D}_0^2} \sum_{s \in \mathcal{C}_g } d_s B_p^s  \, ,
\end{align}
and $\mathcal{D}_0^2=\sum_{s \in \mathcal{C}_0} d_s^2$ is the total quantum dimension of the trivial sector. 

To describe the 1-form symmetry we first fix a decomposition of the Abelian group
\begin{align}
    G 
    \cong \mathbb{Z}_{p_1^{n_1}}\times \dots\times \mathbb{Z}_{p_k^{n_k}}
    \, ,
\end{align}
for primes $p_i$, and their powers $n_i \in \mathbb{N}$. 
Using this decomposition we can express an arbitrary group element and its inverse as
\begin{align}
    g = (g_1,\dots,g_k) \, , && \bar {g} = (-g_1,\dots,-g_k) \, .
\end{align}
where $g_i=0,\dots,p_i^{n_i}-1$ and group composition is given by addition in $\mathbb{Z}_{p_i^{n_i}}$. 
We can now define a generalized clock matrix on the $G$-graded vector space of string types. 
Denoting a basis element from sector $g$ as $\ket{a_g}$, the generalized clock matrix acts via  
\begin{align}
    \widetilde{Z}^g \ket{a_h}
    = \prod_{i=1}^{k} \omega_i^{g_i h_i} \ket{a_h}
    \, ,
\end{align}
where $\omega_i$ is a primitive $p_i^{n_i}$-th root of unity and $g_i h_i$ denotes multiplication in $\mathbb{Z}_{p_i^{n_i}}$. 
This clock operator has commutation relation 
\begin{align}
    \widetilde{Z}^g_e B_p^h = 
    \prod_{i=1}^k \omega_i^{ \sigma^p_e g_i h_i} B_p^h \widetilde{Z}_e^g
    \, ,
\end{align}
with the plaquette representation of $G$, where $e\in \partial p$ and $\sigma^p_e =1$ if the orientation of $e$ matches $p$ and $-1$ otherwise.

The 1-form symmetry is then generated by string operators
\begin{align}
    \widetilde{Z}^g_\gamma : = \prod_{e \cap \gamma} (\widetilde{Z}_e^{g})^{ \sigma^\gamma_e}
    \, ,
\end{align}
where $\gamma$ denotes a closed curve in the dual lattice, and $\sigma_e^\gamma=1$ if $\gamma$ intersects $e$ at a right handed crossing, and $-1$ otherwise. 
When applied to an open curve $\gamma$, running from plaquette $\gamma_-$ to $\gamma_+$, the string operator $\widetilde{Z}^g_\gamma$ creates a $\bar{g}$ boson at $\gamma_-$ and a $g$ boson at $\gamma_+$.

We now utilize this 1-form symmetry, and the Abelian bosons created by open string operators, to construct a lattice model by layering graded string-nets along the $xz$ and $yz$ planes of a cubic lattice and inducing p-string condensation of these bosons in $xy$ planes. 
We consider layers of graded string-net models on the square lattice, where each vertex is resolved into a pair of trivalent vertices, making it equivalent to the honeycomb lattice. 
The decoupled layer model is then described by the Hamiltonian 
\begin{align}
    H_{\text{Decoupled}}= \sum_{\ell_x} H_{SN}^{yz,\ell_x} + \sum_{\ell_y} H_{SN}^{xz,\ell_y} \, ,
\end{align}
where the sums are taken over $yz$ and $xz$ planes of the cubic lattice, respectively. 
This system has one qudit per $\hat{x}$ and $\hat{y}$ edge, two qudits per $\hat{z}$ edge of the cubic lattice, one coming from each layer intersecting at that edge, and several qudits per vertex, coming from the resolved vertices of the 2D square lattice string-net. 
A basis for the qudits on each $\hat{z}$ edge of the cubic lattice is given by a pair of string types, one from each intersecting layer, which we take to share a common orientation. 

To induce p-string condensation we first note that $\widetilde{Z}_{e_{xz}}^{g} \widetilde{Z}_{e_{yz}}^{\bar{g}}$ creates two pairs of $g$ bosons adjacent to $e$ that are equivalent to a small loop of the p-string excitation in the $xy$ plane labelled by $g$. 
Hence adding these operators to the Hamiltonian and taking the limit of large coupling strength induces condensation of these p-strings within $xy$ planes. 
The coupled-layer Hamiltonian is 
\begin{align}
    H(\Delta) = H_{\text{Decoupled}} - \Delta \sum_{e \parallel \hat{z}} \sum_{g \in G} \widetilde{Z}_{e_{xz}}^{g} \widetilde{Z}_{e_{yz}}^{\bar{g}}
    \, ,
\end{align} 
and in the limit of large $\Delta$ it enters the planar p-string condensed phase. 
For $\Delta \rightarrow \infty$ the on-site Hilbert space is projected into the subspace given by $\bigoplus_g \mathcal{C}_g^{xz} \boxtimes \mathcal{C}_g^{yz}$ which is spanned by pairs of strings with matching sector label $\ket{s_g,s'_g}$. At leading order in perturbation theory the p-string condensed Hamiltonian on this Hilbert space is 
\begin{align}
    H_{\text{condensed}} = - \sum_{v} A_v^{xz}+A_v^{yz} - \sum_c B_c
    \, ,
\end{align}
where $A_v^{xz}$ includes the vertex terms for the resolved vertex in the $xz$ layers, and similarly for $A_v^{yz}$. We remark that the $A_v$ terms appear unchanged as they commute with the $\widetilde{Z}_{e}^{\bar{g}}$ operators. 
The cube term $B_c$ is given by 
\begin{align}
    B_c = \frac{1}{|G|^4} \sum_g B_c^g \, ,
\end{align}
where $B_c^g = B_{p xz}^g B_{p' xz}^g B_{q yz}^g B_{q' yz}^g$ with $p,p'$ the $xz$ plaquettes in $\partial c$ and similarly for $q,q'$ and $yz$. 

The emergent excitations of the model are described by the general theory of excitations that arise by applying planar p-string condensation to the $g$ bosons in layers of $\mathcal{Z}(\mathcal{C})$ anyons, see Sec.~\ref{sec:Anyon theory description}. 
\newline

\noindent \textbf{$SU(2)_k$ string-net layers: } 
When the input UFC is given by the $SU(2)_k$ anyon theory introduced in Sec.~\ref{sec:SU(2)_k}, the string types are $\mathbb{Z}_2$-graded into integer and half-integer sectors with the generalized clock operator given by $\widetilde{Z}\ket{j}=(-1)^{2j}\ket{j}$, where $j=0,\frac{1}{2},\dots,\frac{k}{2}$. 
The plaquette terms in the $SU(2)_k$ string-net model can be written as ${B_p=\frac{1}{2}(B_p^+ + B_p^-)}$ where 
\begin{align}
    B_p^{+}= \frac{1}{\mathcal{D}_0^2} \sum_{j \text{ integer}} d_j B_p^j \, ,
    && 
     B_p^{-}= \frac{1}{\mathcal{D}_0^2} \sum_{j \text{ half-integer}} d_j B_p^j 
     \, .
\end{align}
The planar p-string condensation on layers of $SU(2)_k$ string-nets is induced by driving a phase transition with large $\widetilde{Z} \widetilde{Z}$ couplings on every $\hat{z}$ edge. This projects into a subspace where the string types on the $\hat{z}$ edges are forced to both be integer, or both be half-integer.  
The cage operators in the condensed model are given by products $B_c^\pm = B_{p xz}^\pm B_{p' xz}^\pm B_{q yz}^\pm B_{q' yz}^\pm$ using the same notation as above. 

The emergent anyon theory in each string-net layer is described by ${SU(2)_k \boxtimes \overline{SU(2)}_k}$, whose elements we denote by $(i,j)$. 
The $\mathbb{Z}_2$ 1-form symmetry utilized in the p-string condensation is generated by the $(k/2,k/2)$ boson in this anyon theory.  
The anyons that braid nontrivially with this boson, and hence are promoted to lineons in the planar p-string condensed model, are of the form $(\frac{i}{2},j)$ or $(j,\frac{i}{2})$, for $i,j,$ integers, whereas pairs of integers, or half-integers braid trivially with $(k/2,k/2)$ and hence remain planons. The  $(k/2,k/2)$ boson itself is a planon that is equivalent to a composite of fractons.

\subsection{Construction from gauging planar subsystem symmetries}
\label{sec:Gauging}

In this section we describe how planar p-string condensation can be induced by gauging planar subsystem symmetries. 

The planar p-string condensation introduced above can be realized by gauging a planar subsystem symmetry~\cite{VijayPRB2016,WilliamsonPRB2016,kubica2018ungauging,ShirleyGauging2018} along a stack of planes, as introduced in Ref.~\cite{GaugedLayers}, see Ref.~\cite{Qi2020} for a related discussion. 
The particular planar subsystem symmetries are generated by a stack of Abelian string operators, see Fig.~\ref{fig:subsystem}. 
The domain wall of such a planar symmetry corresponds to a p-string, and gauging the symmetry condenses these domain walls, see Fig.~\ref{fig:subsystemdw}. 
This provides insight into the possible anomalies of the subsystem symmetry which prevent it from being gauged, and render the corresponding p-string condensation inconsistent. 
In particular, anomalies of the 1-form symmetries, generated by the string operators involved in the planar symmetries, that arise due to braiding are no obstacle to gauging these symmetries as the string operators involved do not intersect, hence fermionic $\mathbb{Z}_2$ and arbitrary $\mathbb{Z}_N$ anyons (for $N>2$) can be planar p-string condensed. Only anomalies arising from the non on-site nature of the string operators present obstacles to gauging planar symmetries, such as for those generated by a stack of semionic string operators. 

We consider a local Hamiltonian on the cubic lattice $H=\sum_v h_v$ with planar subsystem symmetries in the $xy$ planes of the cubic lattice generated by 
\begin{align}
    \prod_{x,y} U_{x,y,z}(g) \, ,
\end{align}
where each on-site action is given by a product of string operators segments on the intersecting layers ${U(g)= V^{xz}(g) W^{yz}(g)}$. That is, $\prod_x V^{xz}_{x,y,z}(g)$ is an on-site string operator for an Abelian $G$ anyon on an $xz$ layer, and similarly $\prod_y W^{yz}_{x,y,z}(g)$ is a string operator on a $yz$ layer. 
The domain wall obtained by truncating this symmetry corresponds to a p-string excitation formed by a loop of Abelian $g$ anyon. 
We can gauge each planar symmetry following the standard procedure for gauging a global 2D symmetry~\cite{levin2012braiding,Gaugingpaper,williamson2014matrix,SET}, this is known to condense the domain walls, and hence induce planar p-string condensation. Although the symmetries described here are Abelian, the planar gauging can be applied also to non-Abelian symmetries. We describe the general gauging procedure below as it may be useful in future work. 

To gauge the symmetry we first introduce $\mathbb{C}[G]$ gauge spins onto the $\hat{x}$ and $\hat{y}$ edges of the cubic lattice, which are given an orientation. Next we introduce projectors on each vertex that implement a generalized Gauss's law within each plane 
\begin{align}
    P_v^{xy} &= \frac{1}{|G_{xy}|} \sum_{g \in G_{xy}} P_v^{xy}(g) \, ,
    \\
    P_v^{xy}(g) &=  U_v(g) \prod_{e \rightarrow v, e\perp \hat{z}} L_{e}(g)
     \prod_{e \leftarrow v, e\perp \hat{z}} R_{e}(g) \, ,
\end{align}
where $e \rightarrow v~(e\leftarrow v) $ denotes adjacent edges that are oriented towards (away from) the vertex $v$, and $L(g),R(g),$ denote the left and right regular representations respectively. 
We also introduce projectors onto zero flux through each $xy$ plaquette 
\begin{align}
    F_p^{xy} = \sum_{g_1,g_2,g_3,g_4}  \delta(g_1^{\sigma^p_{e_1}} g_2^{\sigma^p_{e_2}}  g_3^{\sigma^p_{e_3}}  g_4^{\sigma^p_{e_4}}  = 1) 
    \, 
    \pi_{e_1 \hat{z}}(g_1) \pi_{e_2 \hat{z}}(g_2) \pi_{e_3 \hat{z}}(g_3) \pi_{e_4 \hat{z}}(g_4) \, ,
\end{align}
where $\pi_{e \hat{z}}(g) = \ket{g}_{e \hat{z}}\bra{g} $ and 
the edges $e_1,e_2,e_3,e_4 \in \partial p$ are in order starting from some vertex in $\partial p$ and following the orientation induced by $p$ with  $\sigma^p_{e_i}=1$ if the orientation of $e_i$ matches and $-1$ otherwise. 

To gauge a local term in the Hamiltonian we extend it onto the gauge qubits and then project onto the subspace of gauge invariant operators as follows
\begin{align}
     \mathcal{G}[ h_v ]  = \mathcal{P}[ h_v \prod_{e\in T_{\mathcal{O}_m}} \pi_{e \hat{z}}(1) ] \, ,
\end{align}
where $T_{h_v}$ is a tree, within an $xy$ plane, that contains the vertices in $S_{h_v}$, the support of $h_v$. 
The projection onto the subspace of gauge invariant operators is 
\begin{align}
    \mathcal{P}[ \mathcal{O} ] = \sum_{\{ g_v \}} \prod_{v \in S_{\mathcal{O}}} P_v^{xy}(g_v)|_{S_{\mathcal{O}}}~\mathcal{O}~ \prod_{v \in S_{\mathcal{O}}}P_v^{xy}(g_v)|_{S_{\mathcal{O}}}^\dagger \, ,
\end{align}
where $S_{\mathcal{O}}$ is the set of sites in the support of $\mathcal{O}$. 
The gauged Hamiltonian is then 
\begin{align}
    H_{\text{gauged}} = \sum_{v} \mathcal{G}[h_v] - \varepsilon \sum_{p} F_p^{xy} - \lambda \sum_{v} P_v^{xy} \, ,
\end{align}
and the Gauss's law constraints becomes strict in the limit of $\lambda \rightarrow \infty$. 

By gauging the planar symmetries, all operators that do not commute with them are projected out. In particular the hopping operators for any anyons in the 2D layers that braid nontrivially with the string operators in each planar symmetry are projected out. This causes these anyons to become stuck between a pair of plains, hence becoming lineons. Any anyons that braid trivially with the planar symmetry string operators remain planons. The gauge charges are equivalent to pairs of anyons that are created by a string operator along $\hat{z}$ that violates the planar symmetry, and hence are planons given by a composite of lineons. 
The gauge fluxes are given by gauged twist defects~\cite{SET,NewSETPaper2017,GaugedLayers}, obtained by terminating a domain wall of the symmetry, and hence correspond to fractons. A pair of such fractons separated by a unit along $\hat{x}$ or $\hat{y}$ is equivalent to one of the $g$ anyons that is being p-string condensed, and hence is a planon if the anyon is bosonic or fermionic, and a lineon if the anyon has a nontrivial self braiding phase.

\begin{figure}[t!]
    \centering
    \subfloat[\label{fig:subsystem}]{\includegraphics[width=0.3\columnwidth]{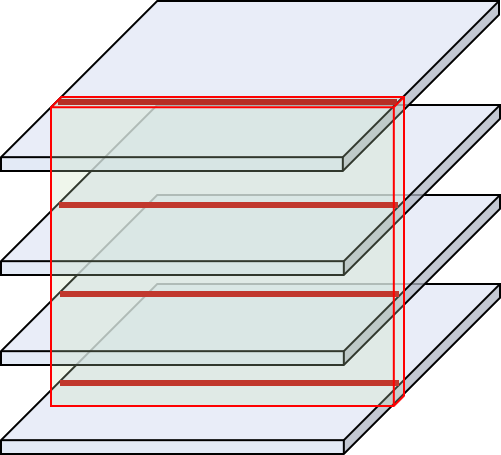}}
    \hspace{.1cm}
    \subfloat[\label{fig:subsystemdw}]{\includegraphics[width=0.3\columnwidth]{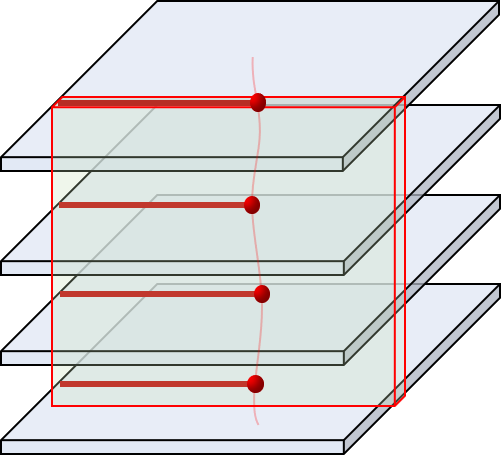}}
    \caption{
    (a) A subsystem symmetry on an $xy$ plane (green), consisting of a product of Abelian anyon string operators (red) where the plane intersects topological layers in the $xz$ planes. 
    \\
    (b) The domain wall created by applying a partial subsystem symmetry is a p-string excitation. Gauging the subsystem symmetry condenses these domain walls within the plane, hence inducing planar p-string condensation. 
    \\
    The connection between subsystem symmetries and p-string excitations depicted in (a), (b), holds similarly when topological layers in $yz$ planes are also included, the simpler case has been depicted for clarity of presentation.
}
\end{figure}

\subsubsection{Honeycomb model example}
\label{sec:Honeycomb}

We now present an example lattice construction based on fermionic planar p-string condensation induced by gauging planar subsystem symmetries on layers of Kitaev's honeycomb model~\cite{Kitaev06a}. 

We consider layers supporting chiral Ising anyons, realized by Kitaev's honeycomb model~\cite{Kitaev06a}, and apply planar p-string condensation to the $\mathbb{Z}_2$ fermions in each layer. This realizes a model that is closely related to the $SU(2)_2$ example from Sec.~\ref{sec:SU(2)_k}, since the anyons only differ by the Frobenius-Schur indicator for the non-Abelian particle~\cite{Rowell2009}, which is $+1$ for Ising and $-1$ for $SU(2)_2$. Hence the resulting fracton models differ only in the Frobenius-Schur indicators of the non-Abelian lineons. 

For the lattice model we consider Kitaev's honeycomb model~\cite{Kitaev06a} in the chiral Ising anyon phase with a perturbation that respects the fermionic 1-form symmetry and opens an energy gap 
\begin{align}
   H(J,\Delta) = - J \sum_{\langle i j \rangle}K_{i j} - \Delta \sum_{\langle ij \rangle\langle ik \rangle}  K_{ij} K_{ik} 
    \nonumber - \Delta \sum_{\langle ij \rangle\langle ik \rangle\langle i\ell \rangle} K_{ij}K_{ik}K_{i \ell} \, ,
\end{align}
where $i,j,k,l,$ denote distinct points and $\langle i j \rangle$ denote edges in the honeycomb lattice. The chirality of the gapped Ising anyon phase is given by $\nu= \sgn \Delta$. The edge operators $K_{ij} $ depend on the orientation of $\langle i j \rangle$ which we denote by $\alpha=x,y,z,$ i.e. 
\begin{align}
    K_{i j} = \sigma_i^{\alpha} \sigma_j^{\alpha} \, ,
\end{align}
where $\sigma^x=X,\sigma^y=Y,\sigma^z=Z.$ 
Abusing notation to only keep track of the edge orientation and after coarse graining to a square lattice with two qubits per site we have 
\begin{align}
K_x=
\begin{array}{c}
\xymatrix@!0{%
IX \ar@{-}[r]  & XI
}
\end{array},
&&
K_{y}= YY \, ,
&&
K_{z}=
\begin{array}{c}
\xymatrix@!0{%
IZ  \ar@{-}[d] 
\\
ZI 
}
\end{array}.
\end{align}
The fermionic 1-form symmetry restricted to the $\hat{x}$-axis is given by 
\begin{align}
    \prod_{i} (ZZ)_{ij}\, ,
\end{align}
truncating this symmetry operator to a finite line creates emergent fermion excitations at the end points. 

We consider a stack of perturbed honeycomb layers (that have been coarse grained to the square lattice) along the $xz$ and $yz$ planes of a cubic lattice such that the $\hat{z}$ axes of the layers align 
\begin{align}
    \sum_{\ell_x} H_{\ell_x}^{yz}(J,\Delta) + \sum_{\ell_y} H_{\ell_y}^{xz}(J,\Delta) \, ,
\end{align}
where the $ H_{\ell_x}^{yz}$ indicates the honeycomb Hamiltonian in a $yz$ plane at $x=\ell_x$. 
This model obeys a large symmetry group given by the product of the fermionic 1-form symmetries within each layer, this contains a 3D 1-form symmetry given by taking products of the fermionic string operators over codimension-1 surfaces~\cite{GaugedLayers}. 
Within the 1-form symmetry group is a subgroup of planar subsystem symmetries along the $xy$ planes of the cubic lattice, generated by products of the fermionic string operators over such a plane 
\begin{align}
    \prod_{i,j} (ZZ)_{ijk}^{yz} (ZZ)_{ijk}^{xz} \, ,
\end{align}
where $yz,xz,$ denote the plane from which the qubits at coordinate $ijk$ originate.  
The domain walls of these planar symmetries are p-strings formed by the fermion excitations in each layer that the p-string intersects. 

To induce planar p-string condensation we gauge the subsystem symmetries defined above. 
This introduces an additional qubit to each $\hat{x}$ and $\hat{y}$ link of the cubic lattice, which we index with half integer coordinates. 
The gauged Hamiltonian is then given by 
\begin{align}
    \sum_{\ell_x} \widetilde{H}_{\ell_x}^{yz}(J,\Delta) + \sum_{\ell_y} \widetilde{H}_{\ell_y}^{xz}(J,\Delta) - \varepsilon \sum_{ijk} F_{ijk} - \lambda \sum_{ijk} G_{ijk} \, ,
\end{align}
where 
\begin{align}
    F_{ijk} = X_{(i+\frac{1}{2})jk} X_{(i+\frac{1}{2})(j+1)k} X_{i(j+\frac{1}{2})k} X_{(i+1)(j+\frac{1}{2})k} \, ,
\end{align}
energetically penalizes nonflat $\mathbb{Z}_2$-gauge connections,  
\begin{align}
    G_{ijk} = (ZZ)_{ijk}^{yz} (ZZ)_{ijk}^{xz} Z_{(i+\frac{1}{2})jk} Z_{(i-\frac{1}{2})jk} Z_{i(j+\frac{1}{2})k} Z_{i(j-\frac{1}{2})k} \, ,
\end{align}
energetically enforces the planar Gauss's law, which becomes strict in the ${\lambda \rightarrow \infty}$ limit, 
and the gauged Hamiltonians within each layer are defined by 
\begin{align}
   \widetilde{H}(J,\Delta) = - J \sum_{\langle i j \rangle} \widetilde{K}_{i j} - \Delta \sum_{\langle ij \rangle\langle ik \rangle}  \widetilde{K}_{ij} \widetilde{K}_{ik} 
    \nonumber - \Delta \sum_{\langle ij \rangle\langle ik \rangle\langle i\ell \rangle} \widetilde{K}_{ij}\widetilde{K}_{ik}\widetilde{K}_{i \ell} \, ,
\end{align}
with minimally coupled local terms 
\begin{align}
\widetilde{K}_x=
\begin{array}{c}
\xymatrix@!0{%
IX \ar@{-}[r] & X  \ar@{-}[r] & XI
}
\end{array},
&&
\widetilde{K}_{y}= YY \, ,
&&
\widetilde{K}_{z}=
\begin{array}{c}
\xymatrix@!0{%
IZ  \ar@{-}[d] 
\\
ZI 
}
\end{array}.
\end{align}
This produces a fracton model whose emergent excitation theory is closely related to the $SU(2)_2$ model described in Sec.~\ref{sec:SU(2)_k} up to the Frobenius-Schur indicator of the nonAbelian lineons being $-1$. 

\subsection{Topological defect network construction} 
\label{sec:Defect Network}

\begin{figure}[t!]
    \centering
    \subfloat[\label{fig:TDNx}]{\includegraphics[width=0.3\columnwidth]{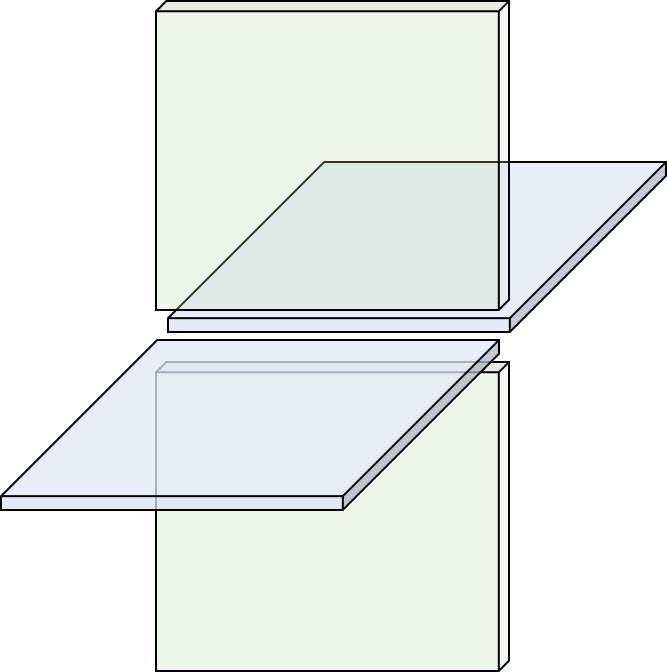}}
    \hspace{.1cm}
    \subfloat[\label{fig:TDNy}]{\includegraphics[width=0.3\columnwidth]{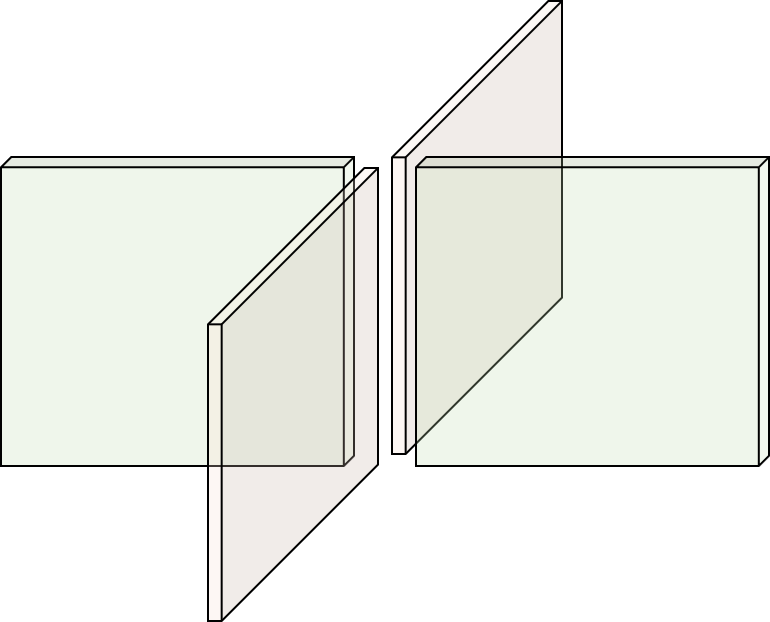}}
    \hspace{.1cm}
    \subfloat[\label{fig:TDNz}]{\includegraphics[width=0.3\columnwidth]{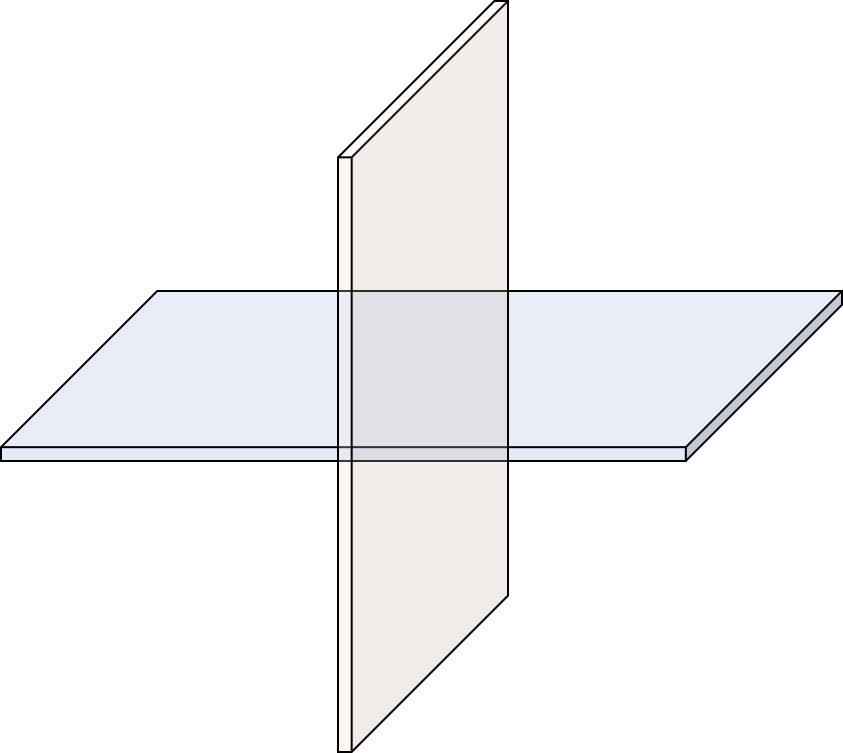}}
    \caption{
    (a) A $\hat{y}$-oriented 1-strata where $xy$-oriented 2-strata supporting $\mathbb{Z}_N$ gauge theory (green) and $yz$-oriented 2-strata supporting a general topological order that contains $\mathbb{Z}_N$ Abelian anyons, meet. 
    \\
    (b) A $\hat{y}$-oriented 1-strata where $xy$-oriented 2-strata supporting $\mathbb{Z}_N$ gauge theory (green) and $xz$-oriented 2-strata supporting a general topological order that contains $\mathbb{Z}_N$ Abelian anyons, meet. 
    \\
    (c) A $\hat{z}$-oriented 1-strata linking $xz$-oriented 2-strata, and $yz$-oriented 2-strata, by a tensor product of identity domain walls. 
}
\end{figure}

The planar p-string condensed models introduced in this paper can be described by the recently introduced topological defect network construction, providing support for the conjecture made in Ref.~\cite{Aasen2020} that all gapped fracton topological orders fit into this framework. 
We follow the procedure used in Ref.~\cite{GaugedLayers} to turn a gauged layer construction into a defect network by introducing layers of gauge theory on the subsystem symmetry planes and gapped boundaries where the planes intersect the initial stacks of topological layers

The defect network construction is given by trivial 3-strata, 2D topological orders described by the anyon theory $\mathcal{M}$ on the $xz$ and $yz$ oriented 2-strata, and $\mathcal{A}$ gauge theory (denoted $\mathcal{Z}(\text{Vec}_{\mathcal{A}})$) on the $xy$ oriented 2-strata. 
The codimension-2 defects on the $\hat{z}$ oriented 1-strata are simply given by the trivial identity domain wall between the pairs of $xz$ 2-strata, and $yz$ 2-strata, meeting there, respectively, see Fig.~\ref{fig:TDNz}. 
This is described by the following Lagrangian algebra of bosons that condenses on the defect 
\begin{align}
    \mathcal{L} = \sum_{a,b \in \mathcal{M}} (a,\bar{a},b,\bar{b}) \, ,
\end{align}
where we have used the folding trick to view the defect as a gapped boundary to vacuum of ${\mathcal{M}_{xz}\boxtimes \mathcal{M}^{\text{rev}}_{xz}\boxtimes\mathcal{M}_{yz}\boxtimes \mathcal{M}^{\text{rev}}_{yz}}$, with the layer subscripts included for guidance. 
Similarly, the defects on the $\hat{x}$ and $\hat{y}$ oriented 1-strata are equivalent to gapped boundaries to vacuum of ${\mathcal{M}\boxtimes \mathcal{M}^{\text{rev}}\boxtimes \mathcal{Z}(\text{Vec}_{\mathcal{A}})\boxtimes\mathcal{Z}(\text{Vec}_{\mathcal{A}})^{\text{rev}}}$ via the folding trick. 
The following Lagrangian algebra describes the appropriate gapped boundary 
\begin{align}
    \mathcal{L} = 
    \sum_{a \in \mathcal{M}} 
    \sum_{b \in \mathcal{Z}(\text{Vec}_{\mathcal{A}})}
    \sum_{\chi \in \widehat{\mathcal{A}}} 
    \sum_{g \in \mathcal{A}} 
    (a_\chi,\bar{g}\otimes\bar{a}_{\bar \chi},\chi \otimes b_g,\bar{b}_{\bar{g}}) \, ,
\end{align}
where we have utilized the $\mathcal{A}$-grading of $\mathcal{A}$ gauge theory by flux sectors $g$, see Figs.~\ref{fig:TDNx} \&~\ref{fig:TDNy}. 
This construction presents an immediate generalization of the construction by replacing the gauge theory layers $\mathcal{Z}(\text{Vec}_{\mathcal{A}})$ with more general $\mathcal{A}$-graded anyon theories. 

\end{widetext}

\bibliographystyle{apsrev4-1}
\bibliography{refs}

\end{document}